\documentclass[twocolumn,amsmath,amssymb,floatfix,prb,epsf,eqsecnum,showpacs]{revtex4}
\usepackage{amsmath,amssymb,natbib,bm,graphicx,url,epsfig}
\usepackage[ansinew]{inputenc}
\usepackage{bm}

\newcommand{\be}{\begin{equation}}
\newcommand{\ee}{\end{equation}}
\newcommand{\bes}{\begin{equation}\begin{split}}
\newcommand{\ees}{\end{split}\end{equation}}
\newcommand{\ba}{\begin{eqnarray}}
\newcommand{\ea}{\end{eqnarray}}

\begin{document}
\title{Interaction effects in 2D electron gas in a random magnetic field:
Implications for composite fermions and quantum critical point}

\author{T. A. Sedrakyan and M. E. Raikh}

\address{Department of Physics, University of Utah, Salt Lake
City, UT 84112}

\date{\today}

\begin{abstract}
We consider a clean two-dimensional interacting electron gas
subject to a random perpendicular magnetic field, $h({\bf r})$.
The field is nonquantizing, in the sense, that  ${\cal N}_h$-a
typical flux into the area $\lambda_{\text{\tiny F}}^2$ in the
units of the flux quantum ($\lambda_{\text{\tiny F}}$ is the de
Broglie wavelength) is small, ${\cal N}_h\ll 1$. If the spacial
scale, $\xi$, of change of $h({\bf r})$ is much larger than
$\lambda_{\text{\tiny F}}$, the electrons move along semiclassical
trajectories. We demonstrate that a weak field-induced curving of
the trajectories affects the interaction-induced electron lifetime
in a singular fashion: it gives rise to the correction to the
lifetime with a very sharp energy dependence. The correction
persists within the interval
 $\omega \sim \omega_0= E_{\text{\tiny F}}{\cal N}_h^{2/3}$
much smaller than the Fermi energy, $E_{\text{\tiny F}}$. It
emerges in the third order in the interaction strength; the
underlying physics is that a small phase volume $\sim
(\omega/E_{\text{\tiny F}})^{1/2}$ for scattering processes,
involving {\em two} electron-hole pairs, is suppressed by curving.
Even more surprising effect that we find is that {\em
disorder-averaged} interaction correction to the density of
states, $\delta\nu(\omega)$, exhibits {\em oscillatory} behavior,
periodic in $\bigl(\omega/\omega_0\bigr)^{3/2}$. In our
calculations of interaction corrections random field is
incorporated via the phases of the Green functions in the
coordinate space. We discuss the relevance of the new low-energy
scale for realizations of a smooth random  field in composite
fermions and in disordered phase of spin-fermion model of
ferromagnetic quantum criticality.
\end{abstract}


\maketitle

\section{Introduction}


Electron-electron interactions are strongly modified when
electrons move diffusively\cite{ZERO}. Resulting enhancement of
the interactions leads, in two dimensions, to a divergent
correction to the density of states\cite{ZERO,AAL80},
$\delta\nu(\omega)$. When electrons move ballistically and are
scattered by point impurities, the anomaly persists, although it
has a different underlying scenario\cite{rudin97}.

Within this scenario, individual impurities (unlike the diffusive
case\cite{ZERO,AAL80}) are responsible for the ballistic zero-bias
anomaly by virtue of the following process.
Static screening of each impurity by the Fermi sea
creates a Friedel oscillation of the electron density with a
period, $\lambda_{\text{\tiny F}}/2$, where $\lambda_{\text{\tiny
F}}$ is the de Broglie wavelength. Then the amplitude of combined
scattering from the impurity and the Friedel oscillation, which it
created, exhibits anomalous behavior\cite{rudin97} when the
scattering angle is either $0$ or $\pi$. Energy, $\omega$, of the
scattered electron, measured from the Fermi level, $E_{\text{\tiny
F}}$, defines the angular interval, $\sim
\left(\omega/E_{\text{\tiny F}}\right)^{1/2}$, within which the
scattering is enhanced. This enhancement translates into
$\delta\nu(\omega) \propto \ln \omega$ correction to the density
of states.

In a diagrammatic language, creation of the Friedel oscillation is
described by a static polarization bubble. We note in passing,
that the {\em same} polarization bubble at finite frequency,
$\omega$, is responsible for the lifetime of electron of energy
$\sim\omega$ with respect to creation of an electron-hole pair.

It is known~\cite{mishchenko02} that, in perfectly clean electron
gas, finite-range interactions do not cause {\em any} anomaly in
$\delta\nu(\omega)$. Then a natural question to ask is whether or
not the anomalous behavior of $\delta\nu(\omega)$ holds when a
weak disorder is not point-like, as in Ref.~\onlinecite{rudin97},
but, instead, smooth. Finding an answer to this question is the
main objective of the present paper. For concreteness we choose a
particular case of 2D electron gas in a smooth random magnetic
field, although our main results apply to the arbitrary smooth
disorder.

Historically, the interest to the problem of 2D electron motion in
a random static magnetic field first  emerged in connection with a
gauge field description of the correlated spin
systems\cite{ioffe,meshkov,lee}. Later this interest was
stimulated by the notion that electron density variations near the
half-filling of the lowest Landau level reduces to random magnetic
field acting on composite fermions\cite{Jain,HLR}. Another
motivation was the possibility to realize an inhomogeneous
magnetic field, acting on 2D electrons,
artificially\cite{Geim90,bending90,Geim92,
Geim94,smith94,mancoff95,gusev96,gusev00,rushforth04}. For {\em
non-interacting} electrons, this motion has been studied
theoretically in
Refs.~\onlinecite{Chalker94,Chalker94',Aronov94,chklovskii94,chklovskii95,
Falko94,Khveshchenko96,Simons99,Shelankov00,Mirlin1,Mirlin2,Mirlin3,baranger01,efetov04}.
In the present paper we
trace how the perturbation of electron motion by a smooth random
field affects the interaction corrections to the single-particle
characteristics of the electron gas.

In Refs.~\onlinecite{ioffe,meshkov,lee} the averaging over static
random field was carried out with the help of the path integral
approach originally employed for diffusively moving electrons in a
noisy environment \cite{AAK} (see also
Refs.~\onlinecite{AW1,AW2}). A crucial fact that ensures the
effectiveness of this approach is that the field is assumed to be
$\delta$-correlated. In fact, the correlation radius must be even
smaller than
$\lambda_{\text{\tiny F}}$. However, in
realizations,\cite{Jain,HLR,Geim90,bending90,Geim92,
Geim94,smith94,mancoff95,gusev96,gusev00,rushforth04} mentioned
above, the spatial scale of change of the random field in much
bigger than $\lambda_{\text{\tiny F}}$. This leads to a completely
different, semiclassical, picture of the electron motion, when
only the paths close to the classical trajectories are relevant.
In the present paper we consider only this limit. Semiclassical
character of motion suggests the way in which to perform the
averaging over disorder realizations.
Namely, the equation of motion can be first solved for {\em a
given} realization, while averaging over realizations is carried
out at the last step.  This order is opposite to
Refs.~\onlinecite{ioffe,meshkov,lee}, where averaging was carried
out in the general expression for the Green function after it was
cast in the form of a path integral.

It might seem counterintuitive that any smooth disorder could
generate a low-frequency scale for the interaction effects.
Indeed, smooth random field (including magnetic) does not produce
Friedel oscillations, which are required for the
anomaly\cite{rudin97} to develop. In a formal language, there are
no static bubbles in the diagrams for the interaction correction
to the self-energy. More precisely, in the smooth random field,
they are exponentially suppressed. We will, however, demonstrate
that the low-frequency scale emerges from {\em dynamic bubbles}
after they are modified by a smooth disorder.

The new low-$\omega$ scale shows up in the virtual processes
involving {\em more than one} electron-hole pair, i.e., two or
more bubbles. This is because the momenta of states, involved in
these processes, are strongly correlated, as was first pointed out
in Ref.~\onlinecite{suhas3}. Namely, these momenta are either
almost parallel or almost antiparallel to each other. It is this
correlation in momenta directions that is affected by the smooth
random magnetic field. By suppressing the correlation, random
field gives rise to the low-$\omega$ feature in
$\delta\nu(\omega)$. Clearly, both the height and the width of the
feature, depend on the magnitude of the random field. The above
argument makes it clear why the low-$\omega$ scale does not emerge
on the level of a single bubble, modified by the random field. The
reason is that the single bubble describes excitation of a single
pair; there is no strong restriction on the momenta directions in
this process.

 Once the mechanism of nontrivial interplay of
smooth disorder and interactions is identified, the following
questions arise: what is the shape of the anomaly in
$\delta\nu(\omega)$, and how it depends on the strength and the
correlation radius of the random field? To address these questions
we develop a systematic approach to the calculation of interaction
corrections in a smooth random field. The key element of our
approach is incorporating the action along the {\em curved}
semiclassical trajectories into the phases of the Green functions.
Our calculation reveal a surprising fact, which could not be
expected on the basis of the above qualitative consideration. It
turns out that {\em disorder-averaged} correction, $\langle
\delta\nu(\omega)\rangle$, exhibits an {\em oscillatory} behavior.
Oscillations emerge when two pairs, participating in one of the
possible processes giving rise to $\delta\nu$, are strongly
correlated with each other. As an example consider the process,
involving creation of the electron-hole pair, rescattering within
the pair, and its subsequent annihilation. In this process,
oscillations come from electron-electron scattering events that
happen at the points, located on a {\em straight line} and at
equal distances. To the best of our knowledge, this is the first
example when disorder does not suppress, but on the contrary, {\em
brings about} the oscillations.

Therefore, as we demonstrate in the present paper, anomaly in the
density of states is created by smooth spatial variation of the
magnetic field, even though this variation does not produce
Friedel oscillations. Although modification of the Friedel
oscillations from a point-like impurity by a smooth random field
is not directly related to our situation with no impurities,
this problem is still useful for gaining a qualitative
understanding. Indeed, the relevant random-field-induced length
scales, in our clean case, emerge in this problem as well. For
this reason we start with the study of suppression of the Friedel
oscillations by the random field, prior to the analysis of the
interaction corrections in the random field.

We are not aware of literature on disorder-induced smearing of the
Friedel oscillations\cite{Friedel}. However, a closely related
issue of smearing of Ruderman-Kittel-Kasuya-Yosida (RKKY)
interaction between the localized spins by the disorder, has a
long history
\cite{deGennes,chatel81,Zyuzin86,Bulaevskii86,Abrahams88,lerner,dobrosavljevich06}.
It is easy to see\cite{deGennes} that a short-range disorder
suppresses exponentially the {\em average} RKKY interaction.
However\cite{chatel81,Zyuzin86,Bulaevskii86}, the average
interaction does not represent the actual value of exchange in a
{\em given realization}. This is due to the fast oscillations of
the exchange with distance. The typical magnitude of the exchange
can be inferred from the averaging of the {\em square} of the RKKY
interaction\cite{chatel81,Zyuzin86,Bulaevskii86}; this average is
suppressed by the disorder only as a power law.

In this paper we demonstrate that the decay of the {\em averaged}
Friedel oscillations in the presence of a smooth disorder is quite
nontrivial. In particular, when the field is strong enough, the
average, in contrast to Ref.~\onlinecite{deGennes}, falls off with
distance as a power law.
We would like to note that recently the notion of averaged Friedel
oscillations became meaningful. This is because the possibility of
visualization of a single-impurity-induced oscillation had been
demonstrated experimentally
\cite{Hasegawa93,Hasegawa06,Hasegawa07,Crommie93,Crommie97,Kempen96,Hofmann97,Hoffmann,Sprunger97}.
The role of averaging can be then played by slow temporal
fluctuations of the environment. Since  experimental advances
\cite{Hasegawa93,Hasegawa06,Hasegawa07,Crommie93,Crommie97,Kempen96,Hofmann97,Hoffmann,Sprunger97}
were reported for correlated systems, recent theoretical
studies\cite{dhlee,giuliani03,Balatsky,Tsvelik07} addressed the
Friedel oscillations created by a {\em single} impurity in such
systems.




The paper is organized as follows. In Section~\ref{Regimes}
possible regimes of electron motion in a random magnetic field are
identified.
 In Section~\ref{main} we
summarize our results on Friedel oscillations and interaction
correction to the density of states for {\em weak} random field,
i.e., for the field, in which the straight-line electron
trajectories are weakly perturbed by the field.  Subsequent
Sections~\ref{Polarization}-\ref{zeroII}
 are devoted to the derivation of the results, outlined in
Section~\ref{main}. Finally, In Section~\ref{Implications} we
translate our results into predictions for experimentally
observable quantities in two prominent situations: composite
fermions in half-field Landau level and electrons interacting with
critical magnetic fluctuations near quantum critical point.
Details of some of the calculations are presented in Appendices
A-F.

\section{Regimes of electron motion in a random magnetic field}
\label{Regimes} Let ${\bf r}\equiv (x,y)$ be the coordinates of
the 2D electron. Random magnetic field along $z$-direction is
characterized by the correlator \be \label{correlator} \langle
h({\bf r})h({\bf r}^{\prime})\rangle= h_0^2\;\text{\large
K}\Bigl(\vert {\bf r}-{\bf r}^{\prime}\vert/\xi\Bigr),
\;\;\;\text{\large K}(0)\equiv 1, \ee where $h_0$ is the r.m.s
magnetic field and $\xi$ is the correlation radius. Throughout the
paper we will assume that the random field is {\em
slow-fluctuating}, in the sense, that $\xi$ is much bigger than
the de Broglie wavelength $\lambda_{\text{\tiny F}}$ , the case
opposite to the limit $\xi \rightarrow 0$ considered in
Refs.~\onlinecite{ioffe,meshkov,lee}. In terms of semiclassical
description, different regimes of motion are classified according
to the {\em classical} electron trajectory, which begins at the
origin and ends at point ${\bf r}$. One should distinguish three
different regimes, as illustrated in Fig.~1.
\begin{figure}[t]
\centerline{\includegraphics[width=90mm,angle=0,clip]{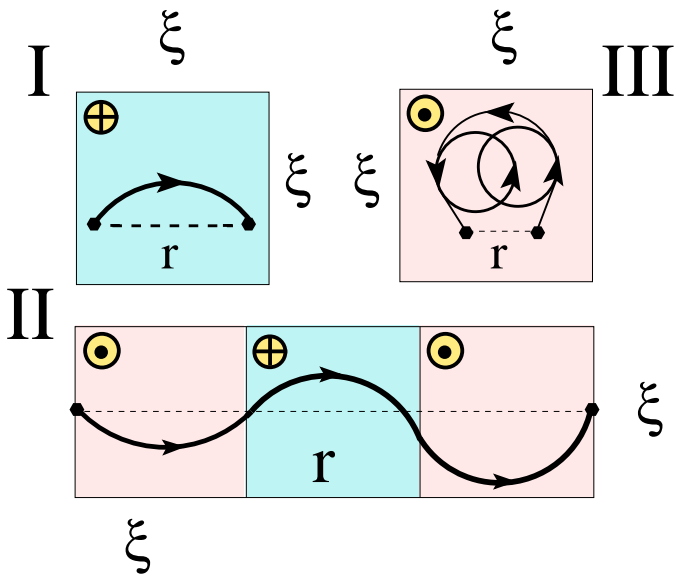}}
\caption{(Color online) Types of semiclassical trajectories
between two points separated by a distance, $r$, in a random
magnetic field: in the regime I the trajectories are of
``arc''-type; in the regime II the trajectories are of
``snake''-type; regime III corresponds to a drifting Larmour
circle.  } \label{trajectories}
\end{figure}

\begin{figure}[t]
\centerline{\includegraphics[width=85mm,angle=0,clip]{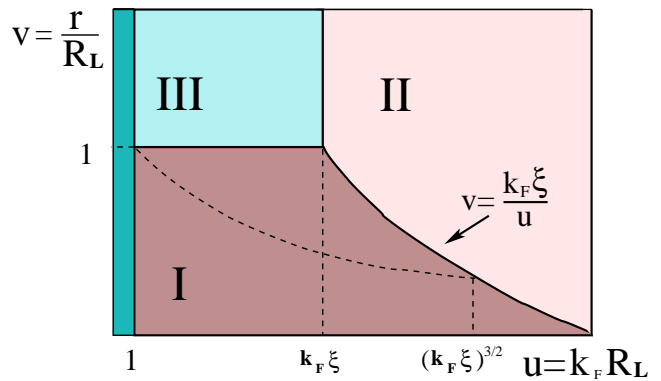}}
\caption{(Color online) Parametric regions for the regimes I, II,
and III. Dashed line, $v=u^{-1/3}$, separates slow and fast {\em
power-law} decays of the {\em averaged} Friedel oscillations
within the regime I: the oscillations fall off as $1/r^2$ to the
left from the dashed line, and as $1/r^{7/2}$ to the right from
the dashed line.} \label{PhaseDiagram}
\end{figure}

\noindent({\em i}) short-distance regime (regime I in Fig.~1). The
trajectory is of the {\em arc}-type. For this regime to realize,
two conditions must be met. Firstly, the change of magnetic field
over the distance, $r$, should be negligible, i.e., $r \ll \xi$.
Secondly, the curving of electron trajectory in the {\em locally
constant} magnetic field must be relative small. The measure of
this curving is $r/R_{\text{\tiny{L}}}$, where
$R_{\text{\tiny{L}}}=\hbar ck_{\text{\tiny{F}}}/eh_0$ is the
Larmour radius in the field, $h_0$, and
$k_{\text{\tiny{F}}}=2\pi/\lambda_{\text{\tiny{F}}}$ is the Fermi
momentum. Thus the short-distance regime corresponds to $r\ll \xi,
R_{\text{\tiny{L}}}$.

\noindent({\em ii}) "weak-field" long-distance regime (regime II
in Fig.~1). The trajectory is of the {\em snake}-type. One
condition for this regime is that magnetic field changes sign many
times within the distance, $r$, i.e., $r\gg \xi$. The other is
that within each interval of length $\sim \xi$ the curving of the
trajectory is weak, i.e., $\xi \ll R_{\text{\tiny{L}}}$.

\noindent({\em iii}) "strong-field" long-distance regime (regime
III in Fig. 1). Electron executes many full Larmour circles before
arriving to the point ${\bf r}$. The conditions for this regime
are $R_{\text{\tiny{L}}} \ll r$ and $R_{\text{\tiny{L}}} \ll \xi$.

Note, that the last two regimes correspond to the "semiclassical"
and "strong" random magnetic field regimes in the language of
Ref.~\onlinecite{Mirlin1}. In order to accommodate all three
regimes within a single diagram, it is convenient to introduce the
dimensionless parameters
\begin{eqnarray}
\label{uandv}
&&u=k_{\text{\tiny{F}}}R_{\text{\tiny{L}}}=\left(c\hbar
k_{\text{\tiny{F}}}^2 /eh_0\right)={\mathcal N}_{h}^{-1},
\nonumber\\
&&v=r/R_{\text{\tiny{L}}}\sim k_{\text{\tiny{F}}} r {\mathcal
N}_{h},
\end{eqnarray}
where ${\mathcal N}_h < 1$ is the flux of the field $h_0$ into the
area $\lambda_{\text{\tiny{F}}}^2$ (in the unites of the flux
quantum). Then the regime I is defined by the lines
$u=k_{\text{\tiny{F}}}\xi$ and $v=k_{\text{\tiny{F}}}\xi/u$, see
Fig.~2. The regime III is separated from the regime I by the line
$v=1$, and from the regime II by the line
$u=k_{\text{\tiny{F}}}\xi$. Finally, the dashed region $u<1$ in
Fig.~2 corresponds to quantizing magnetic field. The diagram
Fig.~2 is compiled for $k_{\text{\tiny{F}}}\xi \gg 1$, so it does
not reflect  white-noise regime, $k_{\text{\tiny{F}}}\xi\ll 1$, of
Refs.~\onlinecite{ioffe,meshkov,lee}.

\section{Main results}
\label{main}
\subsection{Friedel oscillations}

The simplest manifestation of the interplay of external field and
electron-electron interactions shows up in spatial response of the
electron gas to a point-like impurity, or, in other words, in
Friedel oscillations. Denote with $U_{imp}({\bf r})$ the
short-range potential of the impurity. In the presence of
interaction, $V({\bf r}-{\bf r}_1)$, the effective electrostatic
potential in a clean electron gas falls off with $r$ as
$V_{\mbox{\tiny H}}(r)\propto \sin(2k_{\mbox{\tiny F}}r)/r^2$ in a
zero field. In Ref.~\cite{we1} we had demonstrated that in a {\em
constant} magnetic field, $h=h_0$, this behavior modifies to
\begin{equation}
\label{modified1} V_{\mbox{\tiny H}}(r) =-\frac{\nu_0g
V(2k_{\mbox{\tiny F}})}{2\pi r^2} \sin\Biggl[2k_{\mbox{\tiny
F}}r-\frac{(p_0r)^3}{12} \Biggr],
\end{equation}
where the characteristic momentum, $p_0$, is defined as
\begin{equation}
\label{p0}
p_0=\frac{k_{\text{\tiny{F}}}}{(k_{\text{\tiny{F}}}R_{\text{\tiny{L}}})^{2/3}}
=\left(\frac{h_0}{k_{\mbox{\tiny F}}^{1/2}\Phi_0}\right)^{2/3},
\end{equation}
where $\Phi_0$ is the flux quantum. In Eq.~(\ref{modified1})
$\nu_0=m/\pi\hbar^2$ is the free electron density of states,
$V(2k_{\mbox{\tiny F}})$ is the Fourier component of $V({\bf r})$,
and the parameter $g$ is defined as
$g=\int U_{imp}({\bf r})\;d{\bf r}$. Eq.~(\ref{modified}) is valid
within the domain $k_{\mbox{\tiny F}}^{-1} \lesssim  r \lesssim
R_{\mbox{\tiny L}}$, so that $(p_0r)^3/12$ in the argument of sine
does not exceed the the main term, $2k_{\mbox{\tiny F}}r$. As
follows from Eq.~(\ref{p0}), the characteristic length scale,
\begin{eqnarray}
\label{rI} r_{\text{\tiny I}}=\frac{1}{p_0}=k_{\mbox{\tiny
F}}^{1/3}\left(\frac{\Phi_0}{h_0}\right)^{2/3},
\end{eqnarray}
defied by $p_0$, is intermediate between $R_{\mbox{\tiny L}}$ and
$\lambda_{\mbox{\tiny F}}$, so that
\begin{equation}
\label{modified2} R_{\mbox{\tiny L}}\gg r_{\text{\tiny I}}\gg
1/k_{\mbox{\tiny F}}.
\end{equation}

We see from Eq.~(\ref{modified1}) that only  {\em the phase} of
the Friedel oscillations is affected by the constant field, while
the magnitude  still falls off as $1/r^2$. The randomness of
$h(x,y)$ results in randomness of the field-induced phase of the
oscillations. This, in turn, translates into a faster decay of
disorder-averaged oscillations. To quantify the behavior of the
average $\bigl\langle V_{\mbox{\tiny H}}(r)\bigr\rangle$, we
rewrite it the form
\begin{equation}
\label{modified} \bigl\langle V_{\mbox{\tiny H}}(r)\bigr\rangle
=-\frac{\nu_0g V(2k_{\mbox{\tiny F}})}{2\pi r^2}\;F(r)
\sin\Biggl[2k_{\mbox{\tiny F}}r+\phi(r)\Biggr]\;,
\end{equation}
so that $F(r)$ describes the decay of the magnitude of the
disorder-averaged oscillations. For a given distance, $r$, the
character of the phase randomization is different in the regimes I
and II. In regime I, we have $\xi \gg r$, and thus the relevant
scale for the decay of $\bigl\langle V_{\mbox{\tiny
H}}(r)\bigr\rangle$ is $r_{\mbox{\tiny I}}$. In Section~\ref{FO}
we find that in this regime the magnitude, $F_{\text{\tiny I}}$,
and the phase, $\phi_{\text{\tiny I}}$, are the following
functions of the dimensionless ratio $x=r/r_{\text{\tiny I}}$
\begin{eqnarray}
\label{AVERAGED}
&&F_{\text{\tiny I}}(x)=\frac{1}{\left(1+x^6\right)^{1/4}},\\
&&\phi_{\text{\tiny I}}(x)=
-\arctan\left[\frac{\sqrt{1+x^6}-1}{x^3}\right].\nonumber
\end{eqnarray}

In regime II, with snake-like trajectories,
Fig.~\ref{trajectories}, the sign of random field changes many,
$\sim r/\xi \gg 1$, times within the distance, $r$. As
demonstrated in Section~\ref{FO}, in this regime we have
\begin{eqnarray}
\label{FriedelMagnII} F_{\text{\tiny
II}}(x)=\frac{\sqrt{2}x}{\Bigl[1+\frac{4}{9}x^4\Bigr]^{1/2}
\sqrt{\cosh^2x-\cos^2x}},\nonumber\\
\end{eqnarray}
\begin{eqnarray}
\label{FriedelPhaseII} \phi_{\text{\tiny II}}(x)=
&-&\arctan\left[1-\frac{2}{1-\cot x\tanh x}\right]\nonumber\\
&-&\arctan\left[\frac{2}{3}x^2\right],
\end{eqnarray}
where $x=r/r_{\text{\tiny II}}$, with $r_{\text{\tiny II}}$
defined as
\begin{eqnarray}
\label{rII} r_{\text{\tiny II}}=\eta~\Biggl(\frac{k_{\mbox{\tiny
F}}}{\xi}\Biggr)^{1/2}\frac{\Phi_0}{h_0}.
\end{eqnarray}
In Eq.~(\ref{rII}) the numerical factor, $\eta$, depends on the
functional form of the correlator Eq.~(\ref{correlator}) and will
be defined in Section~\ref{FO}.

\begin{figure}[t]
\centerline{\includegraphics[width=85mm,angle=0,clip]{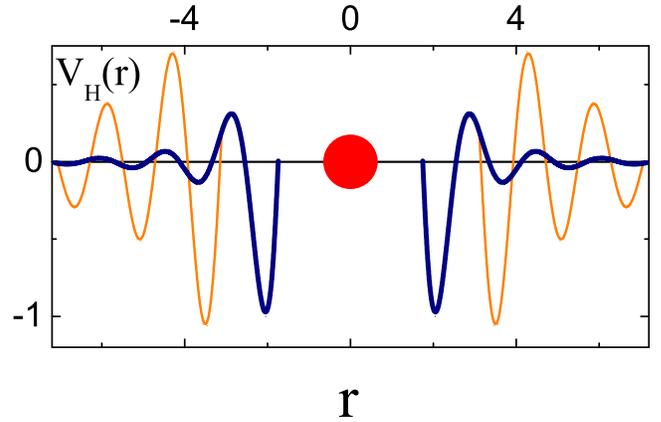}}
\caption{(Color online) Friedel Oscillations of the potential
created by an impurity located at the origin. Thick line: averaged
Friedel oscillations in regime I  is plotted from
Eqs.~(\ref{modified}), (\ref{AVERAGED}). Thin line: oscillations
in the absence of the random field [Eq.~(\ref{modified1}) with
$p_0=0$].} \label{FriedelI}
\end{figure}
In conclusion of this subsection we point out that the actual
character of the decay of Friedel oscillations with distance is
governed by the following dimensionless combination of parameters,
$h_0$, and, $\xi$, in the correlator Eq.~(\ref{correlator}) of the
random field
\begin{eqnarray}
\label{varepsilon}
\varepsilon=\frac{h_0^2\xi^3}{\Phi_0^2k_{\text{\tiny{F}}}}.
\end{eqnarray}
For $\varepsilon \gg 1$, i.e., for strong random field, the {\em
averaged} oscillations decay with $r$ according to
Eq.~(\ref{AVERAGED}) in the regime I. This is because for
$\varepsilon \gg 1$ we have $p_0\xi \gg 1$. In the opposite limit
of a weak random field,  $\varepsilon \ll 1$, we have $p_0\xi \ll
1$, so that the scale $p_0^{-1}$ is irrelevant, and also  {\em no
dephasing} takes place  within the distance, $\xi$. Thus, the
characteristic decay length, $r_{\text{\tiny{II}}} \sim
\xi/\varepsilon^{1/2}$, is much larger than $\xi$. This
automatically guarantees that
$k_{\text{\tiny{F}}}r_{\text{\tiny{II}}}\gg 1$.


\subsection{Tunnel density of states}

Two spatial scales, $r_{\text{\tiny{I}}}$ and
$r_{\text{\tiny{II}}}$, define two energy scales,
\begin{eqnarray}
\label{energies}
\omega_0&=&\frac{v_{\text{\tiny{F}}}}{r_{\text{\tiny{I}}}} \sim
E_{\text{\tiny{F}}}
\left(\frac{h_0}{\Phi_0k_{\text{\tiny{F}}}^2}\right)^{2/3}\sim E_{\text{\tiny{F}}} {\mathcal N}_h^{2/3},\\
\omega_1&=&\frac{v_{\text{\tiny{F}}}}{r_{\text{\tiny{II}}}} \sim
E_{\text{\tiny{F}}}
\left(\frac{\xi^{1/2}h_0}{k_F^{2/3}\Phi_0}\right)\sim
E_{\text{\tiny{F}}} \left(k_{\text{\tiny{F}}}\xi\right)^{1/2}
{\mathcal N}_h.\nonumber
\end{eqnarray}
As shown below, these scales manifest themselves in the anomalous
behavior of the density of states in the {\em third order} in the
electron-electron interaction parameter, $\nu_0V$. More
specifically, in the regime I, the bare density of states,
$\nu_0$, acquires a correction
$\delta\nu_{\text{\tiny{I}}}(\omega)\sim\nu_0(\nu_0 V)^3
\left(\omega_0/E_{\text{\tiny{F}}}\right)^{3/2}{\mathcal
I}\left(\omega/\omega_0\right)$, where $E_{\text{\tiny{F}}}$ is
the Fermi energy. In the regime II the correction has a similar
form $\delta\nu_{\text{\tiny{II}}}(\omega)\sim\nu_0 (\nu_0V)^3
\left(\omega_1/E_{\text{\tiny{F}}}\right)^{3/2}{\mathcal
J}\left(\omega/\omega_1\right)$. Both functions, ${\mathcal I}(z)$
and ${\mathcal J}(z)$, have characteristic magnitude and scale
$\sim 1$. Moreover, they exhibit quite a "lively" behavior. In
particular, a zero-bias anomaly,
$\delta\nu_{\text{\tiny{I}}}(\omega)$, falls off at
$\omega\gg\omega_0$ {\em with aperiodic oscillations}, i.e.,
${\mathcal I}(z)$ has a contribution
$\propto{\sin\left(2^{8/3}\sqrt{3}z\right)}z^{-3/4}\exp\left\{-2^{8/3}z\right\}$
for $z\gg 1$. The origin of the oscillations is the power-law
decay of $F_{\text{\tiny{I}}}(x)$,  given by Eq.~(\ref{AVERAGED}),
and the brunch-point, $x=e^{i\pi/6}$.

The contribution, $\delta\nu_{\text{\tiny{II}}}(\omega)$, also has
a non-monotonic behavior, despite the fact that
$F_{\text{\tiny{II}}}(x)$ falls off exponentially, as
$\exp(-r/r{\text{\tiny{II}}})$ [see Eq.~(\ref{FriedelMagnII})].

It is instructive to trace the evolution of the zero-bias anomaly
upon increasing the magnitude of the random field, $h_0$. This
evolution is governed by parameter, $\varepsilon$,
Eq.~(\ref{varepsilon}). While $\varepsilon$ remains smaller than
one, where the regime II applies, the anomaly is described by the
function ${\mathcal J}(\omega/\omega_1)$ and broadens with $h_0$
as $v_{\text{\tiny{F}}}/r_{\text{\tiny{II}}}(h_0)\propto h_0$.
Upon further increasing $h_0$, when $\varepsilon$ exceeds one, the
crossover to the regime I takes place. Zero-bias anomaly is then
described by ${\mathcal I}(\omega/\omega_0)$; it broadens with
$h_0$ as $v_{\text{\tiny{F}}}/r_{\text{\tiny{I}}}(h_0)\propto
h_0^{2/3}$, and {\em develops oscillations}. The fact that
oscillations in $\delta\nu(\omega)$ {\em emerge} upon
strengthening disorder might seem counterintuitive. This issue
will be discussed in details in Section~\ref{DOS}.

\begin{figure}[t]
\centerline{\includegraphics[width=85mm,angle=0,clip]{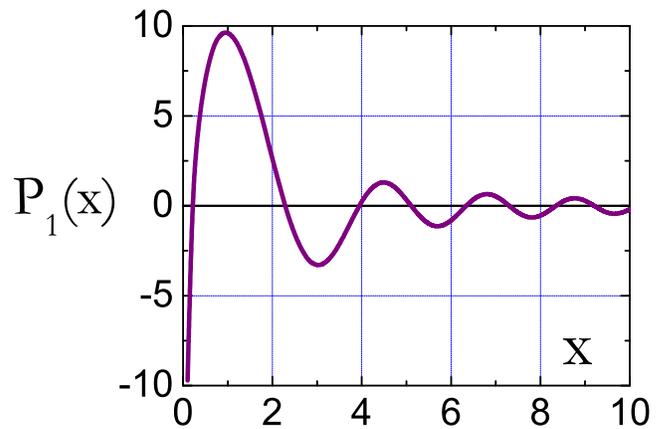}}
\caption{(Color online)
Magnetic-field-induced contribution Eq.~(\ref{last}) to the
ballistic zero-bias anomaly Eq.~(\ref{impurityNU}).
Field-dependent correction,
$\left[\delta\nu(\omega,h)-\delta\nu(\omega,0)\right]/\nu_0$, in
the units
$(\nu_0V)^2(\omega_0/E_{\text{\tiny{F}}})^{1/2}(E_{\text{\tiny{F}}}\tau)^{-1}$
is plotted versus dimensionless energy $2^{2/3}(\omega/\omega_0)$,
where $\omega_0=(2E_{\text{\tiny{F}}})^{1/3}\omega_c^{2/3}\gg
\omega_c$, and $\omega_c$ is the cyclotron frequency.}
 \label{imurity_DOS}
\end{figure}

In a zero magnetic field, an intimate relation between
impurity-induced Friedel oscillations and the zero-bias anomaly
was first established in Ref.~\onlinecite{rudin97}. Namely, it was
demonstrated that for short-range interaction
$\delta\nu(\omega)/\nu_0\sim(\nu_0V/E_{\text{\tiny{F}}}\tau)\ln\omega$,
where $1/\tau=\nu_0 \pi g^2 n_{\text{imp}}$ is the electron
scattering rate by the impurities, and $n_{\text{imp}}$ is the
impurity concentration. This anomaly is of the first order in $V$.
A non-trivial question is whether or not the modification,
Eq.~(\ref{modified1}), in a {\em constant} magnetic field results
in field dependence of the density of states in this order. In
other words, whether or not a weak magnetic field introduces a
cutoff of $\ln\omega$ at small $\omega$. The answer to this
question is negative. In Ref.~\onlinecite{we1} it was demonstrated
that sensitivity of $\delta\nu(\omega)$ to a weak magnetic field
indeed emerges, but in the {\em second} order in $\nu_0V$
(however, still in the first order in $1/\tau$). The
field-dependent correction,
$\left[\delta\nu(\omega,h)-\delta\nu(\omega,0)\right]/\nu_0$, has
a characteristic frequency scale, $\omega=\omega_0$. It is
interesting to note that, at $\omega\gg\omega_0$, this {\em
impurity-induced} correction has an {\em oscillating} character

\begin{eqnarray}
\label{impurityNU}
\frac{\delta\nu(\omega,h)-\delta\nu(\omega,0)}{\nu_0}=
\frac{(\nu_0V)^2}{E_{\text{\tiny{F}}}\tau}\left(\frac{\omega_0}{E_{\text{\tiny{F}}}}\right)^{1/2}
{\text{\Large P}}\left(\frac{\omega}{\omega_0}\right).\nonumber\\
\end{eqnarray}
The dimensionless function, ${\text{\large P}}$, has the following
large-$x$ asymptote
\begin{eqnarray}
\label{calG}{\text{\Large P}}(x)\propto\frac{1}{x^{3/4}}\cos\left[
\frac{8}{3\sqrt{3}} x^{3/2} -\frac{\pi}{4} \right].\;\;\;
\end{eqnarray}
In Fig.~\ref{imurity_DOS} we show the oscillating correction
to the density of states; the form of
the function ${\text{\large P}}\left(\omega/\omega_0\right)$
is addressed in Section~\ref{DOS}. Technically, the derivation of
Eqs.~ (\ref{impurityNU}), (\ref{calG}) is quite analogous to the
derivation of the oscillatory $\delta\nu$ in the random field in
the regime I. For this reason we will outline this derivation in
Section~\ref{DOS}.

\section{Polarization operator in a random magnetic field}
\label{Polarization}

Friedel oscillations, $V_{\mbox{\tiny H}}(r)$, created by a
point-like impurity, and the ballistic zero-bias anomaly
originating from these oscillations\cite{rudin97} are intimately
related to the Kohn anomaly in the polarization operator,
$\Pi(q)$, of a clean electron gas near $q=2k_{\text{\tiny{F}}}$.
In two dimensions, this anomaly behaves as \cite{Stern67}
$(q-2k_{\text{\tiny{F}}})^{1/2}$, which translates into $1/r^2$
decay of the Friedel oscillations and $\propto \ln \omega$
correction to the density of states. Suppression of the Friedel
oscillations, $V_{\mbox{\tiny H}}(r)$, in a random field is a
result of smearing of the Kohn anomaly in the momentum space.
However, since the momentum is not a good quantum number in the
presence of the random field, it is much more convenient to study
the field-induced suppression of $V_{\mbox{\tiny H}}(r)$ directly
in the coordinate space.

\subsection{Evaluation in the coordinate space}

Polarization operator, $\Pi_{\Omega}({\bf r},{\bf r}^{\prime})$,
is defined in a standard way as
\begin{equation}
\label{polar} \Pi({\bf r},{\bf
r}^{\prime},\Omega)=-i\int\frac{d\Omega^{\prime}}{2\pi}G_{\Omega^{\prime}}({\bf
r},{\bf r}^{\prime}) G_{\Omega-\Omega^{\prime}}({\bf
r}^{\prime},{\bf r}).
\end{equation}
Here $G_{\Omega}({\bf r},{\bf r}^{\prime})$ denotes causal Green
function, which coincides with the retarded, $G_{\Omega}^R({\bf
r},{\bf r}^{\prime})$, or advanced $G_{\Omega}^A({\bf r},{\bf
r}^{\prime})$ Green functions for $\Omega>0$ and $\Omega<0$,
respectively. At distances $\vert {\bf r}-{\bf r}^{\prime}\vert\gg
k_{\text{\tiny{F}}}^{-1}$ the polarization operator in coordinate
space represents the sum $\Pi_0(r,\omega)$ and
$\Pi_{2k_{\text{\tiny{F}}}}(r,\omega)$ of slow and rapidly
oscillating parts
\begin{eqnarray}
\label{operator0}
\Pi_0(r,\omega)=-\frac{i\pi\nu_0^2\hbar^4}{2k_{\mbox{\tiny
F}}r}\vert\omega\vert\exp\left\{\frac{i\vert\omega\vert
r}{v_{\mbox{\tiny F}}}\right\},
\end{eqnarray}

\begin{eqnarray}
\label{operator2kF}
\Pi_{2k_{\text{\tiny{F}}}}(r,\omega)=-\frac{\nu_0\hbar^3}{2r^2}
\sin\bigl(2k_{\text{\tiny{F}}}r\bigr)A\Biggl(\frac{2\pi rT}{v_{\mbox{\tiny F}}}\Biggr)\nonumber\\
\times \exp\left\{\frac{i\vert\omega\vert r}{v_{\mbox{\tiny
F}}}\right\}.
\end{eqnarray}
Subindices $0$ and $2k_{\text{\tiny{F}}}$ emphasize that these
parts come from small momenta and momenta close to
$2k_{\text{\tiny{F}}}$ in $\Pi(q)$, respectively.
Eq.~(\ref{operator0}) emerges if one of the Green functions in
Eq.~(\ref{polar}) is retarded and the other is advanced.
Eq.~(\ref{operator2kF}) corresponds to the case when the Green
functions in Eq.~(\ref{polar}) are both advanced or both retarded
\cite{Chubukov1,Chubukov2}. Derivation of Eqs.~(\ref{operator0}),
(\ref{operator2kF}) is presented in Appendix A. In
Eq.~(\ref{operator2kF}) the function,
\begin{eqnarray}
\label{Afunct} A(x)=\frac{x}{\sinh x},
\end{eqnarray}
in $\Pi_{2k_{\text{\tiny{F}}}}$ describes the temperature damping.

\subsection{Qualitative derivation for the constant field}

For a constant magnetic field, $h(x,y)\equiv h_0$, the phase,
$\phi(r)$, in the argument of Eq.~(\ref{modified}) can be inferred
from the following simple qualitative consideration.

\begin{figure}[t]
\centerline{\includegraphics[width=80mm,angle=0,clip]{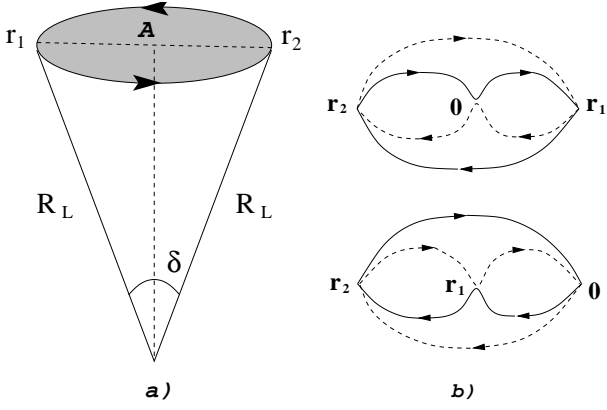}}
\caption{(a)  Origin of the net ``magnetic'' phase
Eq.~(\ref{constantfield}): two arcs, corresponding to the opposite
directions of propagation, define a finite area, ${\cal A}$.
Aharonov-Bohm flux through this area makes the net phase {\em
negative}; (b) Schematic illustration of the scattering processes
giving rise to the additional phases Eqs.~(\ref{product1}),
(\ref{product2}) in the product Eq.~(\ref{product}).}
\label{angle}
\end{figure}

Classical trajectory of an electron in a weak magnetic field is
{\em curved} due to the Larmour motion even at the spatial scales
much smaller than $R_{\mbox{\tiny L}}$. As a result of this
curving, the electron propagator, $G({\bf r}_1,{\bf r}_2)$,
between the points ${\bf r}_1$ and ${\bf r}_2$ contains,  in the
semiclassical limit, a  phase, $k_{\mbox{\tiny F}}{\cal L}$, where
${\cal L}$ is the length of the {\em arc}  of a circle with the
radius $R_{\mbox{\tiny L}}$, that connects the points ${\bf r}_1$
and ${\bf r}_2$, see Fig.~\ref{angle}a. Since the Friedel
oscillations are related to the propagation from ${\bf r}_1$ to
${\bf r}_2$ {\em and back}, it is important that two arcs,
 corresponding to the opposite directions of propagation, define a {\em finite} area,
$\cal A$, so that the product $G({\bf r}_1,{\bf r}_2)G({\bf
r}_2,{\bf r}_1)$ should be multiplied by the Aharonov-Bohm phase
factor, $\exp\left[ih_0{\cal A}/\Phi_0\right]$. Then the phase,
of this product is equal to
\begin{equation}
\label{theta}
2k_{\text{\tiny{F}}}r +\phi(r) =2k_{\mbox{\tiny F}}{\cal L}
-\frac{h_0{\cal A}({\bf r}_1,{\bf r}_2)}{\Phi_0}.
\end{equation}
Simple geometrical relations, see Fig.~\ref{angle}a, yield
\begin{eqnarray}
\label{geometrical} r=\vert {\bf r}_1-{\bf
r}_2\vert=2R_{\mbox{\tiny L}}\sin(\delta/2),\nonumber\\ {\cal
L}=R_{\mbox{\tiny L}}\delta,\;\;\;\;{\cal A}=2 R_{\mbox{\tiny
L}}^2(\delta-\sin\delta).
\end{eqnarray}
Using this relation and assuming $r \ll R_{\mbox{\tiny L}}$, we
find
\begin{eqnarray}
\label{constantfield}
\phi(r)=-\frac{h_0^2r^3}{12k_{\text{\tiny{F}}}\Phi_0^2}=-\frac{(p_0r)^3}{12}.
\end{eqnarray}
At this point, we would like to note, that the conventional
way~\cite{gorkov59} of incorporating magnetic field into the
semiclassical zero-field Green's function amounts to multiplying
it by $\exp\left[(1/\Phi_0)\int {\bf a}\cdot d{\bf l}\right]$,
where the phase factor is the integral of the vector potential,
${\bf a}$, along the {\em straight} line, connecting the points
${\bf r}_1$ and ${\bf r}_2$. Such an incorporation  neglects the
field-induced curvature of the electron trajectories, and thus
does not capture the modification Eq.~(\ref{modified}) of the
Friedel oscillations in magnetic field. Indeed, the magnetic phase
factors, introduced following Ref.~\onlinecite{gorkov59} {\em
cancel out} in the polarization operator.

With phase, $\phi(r)$, given by Eq.~(\ref{constantfield}), Friedel
oscillations in a constant magnetic field acquire the
form\cite{we1} Eq.~(\ref{modified1}). To see this, we notice that,
with accuracy of a factor, $g/2\pi$, the potential,
$V_{\text{\tiny{H}}}(r)$ coincides with
$\Pi_{2k_{\text{\tiny{F}}}}(r,0)$. Then the additional phase
Eq.~(\ref{constantfield}) transforms $\sin(2k_{\text{\tiny{F}}}r)$
into $\sin\left[2k_{\text{\tiny{F}}}r-(p_0r)^3/12\right]$, as in
Eq.~(\ref{modified1}).

In Appendix B we present a rigorous derivation of
Eq.~(\ref{modified1}) starting from {\em exact} electronic states
in a constant magnetic field, as in Ref.~\onlinecite{aleiner95}.

\subsection{Field-induced phase of the Green function:
Analytical derivation in a spatially-inhomogeneous field}

Additional semiclassical phase, $\delta \varphi_{0\rightarrow {\bf
r}}$, of the Green function due to the random magnetic field,
$h(x,y)$, is given by the following generalization of
Eq.~(\ref{constantfield})
\begin{equation}
\label{general1} \delta\varphi_{0\rightarrow {\bf r}}
=\frac{k_{\text{\tiny{F}}}}{2}\int_0^{r}dx\Biggl(\frac{dy}{dx}\Biggr)^2-
\frac{1}{\Phi_0}\int_0^{r}\!dx\; y(x)\;h(x,0),
\end{equation}
where the first term comes from the elongation of the trajectory
in
magnetic field. The second term describes the Aharonov-Bohm flux
into the area restricted by the curve $y(x)$ and the $x$-axis. In
Eq.~(\ref{general1}) we assumed that the field does not change
along the $y$-axis. This is the case when the maximal $y$ is
smaller than the correlation radius, $\xi$, of the random field.
The condition $y<\xi$ is met in the regime of the "arcs" and the
regime of the "snakes", see Figs.~\ref{trajectories},
\ref{PhaseDiagram}.

In Eq.~(\ref{general1}) we have also assumed that the magnitude of
the de Broglie wavelength of the electron does not change along
the trajectory. This can be justified from the equations of motion
\begin{eqnarray}
\label{motion}
m\frac{d^2y}{dt^2}\!\!&=&\!\!\frac{e}{c}\;h(x,0)\frac{dx}{dt}\nonumber\\
m\frac{d^2x}{dt^2}\!\!&=&\!\!-\frac{e}{c}\;h(x,0)\frac{dy}{dt}.
\end{eqnarray}
It follows from Eq.~(\ref{motion}) that the energy of electron
$\frac{m}{2}[(dx/dt)^2+(dy/dt)^2]$ is conserved even if magnetic
field changes with coordinates.

The most important step that allows to find
$\delta\varphi_{0\rightarrow {\bf r}}$ analytically, is that in
the regimes I and II in Fig.~1 we can replace $dx/dt$ by
$v_{\text{\tiny{F}}}$ and set $t=x/v_{\text{\tiny{F}}}$ in the rhs
of Eq.~(\ref{motion}). This allows to replace $d^2y/dt^2$ by
$v_{\text{\tiny{F}}}^2 d^2y/dx^2$. Then the first of the equations
yields
\begin{eqnarray}
\label{motion1}
 mv_{\text{\tiny{F}}}^2\frac{d^2y}{dx^2}=\frac{ev_{\text{\tiny{F}}}}{c}h(x,0).
\end{eqnarray}
Integrating this equation, we obtain
\begin{eqnarray}
\label{trajectory}
\frac{dy}{dx}=\frac{e}{mcv_{\text{\tiny{F}}}}\int_0^x\!dx^{\prime}h(x^{\prime},0)+C.
\end{eqnarray}
The constant, $C$, should be found from the conditions: $y(0)=0$,
and $y(r)=0$, leading to
\begin{eqnarray}
\label{C}
C=-\frac{e}{mcv_{\text{\tiny{F}}}r}\int_0^rdx^{\prime}\int_0^{x^{\prime}}dx^{\prime\prime}
h(x^{\prime\prime},0)\nonumber\\
=-\frac{e}{mcv_{\text{\tiny{F}}}r}\int_0^rdx^{\prime}\Lambda(x^{\prime}),
\end{eqnarray}
where we have introduced an auxiliary function
\begin{eqnarray}
\label{Lambda} \Lambda(x)=\int_0^x\!dx^{\prime}h(x^{\prime},0).
\end{eqnarray}
The meaning of $\Lambda(x)$ is the $y$-projection of the vector
potential. Substituting Eq.~(\ref{C}) back into
Eq.~(\ref{trajectory}), we find
\begin{eqnarray}
\label{deriv}
\frac{dy}{dx}=\frac{e}{mcv_{\text{\tiny{F}}}}\Biggl[\int_0^x\!dx^{\prime}h(x^{\prime},0)\qquad\qquad\qquad\qquad\qquad\nonumber\\
\qquad\qquad\qquad\qquad\qquad-\frac{1}{r}\int_0^r\!dx^{\prime}\int_0^{x^{\prime}}dx^{\prime\prime}
h(x^{\prime\prime},0)\Biggr]\nonumber\\
=\frac{e}{mcv_{\text{\tiny{F}}}}\Biggl[\Lambda(x)-\frac{1}{r}\int_0^r\!dx\Lambda(x)\Biggr].\qquad\qquad\qquad
\end{eqnarray}
With the help of Eq.~(\ref{deriv}) one can express the first term
in additional phase Eq.~(\ref{general1}) in terms of $\Lambda(x)$.
It turns out that the second term in  Eq.~(\ref{general1}) exceeds
twice the first term. To see this, one should multiply the first
of equations Eq.~ (\ref{motion}) by $y(x)$ and integrate over $x$
\begin{eqnarray}
\label{phi}
\int_0^rdx\;y(x)\;\frac{d^2y}{dx^2}=\frac{1}{\Phi_0k_{\text{\tiny{F}}}}\int_0^rdx\;h(x,0)\;y(x).
\end{eqnarray}
The rhs of Eq.~(\ref{phi}) is the second term in
Eq.~(\ref{general1}). The lhs of Eq.~(\ref{phi}) can be related to
the first term in Eq.~(\ref{general1}) upon integration by parts
\begin{eqnarray}
\label{int}
\int_0^r\!dx\;y(x)\frac{d^2y}{dx^2}=-\int_0^rdx\left(\frac{dy}{dx}\right)^2.
\end{eqnarray}
Finally, we get
\begin{eqnarray}
\label{dphi} \delta\varphi_{0\rightarrow {\bf r}}=
-\frac{k_{\text{\tiny{F}}}}{2}\int_0^rdx\left(\frac{dy}{dx}\right)^2\qquad\qquad\qquad\qquad\qquad\\
=-
\frac{1}{\Phi_0^2k_{\text{\tiny{F}}}}\Biggl(\int_0^rdx\;\Lambda^2(x)-\frac{1}{r}
\left[\int_0^rdx\;\Lambda(x)\right]^2\Biggr).\nonumber
\end{eqnarray}
It is convenient to rewrite the final result Eq.~(\ref{dphi})
directly in terms of the random field, $h({\bf r})$. Substituting
Eq.~(\ref{Lambda}) into Eq.~(\ref{dphi}), we obtain

\begin{eqnarray}
\label{directly} \delta\varphi_{0\rightarrow {\bf r}}=
\frac{1}{\Phi_0^2k_{\text{\tiny{F}}}\xi}\int d{\bf r_1}\int d{\bf
r_2} h({\bf r_1}){\cal R}({\bf r_1},{\bf r_2})h({\bf
r_2}),\;\;\;\;\;
\end{eqnarray}
where the dimensionless kernel ${\cal R}({\bf r}_1,{\bf r}_2)$ is
defined as
\begin{eqnarray}
\label{kernel} {\cal R}({\bf r_1},{\bf r_2})=
\xi\;\delta(y_1)\;\delta(y_2)\qquad\qquad\qquad\qquad\\
\times\left[r-\frac{x_1x_2}{r}-x_2\;\theta(x_2-x_1)
-x_1\;\theta(x_1-x_2)\right].\nonumber
\end{eqnarray}
Note, that for the constant field $h(x,y)=h_0$, evaluation of
Eq.~(\ref{directly}) using the kernel Eq.~(\ref{kernel})
reproduces the result Eq.~(\ref{constantfield}), as expected.


\section{Disorder-smeared Friedel oscillations in different
regimes} \label{FO}

Smearing of the Friedel oscillations in the random field, $h({\bf
r})$, originates from the randomness of the phase,
$\varphi_{0\rightarrow {\bf r}}$, which is related to $h({\bf r})$
 via Eqs.~(\ref{directly}), (\ref{kernel}). Quantitatively, the
magnitude, $F(r)$, and the phase, $\phi(r)$, of smeared Friedel
oscillations Eq.~(\ref{modified}) are determined by the following
averages
\begin{eqnarray}
\label{Ups12} \Upsilon_1(r)=\text{Im}\bigl\langle
e^{2i\delta\varphi_{0\rightarrow {\bf r}}}\bigr\rangle_{h({\bf
r})},~\Upsilon_2(r)=\text{Re}\bigl\langle
e^{2i\delta\varphi_{0\rightarrow {\bf
r}}}\bigr\rangle_{h({\bf r})}.\nonumber\\
\end{eqnarray}
Then $F(r)$ and  $\phi(r)$ are related to the functions
$\Upsilon_1(r)$ and $\Upsilon_2(r)$ as
\begin{eqnarray}
\label{Fr}
F(r)=\sqrt{\bigl[\Upsilon_1(r)\bigr]^2+ \bigl[\Upsilon_2(r)\bigr]^2},\nonumber\\
\phi(r)=\arctan\left[\frac{\Upsilon_1(r)}{\Upsilon_2(r)}\right].
\end{eqnarray}
In this Section the averages Eq.~(\ref{Ups12}) will be calculated
separately for the regime of "arcs" and the regime of "snakes".

\subsection{Regime I}
\label{FOI}

In the regime of ``arcs'' we have $r\ll  \xi$, so that the field
is  almost constant within the interval $(0,r)$ and is equal to
its ``local'' value. For this reason, we can perform  the
averaging of $\exp\left[2i\varphi_{0\rightarrow {\bf r}}\right]$
over realizations of the random field, $h(x,y)$, explicitly,
without specifying the form of the correlator, $\text{\large
K}(r/\xi)$. This is because we can first set $h(x,y)\equiv const$
in $\exp\left[2i\varphi_{0\rightarrow {\bf r}}\right]$, and then
make use of the fact that the distribution function of the local
field is Gaussian\cite{we2}. Characteristic spatial scale,
$r_{\text{\tiny{I}}}$, for $F(r)$ and $\phi(r)$ immediately
follows from Eq.~(\ref{directly}) upon setting $h(0,r)=h_0$, and
requiring $2\delta\varphi_{0\rightarrow {\bf r}}=1$. This yields
$r_{\text{\tiny{I}}}=2^{2/3}3^{1/3}/p_0$, where $p_0$ is given by
Eq.~(\ref{p0}).

\subsubsection{Random magnetic field}

As discussed above, we start with Friedel oscillations in a
constant {\em local} magnetic field, $h$, for which we know that
\begin{equation}
\label{RegionII}
F_{\text{\tiny{I}}}(r,h)=1,~~~~~\phi_{\text{\tiny{I}}}(r,h)=
-\epsilon_r\left(\frac{h}{h_0}\right)^2,
\end{equation}
where
$p_0=k_{\text{\tiny{F}}}(\omega_c/E_{\text{\tiny{F}}})^{2/3}$, and
$\omega_c=eh_0/mc$ is the cyclotron frequency in the field, $h_0$.
In Eq.~(\ref{RegionII}) the parameter, $\epsilon_r$, is defined as
\begin{eqnarray}
\label{epsilonr} \epsilon_r=\frac{h_0^2r^3}{12\Phi_0^2
k_{\text{\tiny{F}}}}=\frac{(p_0r)^3}{12}.
\end{eqnarray}
 To find
the form of the averaged Friedel oscillation in the regime I, in
which $p_0\xi \ll 1$, we have to simply substitute the ``local''
value, $h$, of magnetic field into Eq.~(\ref{modified1}), i.e.,
replace $p_0^3$ by $p_0^3h^2/h_0^2$, and perform the gaussian
averaging over the distribution of the local field. This averaging
can be carried out analytically with the use of identity
\begin{eqnarray}
\label{AVERAGING}
\int\limits_{-\infty}^{\infty}\!\!\!\frac{dx}{\sqrt{\pi}}~e^{-x^2}\!\!\cos(\epsilon_r
x^2+\beta)\!=\! \Upsilon_1(\epsilon_r)\cos\beta
-\Upsilon_2(\epsilon_r)\sin\beta,\nonumber\\
\end{eqnarray}
where the functions $\Upsilon_1$ and $\Upsilon_2$ for this case
assume the following forms
\begin{eqnarray}
\label{functionsU1} \Upsilon_{1}\rightarrow
\Biggl(\frac{\pi}{2}\Biggr)^{1/2}\sqrt{\frac{(1+\epsilon_r^2)^{1/2}+
1}{1+\epsilon_r^2}},
\end{eqnarray}

\begin{eqnarray}
\label{functionsU2} \Upsilon_{2}\rightarrow
\Biggl(\frac{\pi}{2}\Biggr)^{1/2}\sqrt{\frac{(1+\epsilon_r^2)^{1/2}-
1}{1+\epsilon_r^2}}.
\end{eqnarray}
Using  Eq.~(\ref{Fr}), we recover from Eqs.~(\ref{functionsU1}),
(\ref{functionsU2}) the final result   Eq.~(\ref{AVERAGED}) for
the magnitude $F_{\text{\tiny{I}}}(r/r_{\text{\tiny{I}}})$ and the
phase $\phi_{\text{\tiny{I}}}(r/r_{\text{\tiny{I}}})$ of the
Friedel oscillations in the regime I.

In terms of variables $u$ and $v$ in the parametric space
Fig.~\ref{PhaseDiagram}, the condition $\epsilon_r=1$ can be
presented as
\begin{eqnarray}
\label{dashedline}
v=\frac{1}{u^{1/3}},\;\;\text{where}\;\;\;u=k_{\text{\tiny{F}}}R_{\text{\tiny{L}}},\;
\;v=\frac{r}{R_{\text{\tiny{L}}}}.
\end{eqnarray}
The dependence Eq.~(\ref{dashedline}) is shown in  Fig.~2 with a
dashed line within the regime I. To the left of this line, we have
$\epsilon_r<1$, so that $1/r^2$ decay of the Friedel oscillations
is unchanged in the random field. To the right of the dashed line,
$\epsilon_r$ is bigger than $1$. Then, the dependence $F(r)\propto
\epsilon_r^{-1/2}$, which follows from
 Eq.~(\ref{AVERAGED}), translates into faster, {\em but still power-law} decay,
$\propto 1/r^{7/2}$, of the Friedel oscillations. Note also, that
the phase of the oscillations also changes as $\epsilon_r$ crosses
over from small to large values. Indeed, as follows from
Eq.~(\ref{AVERAGED}),  we have $\phi(r)\rightarrow
-\pi/4+1/(2r^3)$ in the limit $\epsilon_r \gg 1$.

\subsubsection{Periodic Magnetic Field}

Consider a particular case of a {\em spatially-periodic} magnetic
field $h(x,y)={\tilde h}_0\cos(qx)$. For small enough $q$ the
``local'' description applies. The corresponding condition reads
\begin{eqnarray}
\label{condition1} q\ll {\tilde
p}_0=k_{\text{\tiny{F}}}\Bigl(\frac{{\tilde
h}_0}{k_{\text{\tiny{F}}}^2\Phi_0}\Bigr)^{2/3}.
\end{eqnarray}
Under this condition, the averaged Friedel oscillation can be
found by averaging

Eq.~(\ref{modified1}), in which $p_0$ is replaced by $\tilde p_0
\left(h/\tilde h_0\right)^{2/3}$, over the distribution, $P(h)$,
of the local values of magnetic field rather than over the
gaussian distribution Eq.~(\ref{AVERAGING}). This distribution has
the form
\begin{equation}
\label{periodic} P(h)=\frac{1}{\pi\sqrt{{\tilde h}_0^2 -h^2}},
\end{equation}
so that instead of Eq.~(\ref{AVERAGING}) we have
\begin{eqnarray}
\label{averaging1}\frac{1}{\pi}\int\limits_{-1}^{1}dx\frac{\cos({\tilde
\epsilon}_r x^2+\beta)}{\sqrt{1-x^2}}=
J_0\left({\tilde \epsilon}_r/2\right)\cos\left(\frac{{\tilde \epsilon}_r}{2}+\beta\right)\;\;\;\nonumber\\
={\tilde \Upsilon}_1({\tilde \epsilon}_r)\cos\beta -{\tilde
\Upsilon}_2({\tilde \epsilon}_r)\sin\beta,\;\;\;\;
\end{eqnarray}
where $J_0$ is the Bessel function, $\tilde\varepsilon
_r=\left(\tilde p_0 r\right)^3/12$, and
\begin{eqnarray}
\label{tilde} {\tilde \Upsilon}_1({\tilde
\epsilon}_r)=J_0\left(\frac{{\tilde \epsilon}_r}{2}\right)
\cos\left(\frac{{\tilde \epsilon}_r}{2}\right),\\
{\tilde \Upsilon}_2({\tilde \epsilon}_r)=J_0\left(\frac{{\tilde
\epsilon}_r}{2}\right)\sin\left(\frac{{\tilde
\epsilon}_r}{2}\right),\nonumber
\end{eqnarray}
so that in a periodic field, instead of Eq.~(\ref{AVERAGED}), we
have
\begin{eqnarray}
\label{tilde1} {\tilde F}(r)=\Bigl[{\tilde \Upsilon}_1^2({\tilde
\epsilon}_r)+{\tilde \Upsilon}_2^2({\tilde
\epsilon}_r)\Bigr]^{1/2}= \left\vert
J_0\left(\frac{{\tilde \epsilon}_r}{2}\right)\right\vert,\\
\tilde\phi(r)=-\arctan\left[\frac{{\tilde \Upsilon}_2({\tilde
\epsilon}_r)}{{\tilde \Upsilon}_1({\tilde
\epsilon}_r)}\right]=-\frac{{\tilde \epsilon}_r}{2}.\nonumber
\end{eqnarray}
It is instructive to present the results Eq.~(\ref{tilde1}) in a
different form, by simply showing how the Friedel oscillation
Eq.~(\ref{modified1}) gets modified {\em on average} in the
presence of a periodic magnetic field. Substituting
Eq.~(\ref{tilde1}) into Eq.~(\ref{modified}) we get
\begin{eqnarray}
\label{PERIODICDIRECT} \Bigl\langle V_{\mbox{\tiny
H}}(r)\Bigr\rangle & =&-\frac{\nu_0g V(2k_{\mbox{\tiny F}})}{2\pi
r^2}\;J_0\left(\frac{{\tilde p}_0^3r^3}{24}\right)\qquad\qquad
\nonumber\\
&\times &\sin\Biggl[2k_{\mbox{\tiny F}}r-\frac{({\tilde
p}_0r)^3}{24} \Biggr].
\end{eqnarray}
Eq.~(\ref{PERIODICDIRECT}) is a quite remarkable result. It
suggests that, due to the periodic {\em smooth} magnetic field,
the {\em averaged} Friedel oscillations {\em do not} get smeared.
Rather they acquire an {\em oscillatory envelope},
$J_0\left(\frac{{\tilde p}_0^3r^3}{24}\right)$. This envelope
oscillates with ``period''much larger than the de Broglie wave
length, but {\em much smaller} than the period, $1/q$, of change
of the magnetic field.

Note that this effect provides a unique possibility to measure
experimentally the {\em amplitude} of a periodic modulation. The
reason is the following. The envelop Eq.~(\ref{PERIODICDIRECT})
due to periodic magnetic field (or electric field, i.e., due to
the lateral superlattice) translates into a distinct low-frequency
behavior of the {\em tunnel density of states}. Namely, the tunnel
density of states would exhibit an ``oscillatory''behavior with a
``period'' $\omega \sim {\tilde p}_0v_{\text{\tiny{F}}}$. This
period in $\omega$ depends only on the {\em magnitude} of the
modulation, ${\tilde h}_0 $,
 but not on the spatial  period of modulation, $2\pi/q$.
Therefore, the magnitude of modulation, which, unlike the period,
is hard to measure otherwise, can be inferred from the bias
dependence of the tunneling conductance.


\subsection{Friedel oscillations in a random magnetic field: Regime II}
\label{FOII}

As the magnitude, $h_0$, of the random field decreases, the
character of semiclassical motion changes from arc-like (regime I
in Fig.~\ref{trajectories}) to the snake-like (regime II in
Fig.~\ref{trajectories}). To estimate for the "widths", $\delta
y$, of the snake-like trajectories, we use Eq.~(\ref{deriv}) and
set $x\sim \xi$. This yields
\begin{eqnarray}
\label{ymax} \frac{\delta
y}{\xi}\sim\frac{eh_0\xi}{mcv_{\text{\tiny{F}}}}
\sim\left(\frac{\epsilon}{k_{\text{\tiny{F}}}\xi}\right)^{1/2}.
\end{eqnarray}
Since $k_{\text{\tiny{F}}}\xi\gg 1$ and $\epsilon\ll 1$ in regime
II, we confirm that $\delta y\ll \xi$, i.e., that the snake is
"narrow".

It is clear that at large enough distances, $r$, the magnitude,
$F(r)$, of the averaged Friedel oscillations falls off
exponentially with $r$. The prime question is what is the
characteristic decay length. As stated in Section~\ref{main} this
length, $r_{\text{\tiny{II}}}$, is given by Eq.~(\ref{rII}). Below
we derive this length qualitatively, and then establish the form
of the magnitude, $F_{\text{\tiny{II}}} (r)$, as well as the
phase, $\phi_{\text{\tiny{II}}}(r)$, for the average Friedel
oscillations within the entire domain of $r$ by performing the
functional averaging  of $\exp(2i\delta\varphi_{0\rightarrow {\bf
r}}).$

\subsubsection{Qualitative consideration}

To recover qualitatively the  scale $r_{\text{\tiny{II}}}$ from
Eq.~(\ref{directly}) we consider the following toy model. Let us
divide the interval $(0,r)$ into small intervals of a {\em fixed}
length, $\xi$ (overall, $r/\xi$ intervals). Assume now that the
random field takes only two values, $h_0$ and $-h_0$, each with
probability, $1/2$, within a given interval, $\xi$. Under this
assumption, we find from Eq.~(\ref{Lambda})
$\Lambda(r)=h_0\left[m(r)-n(r)\right]\xi$, where $m(r)$ and $n(r)$
are the numbers of small intervals within the length, $r$,  with
$h=h_0$ and $h=-h_0$, respectively (obviously, $m+n=r/\xi$). From
Eq.~(\ref{dphi}) we get for $\delta\varphi_{0\rightarrow {\bf r}}$
\begin{eqnarray}
\label{via_nm} \delta\varphi_{0\rightarrow {\bf
r}}&=&\frac{h_0^2\xi^2}{\Phi_0^2k_{\text{\tiny{F}}}}
\Biggl\{\int_0^{r}dx\bigl[m(x)-n(x)\bigr]^2\nonumber\\
&-&\frac{1}{r}\left(\int_0^rdx\bigl[m(x)-n(x)\bigr]\right)^2\Biggr\}.
\end{eqnarray}
Second term in Eq.~(\ref{via_nm}) is square of the difference
$\langle m\rangle- \langle n\rangle$ of coordinate ({\em not
statistical}) average values of $m(x)$ and $n(x)$. Rewriting
$m(x)$ as $\langle m\rangle +\delta m(x)$ and $n(x)$ as $\langle
n\rangle +\delta n(x)$, and taking into account that $\delta
m(x)+\delta n(x)=0$, one can cast  Eq.~(\ref{via_nm}) into the
form
\begin{eqnarray}
\label{Via_nm} \delta\varphi_{0\rightarrow {\bf
r}}=\frac{4h_0^2\xi^2}{\Phi_0^2k_{\text{\tiny{F}}}}
\int_0^{r}dx\left[\delta m(x)\right]^2.
\end{eqnarray}
Since the typical value of $\left[\delta m(x)\right]^2$ is
$\langle m(x)\rangle =x/2\xi$, we arrive at the following estimate
$\delta\varphi_{0\rightarrow {\bf r}}\sim h_0^2\xi
r^2/\Phi_0^2k_{\text{\tiny{F}}}$. Equating this additional phase
to unity yields $r=\Phi_0k_{\text{\tiny{F}}}^{1/2}/h_0\xi^{1/2}$,
which coincides with $r_{\text{\tiny{II}}}$ defined by
Eq.~(\ref{rII}) within a numerical factor.

\subsubsection{Evaluation of the functional integral}

Below we present the analytical derivation of
Eqs.~(\ref{FriedelMagnII}), (\ref{FriedelPhaseII}). The averaging
of $\exp\left(2i\delta\varphi_{0\rightarrow{\bf r}}\right)$
required to calculate $F_{\text{\tiny{II}}}(r)$, and
$\phi_{\text{\tiny{II}}}(r)$ from Eqs.~(\ref{Ups12}), (\ref{Fr})
reduces to the functional integral
\begin{eqnarray}
\label{functional} \bigl\langle e^{2i\delta\varphi_{0\rightarrow
{\bf r}}} \bigr\rangle= \frac{\int D \left\{h({\bf
r})\right\}\exp\Bigl[2i\delta \varphi({\bf r})- W\left\{h({\bf
r})\right\}\Bigr]}{\int D\left\{ h({\bf
r})\right\}\exp\Bigl[-W\left\{h({\bf r})\right\}\Bigr]},\nonumber\\
\end{eqnarray}
where $\delta\varphi(r)= \delta\varphi_{0\rightarrow {\bf r}}$ is
given by Eq.~(\ref{dphi}), and $\exp\left(-W\{h\}\right)$  with
$W\left\{h({\bf r})\right\}$ given by
\begin{eqnarray}
\label{W}
W\{h\}=\frac{1}{\xi^4h_0^2}\int_0^{r_2}\int_{-\infty}^{\infty}dx_1\;dy_1
\int_0^{r_2}\int_{-\infty}^{\infty}dx_2\;dy_2\;\qquad\nonumber\\
\times\;h(x_1,y_1)
 h(x_2,y_2)\;\kappa(x_1-x_2,y_1-y_2),\quad\qquad
\end{eqnarray}
is the statistical weight of the realization,  $h(x,y)$. The
dimensionless function $\kappa({\bf r},{\bf r}^{\prime})$ is
related to the correlator Eq.~(\ref{correlator}) in a standard way
\begin{eqnarray}
\label{related} \int d{\bf r}^{\prime}\kappa({\bf r},{\bf
r}^{\prime}) \text{\large K}({\bf r}^{\prime},{\bf
r}^{\prime\prime})=\xi^4\delta({\bf r}-{\bf r}^{\prime\prime}).
\end{eqnarray}
The reason  why the functional integral Eq.~(\ref{functional}) can
be evaluated explicitly is that both $W\{h\}$ and
$\delta\varphi_{0\rightarrow {\bf r}}$ are {\em quadratic} in the
random field, $h(x,y)$. The fact that we integrate over
realizations of $h(x,y)$ defined on the interval which is {\em
finite}, $0<x<r$, in the $x$-direction and infinite in the $y$
direction suggests the following expansion of $h(x,y)$

\begin{eqnarray}
\label{EXPAND}
h(x,y)=h_0\!\sum_{n=-\infty}^{\infty}\int_{-\infty}^{\infty}\!dq{\mathcal
A}_{n,q}e^{iqy/\xi}\exp\left(\frac{2\pi inx}{r}\right).\nonumber\\
\end{eqnarray}
The asymmetry between $x$ and $y$ is quite significant in the
calculation below, namely, for $r\gg \xi$, the characteristic
values of $x$ turn out to be much larger than  the characteristic
values of $y$ is $\sim\delta y\ll\xi$, see Eq.~(\ref{ymax}). This
allows to replace $\text{\large K}(x,y,x^{\prime},y^{\prime})$ in
Eq.~(\ref{related}) by $\gamma \xi\;\text{\large
K}(0,y-y^{\prime})\delta(x-x^{\prime})$, where the dimensionless
constant $\gamma$ is defined by the relation
\begin{eqnarray}
\label{replace} \gamma =\frac{\int\limits_0^{\infty} dx
\int\limits_{-\infty}^{\infty}dy\; \text{\large K}(x,y)}
{\xi\int\limits_{-\infty}^{\infty} dy\;  \text{\large K}(0,y)}=
\left(\frac{\pi}{2}\right)\frac{\int\limits_0^{\infty} dz\;z
\text{\large K}(z)} {\int\limits_{0}^{\infty} dz\;  \text{\large
K}(z)},
\end{eqnarray}
where in the second identity we used the fact that $\text{\large
K}(x,y)$ is isotropic. Substituting Eq.~(\ref{EXPAND}) into
Eq.~(\ref{W}), we obtain
\begin{eqnarray}
\label{W2} W\{h\}=\frac{r}{\gamma\xi}\sum_{n=-\infty}^{\infty}\int
\!dq\;\frac{\vert{\mathcal A}_{n,q}\vert^2} {\tilde{\mathcal
K}(q)},
\end{eqnarray}
where ${\tilde{\mathcal K}(q)}$ is the Fourier transform of the
correlator, more precisely,
\begin{eqnarray}
\label{W1} {\tilde{\mathcal K}(q)}=\frac{1}{\sqrt{2\pi}}\int
\frac{dy}{\xi} e^{iqy/\xi}\text{\large K}(0,y).
\end{eqnarray}
Expression for $\delta\varphi(r)$ in terms of the coefficients,
${\mathcal A}_{n,q}$,
 follows upon substitution of Eq.~(\ref{expansion}) into
Eq.~(\ref{directly})
\begin{eqnarray}
\label{h1}
&&\delta\varphi(r)=\frac{h_0^2}{\Phi_0^2k_{\text{\tiny{F}}}}\sum_{n_1=-\infty}^{\infty}\sum_{n_2=-\infty}^{\infty}
\int dq_1{\mathcal A}_{n_1,q_1}\int dq_2{\mathcal A}_{n_2,q_2}\nonumber\\
&&\int_0^r\!\!dx_1\!\!\int_0^r\!\!dx_2
\Bigl[r-\frac{x_1x_2}{r}-x_2\Theta(x_2-x_1)\nonumber\\
&&-x_1\Theta(x_1-x_2)\Bigr] \exp\left\{\frac{2\pi
i}{r}(n_1x_1+n_2x_2)\right\}.
\end{eqnarray}
Performing the integration, we obtain
\begin{eqnarray}
\label{h2} &&\delta\varphi(r)=\frac{\varepsilon
r^3}{\xi^3}\Biggl\{\frac{1}{12}\int dq\; {\mathcal A}_{0,q}^2 +
\sum_{n>0}c_n\Big\vert \int dq\; {\mathcal A}_{n,q}\Big\vert^2 \nonumber\\
&&+\int dq\; {\mathcal A}_{0,q}\int dq \sum_{n>0}\Bigl[b_n
{\mathcal A}_{n,q}+b_n^{\ast}{\mathcal
A}_{n,q}^{\ast}\Bigr]\Biggr\},
\end{eqnarray}
where numerical coefficients $b_n$ and $c_n$ are defined as
\begin{eqnarray}
\label{bc} b_n=-\frac{1}{2\pi^2n^2}+\frac{i}{2\pi
n},~~~~c_n=\frac{1}{2\pi^2n^2}.
\end{eqnarray}
In writing the result of integration in the form Eq.~(\ref{h2}) we
have used the dimensionless parameter $\varepsilon$ defined by
Eq.~(\ref{varepsilon}).
The meaning of this parameter is the additional phase
Eq.~(\ref{dphi}), acquired by the electron travelling the distance
$\sim \xi$ in a constant magnetic field, $h_0$. Since our
calculation pertains to the limit $r\gg \xi$, the relevant values
of $\varepsilon$ are small.

The functional integration reduces now to the infinite product of
the ratios of integrals over ${\mathcal A}_{n,q}$ and ${\mathcal
A}_{n,q}^{\ast}$. The details of calculation are given in
Appendix~\ref{AppendixC}. Here we present only the final result
for $r\gg \xi$
\begin{eqnarray}
\label{after} \langle e^{2i\delta\varphi(r)}\rangle= \frac{1}
{{1-\frac{2i}{3}\left(\frac{r}{r_{\text{\tiny{II}}}}\right)^2}} \nonumber\\
\times\prod_{n=1}^{\infty}\frac{n^2}{n^2-2i(r/r_{\text{\tiny{II}}})^{2}/\pi^2},
\end{eqnarray}
where the characteristic length, $r_{\text{\tiny{II}}}$, is
defined as
\begin{eqnarray}
\label{r0}
r_{\text{\tiny{II}}}=\frac{2\xi}{\left[\sqrt{2\pi}\gamma\varepsilon
\right]^{1/2}}=\sqrt{\frac{4k_{\text{\tiny{F}}}\Phi_0^2}
{(2\pi)^{1/2}\gamma\xi h_0^2
}}.
\end{eqnarray}
The above definition specifies the numerical coefficient, $\eta$,
in Eq.~(\ref{rII}) of Section~\ref{main} as
$\eta=2/(2\pi)^{1/4}\gamma^{1/2}$. This coefficient depends on the
explicit form of the correlator via the factor $\gamma$, given by
Eq.~(\ref{replace}). It is seen that $r_{\text{\tiny{II}}}\sim
\xi/\varepsilon^{1/2}$ is indeed much larger than $\xi$. This
means that, in the regime II, Friedel oscillations survive well
beyond the correlation radius of random magnetic field. Note also
a distinctive dependence $r_{\text{\tiny{II}}} \propto 1/h_0$ of
the characteristic scale on the magnitude of the random field. In
fact, the infinite product in Eq.~(\ref{after}) can be evaluated
for arbitrary $r/r_{\text{\tiny{II}}}$, using the identity
\begin{eqnarray}
\label{IDENTITY} \frac{\sin x}{x}=\prod_n\Bigl(1-\frac{x^2}{\pi^2
n^2}\Bigr),
\end{eqnarray}
which yields
\begin{eqnarray}
\label{finite} \langle
e^{2i\delta\varphi(r)}\rangle=\frac{1}{{ 1-\frac{2i}{3}\left(\frac{r}{r_{\text{\tiny{II}}}}\right)^2  }}\qquad\qquad\\
\times\frac{(1+i)(r/r_{\text{\tiny{II}}})}{\sin(r/r_{\text{\tiny{II}}})\cosh(r/r_{\text{\tiny{II}}})+i\cos(r/r_{\text{\tiny{II}}})\sinh(r/r_{\text{\tiny{II}}})}.\nonumber
\end{eqnarray}
With the help of Eq.~(\ref{finite}) we can calculate the
magnitude, $F_{\text{\tiny{II}}}(r)$, and the phase,
$\phi_{\text{\tiny{II}}}(r)$, of the Friedel oscillations in the
regime II. Corresponding expressions are given by
Eqs.~(\ref{FriedelMagnII}) and (\ref{FriedelPhaseII}).

\subsubsection{Limiting cases}
It is not surprising that Friedel oscillations in the regime II
are smeared more efficiently than in the regime I. The small-$r$
and the large-$r$ asymptotes of $F_{\text{\tiny{II}}}(r)$ are the
following
\begin{eqnarray}
\label{limit1}
F_{\text{\tiny{II}}}(r)=1-\frac{11}{45}\left(\frac{r}{r_{\text{\tiny{II}}}}\right)^4,
\;\;\;r\ll r_{\text{\tiny{II}}},
\end{eqnarray}
\begin{eqnarray}
\label{limit2}
F_{\text{\tiny{II}}}(r)=3\sqrt{2}\left(\frac{r_{\text{\tiny{II}}}}{r}\right)
\exp\left(-\frac{r}{r_{\text{\tiny{II}}}}\right),\;\;\;r\gg
r_{\text{\tiny{II}}}.
\end{eqnarray}
We see from Eq.~(\ref{limit2}) that Friedel oscillations decay
exponentially as $r$ exceeds $r_{\text{\tiny{II}}}$. This should
be contrasted to Eq.~(\ref{AVERAGED}) for the regime I, where the
$F_{\text{\tiny{I}}}(r)$ falls off slowly, as $r^{-3/2}$, with
$r$. On the qualitative level, the strong difference between the
regimes I and II, that is reflected in the different characters of
decay of $F_{\text{\tiny{I}}}(r)$ and $F_{\text{\tiny{II}}}(r)$,
is that in regime I the random field does not change within the
characteristic spatial interval, $r_{\text{\tiny{I}}}$, while in
regime II the sign of the random field changes many times within
the characteristic spatial interval, $r_{\text{\tiny{II}}}$.

\section{Density of states: Qualitative discussion}
\label{DOSQ} In the previous consideration we had demonstrated
that in two regimes of electron motion in random magnetic field,
i.e., regime of arcs, I, and regime of snakes, II, there are two
length-scales, $r_{\text{\tiny I}}$ and  $r_{\text{\tiny II}}$,
respectively that govern the interaction effects. In this section
we demonstrate that the density of states, $\delta\nu (\omega)$,
exhibits an anomalous behavior within the frequency range $\omega
\sim v_{\text{\tiny F}}/r_{\text{\tiny I}}$ in the regime of arcs,
and $\omega \sim v_{\text{\tiny F}}/r_{\text{\tiny II}}$ in the
regime of snakes.

The process underlying the interaction corrections to the density
of states is creation (and annihilation) of the virtual
electron-hole pairs by an electron moving in the random field. Our
central finding is that, unlike the case of point-like impurities
\cite{rudin97}, the low-$\omega$ structure in the density of
states emerges as a result of electron-electron scattering
processes involving {\em more than one pair}.

\begin{figure}[t]
\centerline{\includegraphics[width=85mm,angle=0,clip]{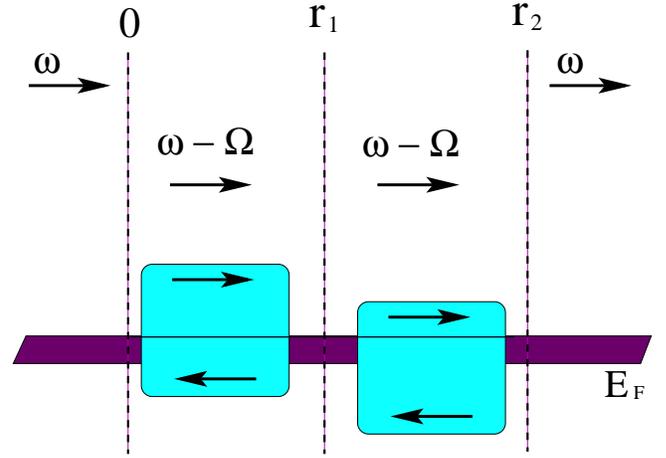}}
\caption{(Color online) Third-order process describing creation of
a pair by initial electron at point ${\bf r}=0$, rescattering
within the pair at point ${\bf r}={\bf r}_1$, and annihilation of
the pair at point ${\bf r}={\bf r}_2$. Diagram corresponding to
this process is shown later in the text (first diagram in
Fig.~\ref{9diagrams}).} \label{three}
\end{figure}

We start with a three-scattering process in the regime of arcs,
and demonstrate qualitatively how the frequency scale,
$v_{\text{\tiny F}}/r_{\text{\tiny I}}$, emerges. Three-scattering
process involves {\em two} virtual pairs. Consider first this
process in the absence of the random field. It is illustrated in
Fig.~\ref{three}. In analysis of this process
\cite{suhas1,suhas2,suhas3} it was established that the directions
of momenta of the participating electrons are strongly correlated,
namely, they are either almost parallel or almost antiparallel.
Quantitative estimate for the degree of alignment of the momenta
can be obtained from inspection of Fig.~\ref{three}. If the
scattering acts take place at points $0$, ${\bf r}_1$, and ${\bf
r}_2$, then the corresponding matrix element contains a phase
factor
\begin{equation}
\label{afactor} \exp\left[2ik_{\text{\tiny
F}}\left(r_1-r_2+\vert{\bf r}_1-{\bf r}_2\vert\right)\right].
\end{equation}
This phase factor does not oscillate, if the angle between the
vectors ${\bf r}_1$ and ${\bf r}_2$ is smaller than
$\left(1/k_{\text{\tiny F}}r\right)^{1/2}$,  where $r$ is the
typical length of ${\bf r}_1$, ${\bf r}_2$.

The above angular restriction constitutes the origin of a
zero-bias anomaly in the regime of arcs.  Zero-bias anomaly
emerges as a result of the suppression of the three-scattering
process in the field, $h_0$. This suppression is due to curving of
the electron trajectory by the angle $\sim r/R_{\text{\tiny L}}$,
see Fig.~\ref{angle}, and it occurs when the curving angle exceeds
the allowed angle of alignment. Therefore, upon equating
$\left(1/k_{\text{\tiny F}}r\right)^{1/2}$ to $ r/R_{\text{\tiny
L}}$, we find $r=r_{\text{\tiny I}}$, which leads us to the
conclusion that $\omega\sim v_{\text{\tiny F}}/r_{\text{\tiny I}}$
is the energy scale at which $\delta\nu(\omega)$ exhibits a
feature. Note that, in considering the Friedel oscillations, we
inferred the scale $r_{\text{\tiny I}}$ from a different
condition, namely, that the additional phase, $\sim
(p_0r_{\text{\tiny I}})^3$, due to the {\em elongation} of a
trajectory in magnetic field is $\lesssim 1$. Thus we conclude
that, in the regime of arcs, the same spatial scale,
$r_{\text{\tiny I}}$, which governs the ``dephasing'' of
$\Pi_{2k_{\text{\tiny{F}}}}(r)$ (a polarization bubble) also
governs the suppression of the three-scattering process, which
involves {\em three loops }.

The above analysis of phases in the matrix element of the
three-scattering process can be extended to the regime of snakes.
This analysis yields that three-scattering process is efficient at
distances $r\lesssim r_{\text{\tiny II}}$, see Eq.~(\ref{rII}).
Analysis of phases similar to the phase, given by
Eq.~(\ref{afactor}), also suggests that  {\em two-scattering}
processes are {\em insensitive} to the magnetic field. This
insensitivity can be explained as follows. Calculation of the
contribution to the density of states from the three-scattering
process with matrix element Eq.~(\ref{afactor}) involves
integration over positions of ${\bf r}_1$ and ${\bf r}_2$, {\em
with respect to the origin}, ${\bf r}=0$, which reveals the
angular restriction on their orientations. Similar integration for
a two-scattering process involves only the orientation of the
interaction point, ${\bf r}$, with respect to the origin. Then the
angular restriction, and its lifting by magnetic field, does not
emerge. In the next subsection the above qualitative arguments are
supported by a rigorous calculation.

\section{Density of states: Analytical derivation}
\label{DOS}
\subsection{Absence of a zero-bias anomaly in the second order in
the interaction strength}

\begin{figure}[t]
\centerline{\includegraphics[width=75mm,angle=0,clip]{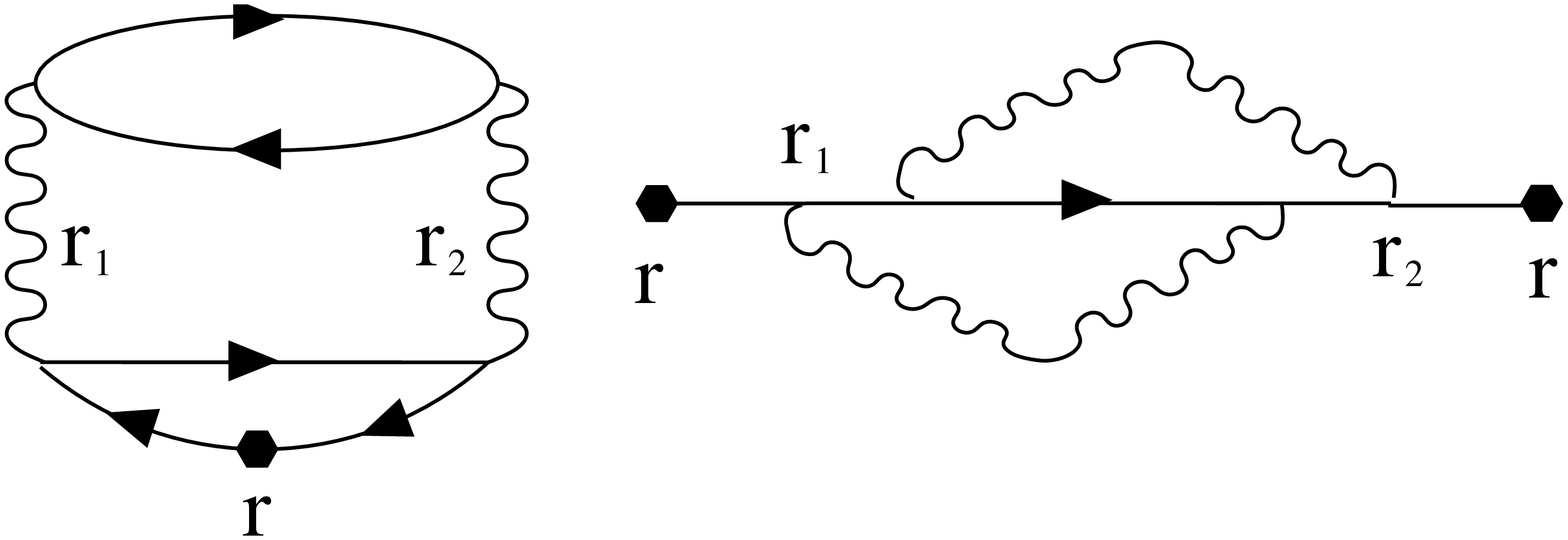}}
\caption{Diagrams for the second-order corrections
Eq.~(\ref{deltanu1}) (left) and Eq.~(\ref{deltanu2}) (right) to
the density of states.} \label{oneloop}
\end{figure}

We start from general expression for the average density of states
\begin{eqnarray}
\label{gen1} \delta\nu(\omega)=-\frac{1}{\pi}\;\Bigl\langle
{\text{Im}}\;G_{\omega}({\bf r},{\bf r})\Bigr\rangle_{h(x,y)},
\end{eqnarray}
where $\langle\dots\rangle$ denotes disorder averaging defined by
Eq.~(\ref{functional}). In the second order in interaction
strength, the random-field-induced correction to the density of
states are determined by two diagrams shown in Fig.~\ref{oneloop}.
The corresponding analytical expressions read
\begin{eqnarray}
\label{deltanu1} \delta\nu_1(\omega)\!&=&\!4\;
{\text{Im}}\frac{2}{\pi }\int\frac{d\Omega}{2\pi}\int d{\bf
r}\;d{\bf r}_1d{\bf r}_2 \;G_{\omega}({\bf r},{\bf
r}_1)\nonumber\\
&\times&\!G_{\Omega}({\bf r}_1,{\bf
r}_2)\Bigl\{V^2\!\left(2k_{\text {\tiny
F}}\right)\Pi_{2k_{\text{\tiny{F}}}}({\bf r}_1,{\bf
r}_2,\omega-\Omega)\nonumber\\&+&\!V^2(0)\Pi_{0}({\bf r}_1,{\bf
r}_2,\omega-\Omega)\Bigr\} G_{\omega}({\bf r}_2,{\bf r}),
\end{eqnarray}

\begin{eqnarray}
\label{deltanu2} \delta\nu_2(\omega)\!\!&=&\!\!-2\;
{\text{Im}}\frac{2}{\pi}\int\frac{d\Omega}{2\pi}\int d{\bf
r}\;d{\bf r}_1d{\bf r}_2\;G_{\omega}({\bf r},{\bf
r}_1)\nonumber\\
&\times&\!\!G_{\Omega}({\bf r}_1,{\bf r}_2)G_{\omega}({\bf
r}_2,{\bf r})\Bigl\{V(0)\bigl[2V(2k_{\text {\tiny
F}})-V(0)\bigr]\nonumber\\
&\times&\!\!\!\Pi_{2k_{\text{\tiny{F}}}}({\bf r}_1,{\bf
r}_2,\omega-\Omega)
+V^2(0)\Pi_{0}({\bf r}_1,{\bf r}_2,\omega-\Omega)\Bigr\},\nonumber\\
\end{eqnarray}
where $V(0)$ and $V(2k_{\text {\tiny F}})$ are the Fourier
components of the interaction potential $V({\bf r})$ with momenta
zero and $2k_{\text {\tiny F}}$, respectively. Three Green
functions in Eqs.~(\ref{deltanu1}), (\ref{deltanu2}) describe the
propagation of electron between the points $\left({\bf r},{\bf
r}_1\right)$, $\left({\bf r}_1,{\bf r}_2\right)$, and $\left({\bf
r}_2,{\bf r}\right)$, Fig.~\ref{oneloop} . Polarization bubble
describes the creation of electron-hole pair at point ${\bf r}_1$
and annihilation at point ${\bf r}_2$. Difference in signs in
Eqs.~ (\ref{deltanu1}), (\ref{deltanu2}) is due to the fact that
the first diagram contains two closed fermionic loops, whereas the
second diagram contains only one. Numerical factors $4$ and $2$ in
Eqs.~(\ref{deltanu1}), (\ref{deltanu2}) come from summation over
the spin indices. The difference between them is due two the fact
the spin of electron-hole pair is not fixed in the first diagram,
but it is fixed in the second diagram. The factor $2$ in the
product $2V(0)V(2k_{\text {\tiny F}})$ in Eq.~(\ref{deltanu2}) is
related to the annihilation of the electron-hole pair, since the
hole is annihilated with {\em initial} electron. Then the momentum
transfer can be $2k_{\text {\tiny F}}$ in the course of creation
and zero in the course of annihilation, and vice versa.

\begin{figure}[t]
\centerline{\includegraphics[width=85mm,angle=0,clip]{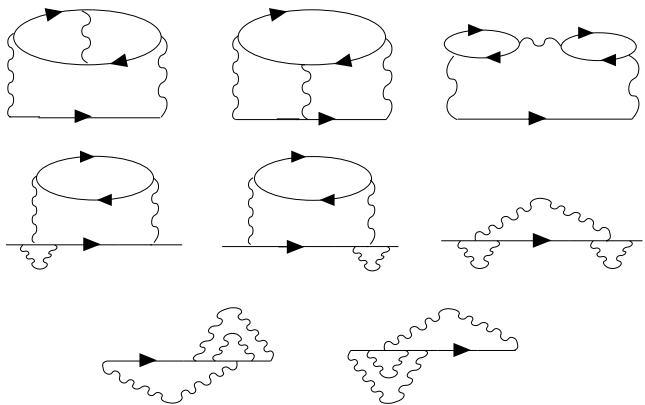}}
\caption{Third-order diagrams contributing to the zero-bias
anomaly in the density of states. Random field enters via the
phases of the Green functions.} \label{9diagrams}
\end{figure}


It is important to emphasize that the Green functions and
polarization operators in Eqs.~(\ref{deltanu1}), (\ref{deltanu2})
contain the information about the random field, $h(x,y)$, via
their additional phases: $\varphi_{{\bf r}_1\rightarrow{\bf r}_2}$
in $G_{\omega}({\bf r}_1,{\bf r}_2)$ and $2\varphi_{{\bf
r}_1\rightarrow{\bf r}_2}$ in $\Pi_{2k_{\text{\tiny{F}}}}({\bf
r}_1,{\bf r}_2,\omega)$.  The phase, $\varphi_{{\bf
r}_1\rightarrow{\bf r}_2}$, always enters in combination with a
main term, $k_{\text {\tiny F}}\vert{\bf r}_1-{\bf r}_2\vert$.
Obviously, $\Pi_0({\bf r}_1,{\bf r}_2,\omega)$ does not contain a
field-induced phase. Thus, only the terms containing
$\Pi_{2k_{\text{\tiny{F}}}}$ in Eqs.~(\ref{deltanu1}),
(\ref{deltanu2}) should be considered.

Now it is easy to see that $\delta\nu_1$ and $\delta\nu_2$ {\em do
not} exhibit a field-induced anomaly at small $\omega$. This is
because the field dependence is {\em cancelled out} in the
integrands of Eqs.~(\ref{deltanu1}), (\ref{deltanu2}). To see
this, we first note that the integration over ${\bf r}$ in
Eqs.~(\ref{deltanu1}), (\ref{deltanu2}) can be easily performed
using the fact that $\int d{\bf r}G_{\omega}({\bf r}_1,{\bf
r})G_{\omega}({\bf r},{\bf r}_2)$ is equal to the derivative,
$\partial G_{\omega}({\bf r}_1,{\bf r}_2)/\partial \omega$. Then
we note that the contribution to $\delta\nu_1$, $\delta\nu_2$
comes only from
 ``slow'' terms, in the product
of two Green functions, $G_{\omega}({\bf r}_1,{\bf r}_2)$,
$G_{\Omega}({\bf r}_1,{\bf r}_2)$, and
$\Pi_{2k_{\text{\tiny{F}}}}$. These slow terms do not contain
rapidly oscillating factors $\exp\{2ik_{\text {\tiny F}}\vert{\bf
r}_1-{\bf r}_2\vert\}$. On the other hand, cancellation of the
rapid terms in the product {\em automatically} results in the
cancellation of the field-dependent terms.


As it was explained in qualitative discussion, the situation
changes in the third order in the interactions. Corresponding
expression for $\delta\nu(\omega)$ is derived in the next
Subsection.


\subsection{General expression for the third-order interaction correction
to the density of states.}

Relevant diagrams for the third-order correction to the density of
states are shown in Fig.~\ref{9diagrams}. The same 8 diagrams were
considered in Ref.~\onlinecite{suhas3} in the momentum space. In
Ref.~\onlinecite{suhas3} the analysis of these diagrams was
restricted to small momenta. In our coordinate representation this
means that only $\Pi_0({\bf r})$ parts of the polarization
operators was kept, whereas $\Pi_{2k_{\text{\tiny{F}}}}({\bf r})$
parts were neglected. As explained above, to reveal the
sensitivity to the random field, we will keep {\em only} the
$\Pi_{2k_{\text{\tiny{F}}}}({\bf r})$ parts. Then the correction
to the Green function corresponding to the sum of eight diagrams
in Fig.~\ref{9diagrams} acquires the form
\begin{eqnarray}
\label{deltaG} &\delta&\!\!\!\!\nu (\omega)=2V(0)V(2k_{\text
{\tiny F}})\Bigl[2V(2k_{\text {\tiny
F}})-V(0)\Bigr]\\
&\times&{\text{Im}}\frac{i}{2\pi^2}\int\frac{d\Omega}{2\pi}\int
d{\bf r}\;d{\bf r}_1d{\bf r}_2\;G_{\omega}({\bf r},{\bf r}_1)\;
G_{\Omega}({\bf r}_1,{\bf r}_2) \nonumber\\&\times&
\Pi_{2k_{\text{\tiny{F}}}}({\bf r}_1,0,\omega-\Omega)\;
\Pi_{2k_{\text{\tiny{F}}}}(0,{\bf
r}_2,\omega-\Omega)\;G_{\omega}({\bf r}_2,{\bf r}).\nonumber
\end{eqnarray}
All the diagrams reduce to the same integrals. Concerning  the
difference in numerical coefficients, it comes from the number of
closed fermionic loops and the spin degrees of freedom. Taking
this into account interaction coefficient corresponding to the
first two diagrams will be $2\cdot
(-2)^2V^3(2k_{\text{\tiny{F}}})$. Coefficient of the third diagram
is $(-2)^3V^3(2k_{\text{\tiny{F}}})$. Thus we see, that the
contributions $\propto V^3(2k_{\text{\tiny{F}}})$ cancel each
other.

The first and the second diagrams in the second row are equal to
each other, and each of them has a coefficient
$(-2)^2V(0)V^2(2k_{\text{\tiny{F}}})$. Coefficient of the last
diagram in the second row is $(-2)V^2(0)V(2k_{\text{\tiny{F}}})$,
since it has only one closed fermionic loop. Finally, the first
diagram in third row has only one closed fermionic loop and is
equal to the second diagram on the third row. Each of these
diagrams contributes with the coefficient
$(-2)V(0)V^2(2k_{\text{\tiny{F}}})$.

On the physical level, 8 diagrams in Fig.~\ref{9diagrams} describe
different electron-electron three-scattering processes. For
example, the first diagram corresponds to creation of
electron-hole pair by the initial electron followed by
rescattering {\em within a created pair} and, finally, its
annihilation. Three stages of this process are illustrated in
Fig.~\ref{three}. However, creation, rescattering, and
annihilation of a pair can follow a different scenario, namely,
the rescattering process can  involve {\em the initial electron}.
This scenario is captured by the second diagram in the first row
in  Fig.~\ref{9diagrams}.


\begin{figure}[t]
\centerline{\includegraphics[width=90mm,angle=0,clip]{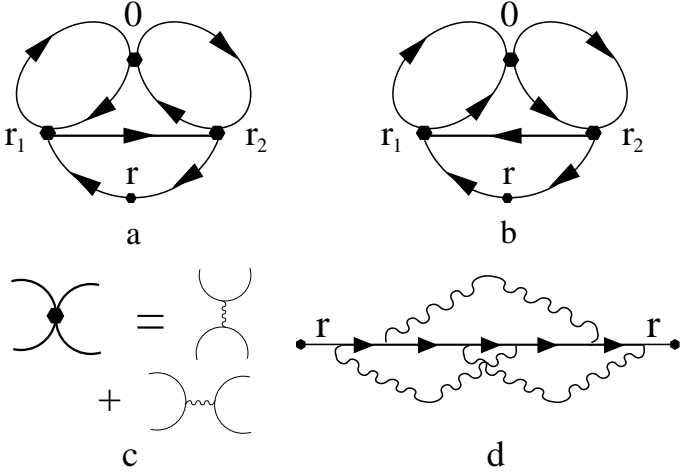}}
\caption{(a) Eight diagrams for $\delta G_{\omega}({\bf r},{\bf
r})$, that are shown in Fig.~\ref{9diagrams}, are combined into
one generalized diagram. Electron-electron scattering processes
take place at points $0$, ${\bf r}_1$, and ${\bf r}_2$; (b) Eight
third-order diagrams that {\em do not} contribute to the zero-bias
anomaly are combined into one generalized diagram;
 (c) Two types of four-leg interaction vertices are combined into big dots;
(d) An example of a third-order diagram of type (b). }
\label{two_loops}
\end{figure}

At this point, we note that diagrams in Fig.~\ref{9diagrams} do
{\em not} exhaust all possible three-scattering processes. In
fact, all diagrams in Fig.~\ref{9diagrams} have identical
structure, in the sense, that they can be combined into a single
{\em generalized} diagram, as shown in Fig.~\ref{two_loops}a.
There are also eight other diagrams combined into a single
generalized diagram, as shown in Fig.~\ref{two_loops}b that are
not sensitive to the random field. This is because, in the absence
of the random field, the phase factor corresponding to
Fig.~\ref{two_loops}b is large, namely, $2\cdot
2k_{\text{\tiny{F}}}(r_1+r_2)$.

The crucial difference between the contributions
Eqs.~(\ref{deltanu1}), (\ref{deltanu2}) and Eq.~(\ref{deltaG}) is
that the cancellation of the rapid-oscillating terms in in the
integrand of  Eq.~(\ref{deltaG}) {\em preserves} the
field-dependence. To see this, we first replace $\int d{\bf r}
G_{\Omega}({\bf r}_1,{\bf r})G_{\Omega}({\bf r},{\bf r}_2)$ by
$\partial G_{\Omega}({\bf r}_1,{\bf r}_2)/\partial \Omega$, as
discussed above, and then consider the phase of the product
\begin{eqnarray}
\label{product} G_{\Omega}({\bf r}_1,{\bf r}_2)~G_{\omega}({\bf
r}_1,{\bf r}_2)\qquad\qquad\qquad\qquad\qquad \nonumber\\ \times
\Pi_{2k_{\text{\tiny{F}}}}(0,{\bf
r}_2,\omega-\Omega)~\Pi_{2k_{\text{\tiny{F}}}}({\bf
r}_1,0,\omega-\Omega).
\end{eqnarray}
Fig.~\ref{angle}b illustrates this product graphically. It is seen
from Fig.~\ref{angle}b  that, when the fast oscillating terms
$\exp\{2ik_{\text {\tiny F}}\vert{\bf r}_1- {\bf r}_2\vert\}$,
$\exp\{2ik_{\text {\tiny F}} r_1\}$, and $\exp\{2ik_{\text {\tiny
F}} r_2\}$ cancel each other  out, the additional phase enters
into the product either in combination
\begin{eqnarray}
\label{product1}
2\delta\varphi_{\Sigma}^{(+)}=2\delta\varphi_{{\bf r}_1\rightarrow
0} + 2\delta\varphi_{{\bf r}_2\rightarrow 0}-2\delta\varphi_{{\bf
r}_1\rightarrow {\bf r}_2},
\end{eqnarray}
or in combination (see Fig.~\ref{angle}b)
\begin{eqnarray}
\label{product2}
2\delta\varphi_{\Sigma}^{(-)}=2\delta\varphi_{{\bf r}_1\rightarrow
0}- 2\delta\varphi_{{\bf r}_2\rightarrow 0} +2\delta\varphi_{{\bf
r}_1\rightarrow {\bf r}_2}.
\end{eqnarray}

Since additional phases defined by Eqs.~(\ref{directly}),
(\ref{kernel}) are {\em cubic} in distance, the combinations
Eq.~(\ref{product1}) and Eq.~(\ref{product2}) are {\em nonzero}.
This is in contrast to the two-scattering processes, where the
cancellation occurs {\em identically} for arbitrary dependence of
$\delta\varphi(r)$ on $r$. In turn, non-cancellation of additional
phases in Eqs.~(\ref{product1}), (\ref{product2}) means that the
random field causes a zero-bias anomaly, more specifically, a
feature in $\delta\nu(\omega)$ at small $\omega$.

The final form of $\delta\nu(\omega)$ emerges upon integration of
Eq.~(\ref{deltaG})  over azimuthal angles of ${\bf r}_1$ and ${\bf
r}_2$, which  can be performed analytically, using the relation

\begin{eqnarray}
\label{int-azimut} \left\langle e^{i{\bf p}\left({\bf r}_1+ {\bf
r}_2\right)} \right\rangle_{\varphi_{\bf p},\varphi_{{\bf
r}_1},\varphi_{{\bf r}_2}}= \frac{\sin\left[p\left(r_1\pm
r_2\right)+\pi/4\right]}{p(r_1r_2)^{1/2}}.
\end{eqnarray}
Upon combining rapidly oscillating terms in the integrand of
Eq.~(\ref{deltaG}) into ``slow''
terms, we obtain
\begin{eqnarray}
\label{dos} \delta\nu (\omega)=
\delta\nu^{(+)}(\omega)+\delta\nu^{(-)}(\omega),
\end{eqnarray}
where
\begin{eqnarray}
\label{nuPLUS}
\frac{\delta\nu^{(+)}(\omega)}{\nu_0}=-\frac{(\nu_0V)^3}{2E_{\text
{\tiny F}}\pi^{3/2}k_{\text {\tiny F}}^{1/2}}
\!\int_{r_2>r_1}\frac{d{r_1}d{r_2}}{(r_1r_2)^{3/2}}\nonumber\\
\times (r_1+r_2)^{1/2}\int_0^{\omega}\!\!d\Omega
\sin\left[v_{\text {\tiny F}}^{-1}(\omega-\Omega)(r_1+r_2)\right]
\nonumber\\\times\sin\left\{ 2\delta\varphi_{\Sigma}^{(+)}
+\frac{\pi}{4}-\frac{(\omega+\Omega)}{v_{\text {\tiny
F}}}(r_1+r_2)\right\},
\end{eqnarray}
and
\begin{eqnarray}
\label{nuMINUS}
\frac{\delta\nu^{(-)}(\omega)}{\nu_0}=-\frac{(\nu_0
V)^3}{2E_{\text {\tiny F}}\pi^{3/2}k_{\text {\tiny F}}^{1/2}}
\!\int_{r_2>r_1}\frac{d{r_1}d{r_2}}{(r_1r_2)^{3/2}}\nonumber\\
\times (r_2-r_1)^{1/2}\int_0^{\omega}\!\!d\Omega
\sin\left[v_{\text {\tiny F}}^{-1}(\omega-\Omega)(r_1+r_2)\right]
\nonumber\\\times\sin\left\{2 \delta\varphi_{\Sigma}^{(-)}
+\frac{\pi}{4}+\frac{(\omega+\Omega)}{v_{\text {\tiny
F}}}(r_2-r_1)\right\},
\end{eqnarray}
where we had assumed that the interaction is short-ranged and set
$V(0)=V(2k_{\text {\tiny F}})=\nu_0V$. Two contributions in
Eq.~(\ref{dos}) correspond to the locations of the points ${\bf
r}_1$ and ${\bf r}_2$ on the opposite and the same sides from the
origin, respectively, see Fig.~\ref{angle}b.

We note that the phases $\delta\varphi_{\Sigma}^{(+)}$,
$\delta\varphi_{\Sigma}^{(-)}$, which enter into the argument of
sine in Eqs.~(\ref{nuPLUS}), (\ref{nuMINUS}), are {\em quadratic}
in the random field, $h(x,y)$, as seen from Eqs.~(\ref{directly}),
(\ref{kernel}). This suggests that the averaging over realizations
of $h(x,y)$ can be carried out  analytically {\em in the
integrands} of Eqs.~(\ref{nuPLUS}), (\ref{nuMINUS}). Similarly to
the case of Friedel oscillations, it is convenient to perform this
averaging separately for the regimes I and II. This is done in
Sections~\ref{zeroI}, \ref{zeroII} below. In the remainder of this
Section we will evaluate the interaction correction,
$\delta\nu(\omega)$, for two particular cases: (i) constant
magnetic field, $h(x,y)\equiv h_0$, in a clean electron gas, and
(ii) $h(x,y)\equiv h_0$ in electron gas with small concentration
of  point-like impurities.

\subsection{Case of Constant Magnetic Field: Oscillations of $\delta\nu(\omega)$}
\begin{figure}[t]
\centerline{\includegraphics[width=85mm,angle=0,clip]{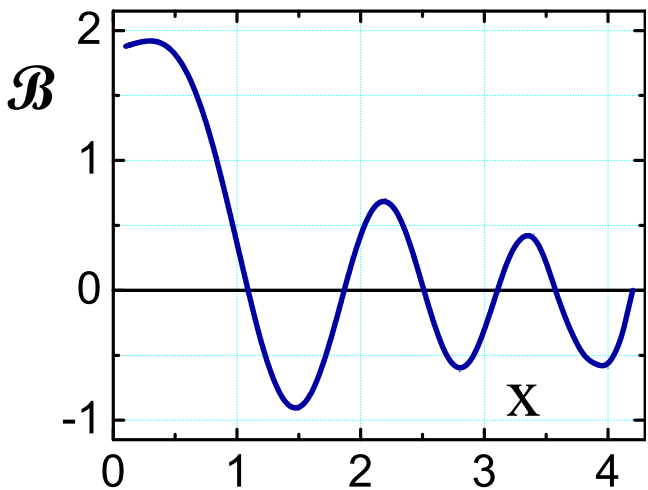}}
\caption{(Color online) Dimensionless correction Eq.~(\ref{calB})
to the tunnel density of states in a weak constant magnetic field
is plotted vs. dimensionless energy $x=2^{2/3}\omega/\omega_0$.
The plot is obtained upon numerical integration in
Eqs.~(\ref{nuPLUS}), (\ref{nuMINUS}). }
 \label{Oscillations}
\end{figure}

In a constant magnetic field $h(x,y)\equiv h_0$ the characteristic
scale of frequency in Eqs.~(\ref{nuPLUS}), (\ref{nuMINUS}) is
$\omega_0=v_{\text {\tiny F}}/r_{\text {\tiny I}}$. This was
stated in Section~\ref{main}. Now this scale of frequencies
emerges naturally upon substituting in Eqs.~(\ref{nuPLUS}),
(\ref{nuMINUS}) the phases $2\delta\varphi_{\Sigma}^{(+)}$,
$2\delta\varphi_{\Sigma}^{(+)}$, calculated from
Eq.~(\ref{directly}) in a constant magnetic field
\begin{eqnarray}
\label{twophases}
2\delta\varphi_{\Sigma}^{(\pm)}=\mp\frac{p_0^3}{4}r_1r_2(r_1\pm
r_2),
\end{eqnarray}
where $p_0$ is defined by Eq.~(\ref{p0}). The integrals in
Eqs.~(\ref{nuPLUS}), (\ref{nuMINUS}) converge at distances $r_1,
r_2 \sim p_0^{-1}=r_{\text {\tiny I}}$. As a result, $\delta
\nu^{(+)}$ and $\delta \nu^{(-)}$ are certain universal functions
of $\omega r_{\text {\tiny I}}/v_{\text {\tiny
F}}=\omega/\omega_0$. The plot of
$\delta\nu^{(+)}+\delta\nu^{(-)}$ vs. dimensionless ratio
$x=2^{2/3}\omega/\omega_0$ is presented in
Fig.~\ref{Oscillations}. To isolate the frequency dependence, in
addition to $x$, we had introduced the dimensionless variables
$r_1/r_{\text {\tiny I}}$ and $r_2/r_{\text {\tiny I}}$ after
which $\delta \nu(\omega)$ acquires the form
\begin{eqnarray}
\label{calB}
\frac{\delta\nu(\omega)}{\nu_0}=-\frac{(\nu_0V)^3}{2^{2/3} (\pi
k_{\text {\tiny F}}r_{\text {\tiny I}})^{3/2}}{\mathcal B}(x).
\end{eqnarray}
The integral over $\Omega$ in Eqs.~(\ref{nuPLUS}), (\ref{nuMINUS})
can be evaluated analytically. The remaining dimensionless double
integrals were calculated numerically. While the characteristic
scale, $x\sim 1$, of change of the function ${\mathcal B}(x)$
follows from qualitative consideration, Fig.~\ref{Oscillations}
indicates that ${\mathcal B}(x)$ also exhibits sizable
oscillations. These oscillations come only from the contribution
$\delta\nu^{(+)}$. They owe their existence to the peculiar
structure of the argument of sine in Eq.~(\ref{nuPLUS}). Namely,
this argument has saddle points  with respect to {\em both} $r_1$
and $r_2$ at $r_1=r_2=2^{1/3}r_{\text {\tiny
I}}(\omega/\omega_0)^{1/2}/3^{1/2}$ Oscillatory behavior of
${\mathcal B}(x)$ is governed by the value of the argument at the
saddle  point, which is $\sim (\omega/\omega_0)^{3/2}$. Strictly
speaking, the saddle point determines the value of the integral
only when $\omega \gg \omega_0$. However, numerics shows that
oscillations in Fig.~\ref{Oscillations}, set in starting already
from $x\sim 1$. These oscillations reflect the distinguished
contribution from the three-scattering process, shown in
Fig.~\ref{angle}b, in which scattering events occur at
$r_1=r_2=2^{1/3}r_{\text {\tiny
I}}(\omega/\omega_0)^{1/2}/3^{1/2}$.


Eq.~(\ref{calB}) and Fig.~\ref{Oscillations} constitute an
experimentally verifiable prediction. Correction Eq.~(\ref{calB})
describes the the feature in the tunneling conductance of a clean
two-dimensional electron gas  as a function of bias that emerges
in a weak magnetic field, $h_0$. It follows from prefactor in
Eq.~(\ref{calB}) that the magnitude of $\delta\nu$ scales with
$h_0$ as $r_{\text {\tiny I}}^{-3/2}\propto h_0$. We emphasize
that the correction $\delta\nu (\omega)$ remains distinguishable
even when the structure in the density of states due to the Landau
quantization is completely smeared out, {\em e.g.}, due to finite
temperature. This follows from the above relation between
$\omega_0$ and the cyclotron frequency, $\omega_c$, namely,
$\left(\omega_c/\omega_0\right) \sim \left(\omega_c/E_{\text
{\tiny F}}\right)^{1/3}\ll 1$.

In discussing the relevance to the experiment one should have in
mind that realistic samples always contain certain degree of
disorder. Therefore, the question remains as to whether the
oscillations of $\delta \nu(\omega)$ in a constant magnetic field
survive in the presence of the short-range impurities. This
question is non-trivial, since impurities themselves give rise to
the singular correction to $\delta \nu(\omega)$ (zero-bias
anomaly) even in a zero field. Then the above question can be
reformulated as: whether the field-induced oscillations are
distinguishable on the background of a zero-bias anomaly. It turns
out that, by introducing the Friedel oscillations, point-like
impurities actually enhance the oscillatory part of
$\delta\nu(\omega)$. This question is addressed in the next
subsection.

\begin{figure}[t]
\centerline{\includegraphics[width=75mm,angle=0,clip]{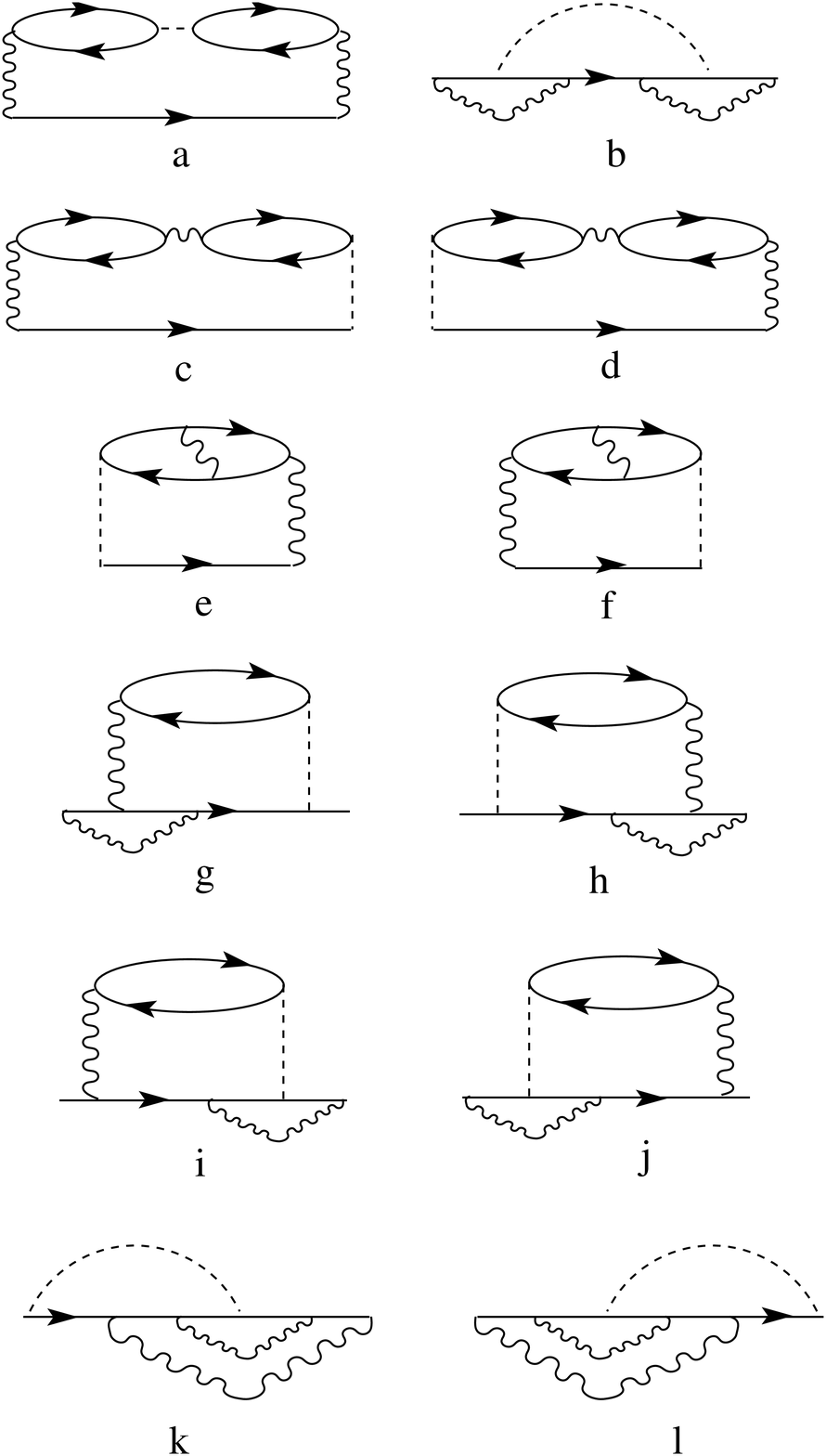}}
\caption{Second-order diagrams contributing to the oscillating
part (see Fig.~\ref{imurity_DOS}) of the ballistic zero-bias
anomaly in a weak {\em constant} magnetic field. Magnetic field
enters through the phases Eq.~(\ref{constantfield}) of the Green
functions. Dashed line represents the impurity scattering. All
$12$ diagrams (a)-(l) contain two static polarization operators.}
\label{diagrams_1}
\end{figure}

\subsection{Ballistic Zero-Bias Anomaly in a Constant Magnetic Field}

\begin{figure}[t]
\centerline{\includegraphics[width=75mm,angle=0,clip]{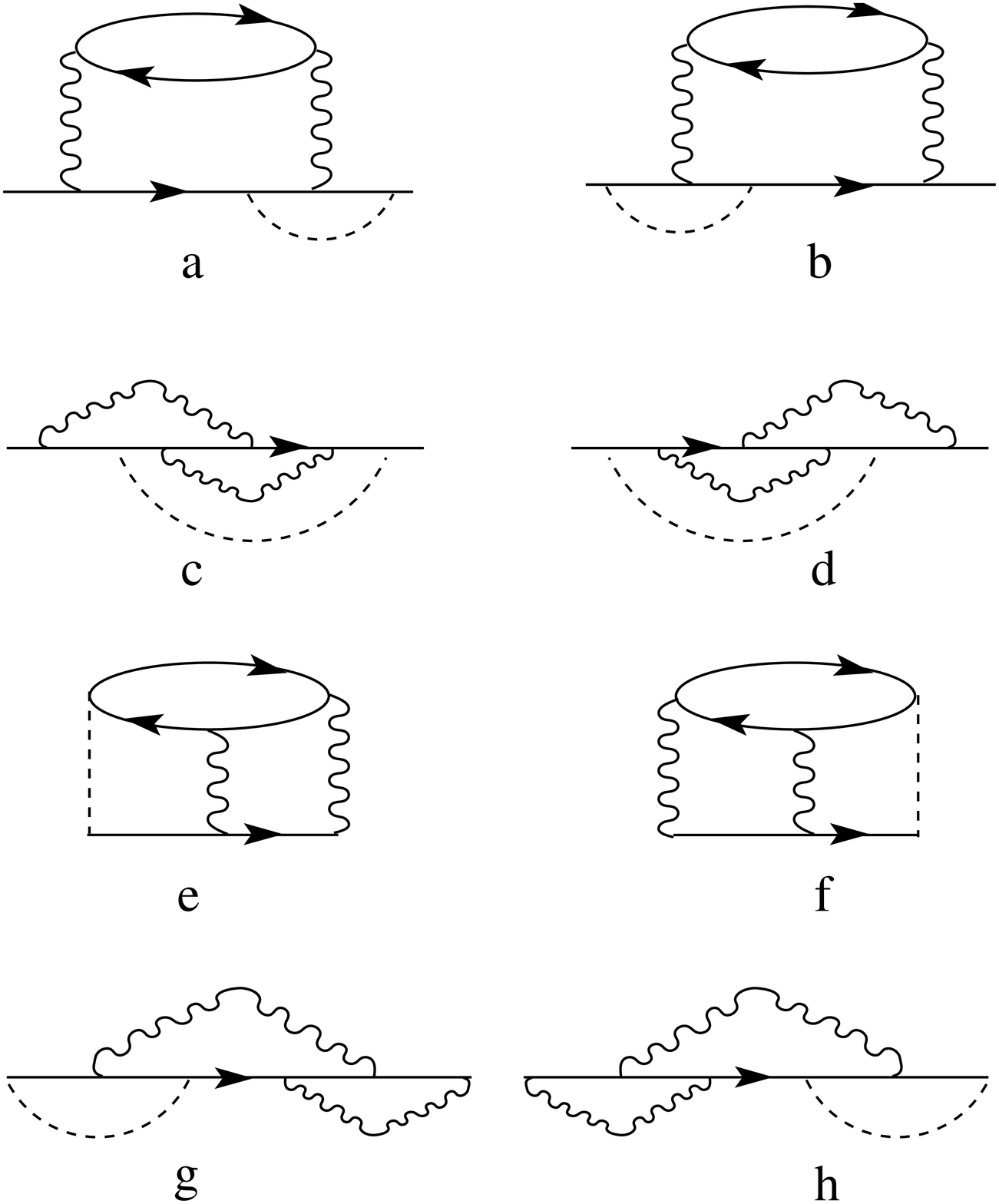}}
\caption{$8$ out of total $24$ second-order diagrams for ballistic
zero-bias anomaly in a weak {\em constant} magnetic field, which
contain {\em one} dynamic polarization operator.}
\label{diagrams_2}
\end{figure}

\begin{figure}[t]
\centerline{\includegraphics[width=75mm,angle=0,clip]{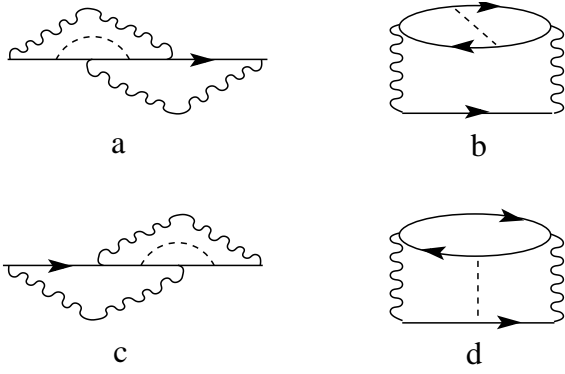}}
\caption{$4$ out of total $24$ second-order
 diagrams for the
ballistic zero-bias anomaly in a weak {\em constant} magnetic
field, which contain a polarization loop crossed by the impurity
line.} \label{diagrams_3}
\end{figure}

Conventional ballistic zero-bias anomaly\cite{rudin97}, caused by
point-like impurities, is described by two second-order diagrams,
shown Fig.~\ref{oneloop}, in which one of two interaction lines is
replaced by an impurity line. As was shown in
Ref.~\onlinecite{rudin97}, these diagrams with one interaction
line and one impurity line yield a singular correction, $\delta
\nu(\omega)/\nu_0 \sim \left(\nu_0V/E_{\text {\tiny
F}}\tau\right)\ln(\omega)$, to the density of states. Here
$\tau^{-1}$ is the scattering rate proportional to the impurity
concentration. Qualitatively, the singular correction originates
from the combined scattering of electron by the impurity and the
Friedel oscillation $\propto \sin(2k_{\text {\tiny F}}r)/r^2$,
created {\em by the same impurity}. This Friedel oscillation is
represented by the polarization loop in Fig.~\ref{oneloop}. In the
presence of the impurity, this loop describes {\em static}
response of the electron gas, and thus the polarization operator,
$\Pi_{2k_{\text {\tiny F}}}(\omega,r)$, corresponding to the loop
should be taken at $\omega=0$. As  was mentioned in
Section~\ref{main}, a weak perpendicular magnetic field, $h$,
leaves the logarithmic correction unchanged. To reveal the
sensitivity to $h$, one should calculate $\delta \nu$  to the next
(second) order in $V$. Corresponding diagrams with one impurity
and two interaction lines are shown in Figs.~\ref{diagrams_1},
\ref{diagrams_2}, and \ref{diagrams_3}. It is easy to see that
there are overall $24$ different diagrams. Indeed, the generalized
diagram, Fig.~\ref{two_loops}(a), for the third-order {\em
interaction} correction contains three generalized four-leg
vertices shown in Fig.~\ref{two_loops}(c). Hence,
Fig.~\ref{two_loops}~(a) represents $2^3=8$ different diagrams. In
each of these $8$ diagrams, the impurity line can replace
interaction line in three places, generating one of $24$ different
diagrams that are shown in Figs.~\ref{diagrams_1},
\ref{diagrams_2}, and \ref{diagrams_3}. All these diagrams are
divided into three groups according to their dependence on
$\omega$. Namely, {\em all} $12$ diagrams in Fig.~\ref{diagrams_1}
have the same $\omega$-dependence. Similarly, the
$\omega$-dependence of {\em all} $8$ diagrams in
Fig.~\ref{diagrams_2} is the same. This also applies to $4$
diagrams in Fig.~\ref{diagrams_3}. However, the corresponding
$\omega$-dependencies are slightly different from each other. The
origin of this difference can be traced from comparison of
diagrams Fig.~\ref{diagrams_1}~(a), Fig.~\ref{diagrams_2}~(a), and
Fig.~\ref{diagrams_3}~(b). Diagram Fig.~\ref{diagrams_1}~(a)
contains two polarization loops separated by the impurity line. As
a result, the expression corresponding to this diagram, contains
two {\em static} polarization operators, $\Pi_{2k_{\text {\tiny
F}}}(0,r)$. Diagram Fig.~\ref{diagrams_2}~(a) contains {\em one}
finite-$\omega$ polarization loop, $\Pi_{2k_{\text {\tiny
F}}}(\omega,r)$. Finally, the diagram Fig.~\ref{diagrams_3}~(b)
does not contain polarization operators at all, but rather a
different object, namely, a polarization loop {\em crossed} by the
impurity line. Important is that the expression, corresponding to
this object
\begin{eqnarray}
\label{G4}
&&\!\!\!\!\!\!\!\!\!\!\!\!{\prod}(\omega-\Omega,\vert{\bf
r}_1-{\bf
r}_2\vert)=-i\int\frac{d\Omega_1}{2\pi}G_{\Omega_1}(0,{\bf
r}_1)\nonumber\\
&&\!\!\!\!\!\!\!\!\!\!\!\!\!\times G_{\omega-\Omega+\Omega_1}({\bf
r}_1,0)G_{\omega-\Omega+\Omega_1}(0,{\bf r}_2) G_{\Omega_1}({\bf
r}_2,0),
\end{eqnarray}
contains a ``fast'' part, ${\prod}_{2k_{\text {\tiny
F}}}(\omega,r)$, which oscillates as $\exp\left(2ik_{\text {\tiny
F}}\vert{\bf r}_1-{\bf r}_2\vert\right)$, i.e., in the same way as
polarization operator.


The full analytical expression corresponding to the diagram
Fig.~\ref{diagrams_1}~(a) reads
\begin{eqnarray}
\label{constG1} \!\!\!\!\!\!\!&&\delta\nu_{1} (\omega,h)=
{\text{Im}}\frac{4V^2(2k_{\text {\tiny F}})}{2\pi^2 \nu_0
\tau}\int d{\bf r}d{\bf r}_1d{\bf
r}_2\;G_{\omega}({\bf r},{\bf r}_1)\nonumber\\
\!\!\!\!\!\!\!&& \times G_{\omega}({\bf r}_1,{\bf
r}_2)\Pi_{2k_{\text {\tiny F}}}(0,{\bf r}_1)\Pi_{2k_{\text {\tiny
F}}}(0,{\bf r}_2)G_{\omega}({\bf r}_2,{\bf r})\nonumber\\
&&={\text{Im}}\frac{6V^2(2k_{\text {\tiny F}})}{\pi^2 \nu_0
\tau}\int d{\bf r}_1\;d{\bf
r}_2\;\partial_{\omega}G_{\omega}({\bf r}_1,{\bf r}_2)\nonumber\\
\!\!\!\!\!\!\!&& \times G_{\omega}({\bf r}_1,{\bf
r}_2)\Pi_{2k_{\text {\tiny F}}}(0,{\bf r}_1)\Pi_{2k_{\text {\tiny
F}}}(0,{\bf r}_2),
\end{eqnarray}
where  in the second identity  we had performed integration over
${\bf r}$.

Analytical expression for the diagram Fig.~\ref{diagrams_2}~(a)
has the form
\begin{eqnarray}
\label{constG2} \!\!\!\!\!\!\!&&\delta\nu_{2} (\omega,h)=
-{\text{Im}}\frac{2V^2(2k_{\text {\tiny F}})}{2\pi^2\nu_0
\tau}\!\!\int\!\! d{\bf r}\;d{\bf r}_1d{\bf r}_2 \nonumber\\
&&\times\; G_{\omega}({\bf r},{\bf r}_1) G_{\omega}({\bf
r}_1,0)G_{\omega}({\bf r}_2,{\bf r})
\\
&&\times\int\frac{d\Omega}{2\pi}G_{\Omega} (0,{\bf
r}_1)G_{\Omega}({\bf r}_1,{\bf r}_2)\Pi_{2k_{\text {\tiny
F}}}(\omega-\Omega,\vert{\bf r}_1-{\bf r}_2\vert).\nonumber
\end{eqnarray}
Finally, the expression for the diagram \ref{diagrams_3}~(b) is
the following
\begin{eqnarray}
\label{constG3} &&\!\!\!\!\!\!\!\delta\nu_{3} (\omega,h)=
-{\text{Im}}\frac{2V^2(2k_{\text {\tiny F}})}{2\pi^2\nu_0
\tau}\!\!\int\!\! d{\bf r}\;d{\bf r}_1d{\bf r}_2 \;G_{\omega}({\bf
r},{\bf r}_1)G_{\omega}({\bf
r}_2,{\bf r})\nonumber\\
&&\times\int\frac{d\Omega}{2\pi}\int \frac{d\Omega_1}{2\pi}
G_{\Omega}({\bf r}_1,{\bf r}_2)G_{\Omega_1}(0,{\bf
r}_1)\nonumber\\&&\times G_{\omega-\Omega+\Omega_1}({\bf
r}_1,0)G_{\omega-\Omega+\Omega_1}(0,{\bf r}_2) G_{\Omega_1}({\bf
r}_2,0).
\end{eqnarray}
Upon integration over ${\bf r}$, it can be expressed through
${\prod}_{2k_{\text {\tiny F}}}(r)$, defined by Eq.~(\ref{G4}), as
\begin{eqnarray}
\label{GG3} &&\!\!\!\!\!\!\!\delta\nu_{3}
(\omega,h)=-\frac{V^2(2k_{\text {\tiny F}})}{\pi^4\nu_0
\tau}\!\!\int\!\!d{\bf r}_1d{\bf
r}_2\;\partial_{\omega}{\text{Im}}G_{\omega}({\bf r}_1,{\bf
r}_2)\nonumber\\
&&\times\int_0^{\omega} \frac{d\Omega}{2\pi}
{\text{Im}}G_{\Omega}({\bf r}_1,{\bf
r}_2){\text{Im}}\;{\prod}_{2k_{\text {\tiny
F}}}(\omega-\Omega,\vert{\bf r}_1-{\bf
r}_2\vert).\nonumber\\
\end{eqnarray}

Despite all $12$ diagrams in Fig.~\ref{diagrams_1} have the same
frequency dependence, their prefactors represent different
combinations of $V^2(2k_{\text {\tiny F}})$, $V^2(0)$, and
$V(2k_{\text {\tiny F}})V(0)$. The same applies to $8$ diagrams in
Fig.~\ref{diagrams_2} and to $4$ diagrams in
Fig.~\ref{diagrams_3}. Taking into account the numerical factors
in these combinations amounts to the following replacements: in
$\delta\nu_1$
\begin{eqnarray}
\label{truefactor1}
 4 V^2(2k_{\text {\tiny
F}}) \rightarrow  3V^2(0),
\end{eqnarray}
in $\delta\nu_2$
\begin{eqnarray}
\label{truefactor2} -2V^2(2k_{\text {\tiny F}}) \rightarrow
4\Bigl[V(0) V(2k_{\text {\tiny
F}})-V^2(2k_{\text {\tiny F}})-V^2(0)\Bigr],\nonumber\\
\end{eqnarray}
and in $\delta\nu_3$
\begin{eqnarray}
\label{truefactor3} -2V^2(2k_{\text {\tiny F}})  \rightarrow
2\Bigl[V(0) V(2k_{\text {\tiny
F}})-V^2(2k_{\text {\tiny F}})-V^2(0)\Bigr].\nonumber\\
\end{eqnarray}
These replacements must be taken into account when calculating the
full correction $\delta\nu(\omega)$ from $\delta\nu_1$,
$\delta\nu_2$, and $\delta\nu_3$.

Below we demonstrate that all three contributions $\delta\nu_1$,
$\delta\nu_2$, and $\delta\nu_3$ are {\em oscillatory} functions
of $\omega$. Detailed derivation will be presented only for
$\delta\nu_1$.

Analogously to the derivation of Eqs.~(\ref{nuPLUS}),
(\ref{nuMINUS}), we can perform the integration over the azimuthal
angles of ${\bf r}_1$ and ${\bf r}_2$ analytically using
Eq.~(\ref{int-azimut}). Then, extracting a ``slow'' term from the
product of trigonometrical functions, we obtain
$\delta\nu_1(\omega)= \left[{6\nu_0^3V^2(2k_{\text {\tiny
F}})}/{E_{\text {\tiny F}}\tau}\right] \left(\omega_0/E_{\text
{\tiny F}}\right)^{1/2} \text{\large
P}_1(2^{2/3}\omega/\omega_0)$, with
$\omega_0=2^{1/3}\omega_c^{2/3}E_{\text {\tiny F}}^{1/3}$, where
the function $\text{\large P}_1(x)$ is defined as
\begin{eqnarray}
\label{Pfunction} \text{\large P}_1(x)=\text{\large
P}^{+}_1(x)+\text{\large P}^{-}_1(x),
\end{eqnarray}
where
\begin{eqnarray}
\label{formulaP+} &&\text{\large P}^{+}_1(x)=\sigma
\int_{\rho_2>\rho_1}\frac{d\rho_1d\rho_2}{(\rho_1\rho_2)^{3/2}}\Biggl\{
(\rho_1+\rho_2)^{1/2}\nonumber\\
&&\times\Bigl\{\cos\left[x(\rho_1+\rho_2)-\frac{\pi}{4}-\rho_1\rho_2(\rho_1+\rho_2)\right]\nonumber\\
&&-\cos\left[x(\rho_1+\rho_2)-\frac{\pi}{4}\right]\Bigr\}\Biggr\},
\end{eqnarray}
\begin{eqnarray}
\label{formulaP-} &&\text{\large P}^{-}_1(x)=-\sigma
\int_{\rho_2>\rho_1}\frac{d\rho_1d\rho_2}{(\rho_1\rho_2)^{3/2}}
\Biggl\{(\rho_2-\rho_1)^{1/2}\nonumber\\
&&\times\Bigl\{\cos\left[x(\rho_2-\rho_1)+\frac{\pi}{4}+\rho_1\rho_2(\rho_2-\rho_1)\right]\nonumber\\
&&-\cos\left[x(\rho_2-\rho_1)+\frac{\pi}{4}\right]\Bigr\}\Biggr\}.
\end{eqnarray}
Here the constant factor, $\sigma$, is given by $\sigma =(3 \cdot
2^{1/6})/\pi^{3/2}$. In Appendix~\ref{AppendixD1} we demonstrate
how the function $\text{\large P}_1(x)$ can be cast in the form
that is convenient for numerical evaluation and extracting
asymptotes. This form is given by the following double integral
\begin{eqnarray}
\label{last} \text{\large P}_1(x)=4\sigma
\int_0^{\infty}\frac{dz}{z^{3/2}}
\int_{-4}^0\frac{dv}{\sqrt{v+4}}\qquad\qquad\\
\times\Biggl(\cos\left[xz+\frac{\pi}{4}+\frac{z^3}{v}\right]
-\cos\left[xz+\frac{\pi}{4}\right]\Biggr).\nonumber
\end{eqnarray}
The fact that $\text{\large P}_1(x)$ oscillates at large $x\gg 1$
follows from the observations that (i) first cosine in the
brackets in Eq.~(\ref{last}) has a saddle point $z=(x\vert
v\vert/3)^{1/2}$, and (ii) the major contribution to the integral
over $v$ comes from the lower limit $v=-4$ (corresponding steps
are outlined in Appendix~\ref{AppendixD}). This yields
\begin{eqnarray}
\label{xgg1} \text{\large P}_1(x)\mbox{\Large$|$}_{x\gg
1}=\frac{2^{5/3}3^{9/4}}{\pi^{1/2}}\;\frac{1}{x^{7/4}}
\sin\left[4\left(\frac{x}{3}\right)^{3/2}+\frac{\pi}{4}\right].\nonumber\\
\end{eqnarray}
The argument $x^{3/2}$ in the cosine in Eq.~(\ref{xgg1}) can be
presented as $\omega^{3/2}/\left(2^{1/2}\omega_cE_{\text {\tiny
F}}^{1/2}\right)$, so that the ``period'' in $\omega$ is much
bigger than the cyclotron energy, $\omega_c$, as was discussed
above.

The analysis of the contributions $\delta\nu_2(\omega)$ and
$\delta\nu_3(\omega)$ can be carried out in a similar way. They
exhibit the same oscillations as Eq.~(\ref{xgg1}). The difference
is that, due to integration over $\Omega$ in Eqs.~(\ref{constG2})
and (\ref{GG3}), both $\delta\nu_2(\omega)$ and
$\delta\nu_3(\omega)$ contain an extra factor $\omega/\omega_0$,
see Eq.~ (\ref{calG}), and thus their contribution to the net
correction $\delta\nu$ is dominant at $\omega \gg \omega_0$.

\section{ Zero-bias anomaly in the averaged density of states in regime I }
\label{zeroI}

With the help of the identity Eq.~(\ref{AVERAGING}) the integrand
in the average $\delta\nu(\omega)$ can be expressed in terms of
functions $U_{1,2}\left[r_1r_2(r_2\pm r_1)p_0^3/4\right]$, where
the functions $U_{1,2}$ are defined as

\begin{eqnarray}
\label{U1} U_{1}(x)=
\Biggl(\frac{\pi}{2}\Biggr)^{1/2}\sqrt{\frac{(1+x^2)^{1/2}+
1}{1+x^2}},
\end{eqnarray}

\begin{eqnarray}
\label{U2} U_{2}(x)=
\Biggl(\frac{\pi}{2}\Biggr)^{1/2}\sqrt{\frac{(1+x^2)^{1/2}-
1}{1+x^2}}.
\end{eqnarray}

\begin{figure}[t]
\centerline{\includegraphics[width=85mm,angle=0,clip]{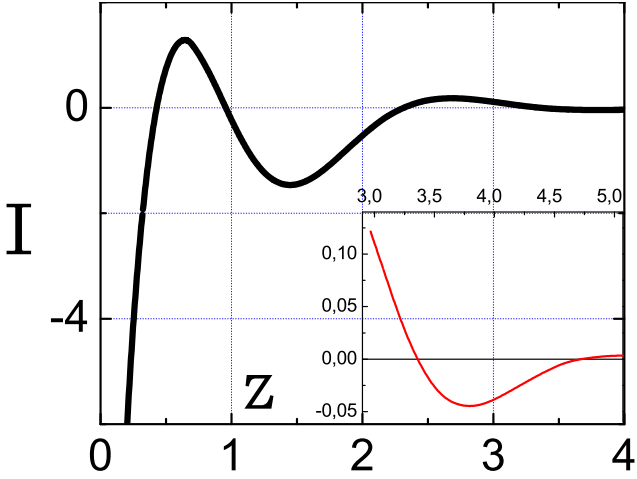}}
\caption{(Color online) Dimensionless function, ${\mathcal I}(z)$,
describing the shape of a zero-bias anomaly in regime I, is
plotted from Eq.~(\ref{averaged}) versus dimensionless energy,
$z=\omega/\omega_0$. Inset in the lower-right corner:
 enlarged plot of ${\mathcal I}(z)$ in the domain $3<z<5$.}
 \label{RegimeI}
\end{figure}

Upon introducing dimensionless variables
$\rho_{1,2}=p_0r_{1,2}/2^{2/3}$, we present the final result in
the form
\begin{eqnarray}
\label{CALI} \frac{\delta \nu(\omega)}{\nu_0}={\mathcal C}\;
{\mathcal I}\left(\frac{\omega}{\omega_0}\right),
\end{eqnarray}
 with
\begin{eqnarray}
\label{wm}
\omega_0=v_{\text{\tiny{F}}}p_0=2E_{\text{\tiny{F}}}\left(\frac{h_0}{k_{\text{\tiny{F}}}^2\Phi_0}\right)^{2/3},
\end{eqnarray}
and with constant, ${\mathcal C}$, defined as
\begin{eqnarray}
\label{A} {\mathcal
C}=-\frac{(\nu_0V)^3}{2\pi}\left(\frac{h_0}{k_{\text{\tiny{F}}}^2\Phi_0}\right)=
-\frac{(\nu_0V)^3}{4\sqrt{2}\pi}\left(\frac{\omega_0}{E_{\text{\tiny{F}}}}\right)^{3/2}.
\end{eqnarray}
The dimensionless function, ${\mathcal I}(z)$, describing the
shape of the anomaly, is given by the following double integral
over $\rho_1$, $\rho_2$
\begin{eqnarray}
\label{averaged} {\mathcal I}(z)= {\mathcal I}^{+}(z)+{\mathcal I}^{-}(z)\qquad\qquad\qquad\qquad\\
=\int\limits_{\rho_2>\rho_1}\!
\frac{d{\rho_1}d{\rho_2}}{(\rho_1\rho_2)^{3/2}}\!\!\int\limits_0^z\!dz^{\prime}
\sin\Bigl[(z-z^{\prime})(\rho_1+\rho_2)\Bigr]\qquad\nonumber\\
\times\Bigl\{S_{+}(\rho_1,\rho_2)\!+\!C_{+}(\rho_1,\rho_2)\!+\!S_{-}(\rho_1,\rho_2)\!+
\!C_{-}(\rho_1,\rho_2)\Bigr\},\nonumber
\end{eqnarray}
where the functions $S_{+}$, $S_{-}$, $C_{+}$, and $C_{-}$ are
defined as
\begin{eqnarray}
\label{functions1}
S_{\pm}(\rho_1,\rho_2)={(\rho_1\pm\rho_2)^{1/2}}
\sin\left[\frac{\pi}{4}\mp (z+z^{\prime})(\rho_1\pm \rho_2)\right]\nonumber\\
\times\Biggl\{ U_1\biggl(\rho_1\rho_2(\rho_1 \pm \rho_2)
\biggr)-\sqrt{\pi}\Biggr\},\qquad\qquad\qquad\qquad\\
C_{\pm}(\rho_1,\rho_2)={(\rho_1\pm\rho_2)^{1/2}}
\cos\left[\frac{\pi}{4} \mp (z+z^{\prime})(\rho_1\pm \rho_2)\right]\nonumber\\
\times
U_2\biggl(\rho_1\rho_2(\rho_1\mp\rho_2)\biggr).\qquad\qquad\qquad\qquad\qquad\qquad
\end{eqnarray}
In definitions of $S_{+}$ and $S_{-}$ we had subtracted from the
function $U_1(\alpha)$ the zero-field value $U_1(0)=\sqrt{\pi}$.
Integration over $z^{\prime}$ in Eq.~(\ref{averaged}) can be
easily carried out analytically. The remaining integrals over
$\rho_1$, $\rho_2$ were evaluated numerically. Direct numerical
integration encounters difficulties due to very fast oscillations
of the integrand in Eq.~(\ref{averaged}). These difficulties can
be overcome by a proper change of variables in the integrand. This
procedure is described in Appendix D. The resulting shape of the
zero-bias anomaly is shown in Fig.~\ref{RegimeI}. The small-$z\ll
1$ behavior of  ${\mathcal I}(z)$ is $8\ln z$, i.e., it diverges
logarithmically. The cutoff is chosen from the condition that
${\mathcal I}(z)$ approaches zero at large $z$. Note, that
${\mathcal I}(z)$ exhibits a pronounced feature around $z=1$. The
origin of this feature lies in strong oscillations of the
integrand in Eq.~(\ref{dos}). The ``trace'' of these oscillations
{\em survives} after averaging over the magnitude of the random
field. In fact, the oscillations persist beyond $z=3$. This is
reflected in the $z\gg 1$ asymptote of the function ${\mathcal
I}(z)$,
\begin{eqnarray}
\label{oscillation} {\mathcal I}^{+}(z)\mbox{\Large$|$}_{z\gg
1}\approx-2^{3/4}\sqrt{\pi}\;\frac{\sin\left(2^{8/3}\sqrt{3}z\right)}{z^{3/4}}\exp\left\{-2^{8/3}z\right\}.\;\;\nonumber\\
\end{eqnarray}
To derive this asymptote, it is more convenient to first take the
limit of large $\omega$ in Eq.~(\ref{dos}) and perform the
averaging over the random field only {\em as a last step}. In the
limit $\omega \gg \omega_0$ following simplifications of Eq.
(\ref{dos}) become possible. Firstly, the second term in the
square brackets can be neglected, since it does not produce
oscillatory contribution to $\delta\nu$. Secondly, one can set
$\Omega=0$ in the integrand, so that the integration over $\Omega$
reduces to multiplying by $\omega$. Lastly, upon converting the
product of sines into the difference of cosines, one finds that
the $\omega$-dependence is present only in the term, corresponding
to the difference of arguments. As a result, the oscillatory part
of $\delta\nu(\omega)$ at $\omega \gg \omega_0$ acquires the form
\begin{eqnarray}
\label{FORM} &&\!\!\!\!\!\!\left\langle\frac{\delta\nu
(\omega)}{\nu_0}\right\rangle\!=\!-\frac{
(\nu_0V)^3\omega\omega_0^{1/2}}{2^{15/6}\pi^{3/2}E_{\text{\tiny{F}}}^{3/2}}
\!\!\!\int\limits_{\rho_2>\rho_1}\!\!
\!\frac{d{\rho_1}d{\rho_2}}{(\rho_1\rho_2)^{3/2}}
(\rho_1+\rho_2)^{1/2}\nonumber\\
&&\!\!\!\!\!\!\Biggl\langle\!\left(\frac{h}{h_0}\right)
\cos\Biggl[\rho_1\rho_2(\rho_1+\rho_2)+\frac{\pi}{4}-\frac{2^{5/3}\omega}{\omega_0}\left(\frac{h_0}{h}\right)^{2/3}\nonumber\\
&&\;\;\;\;\;\;\;\;\;\;\;\;\;\;\;\;\;\;\times(\rho_1+\rho_2)\Biggr]\!\Biggr\rangle_{h(x,y)}.
\end{eqnarray}
The steps leading from this expression to the asymptote
Eq.~(\ref{oscillation}) are outlined in Appendix E.

\section{Zero-bias anomaly in averaged density of states in regime II}
\label{zeroII}
\subsection{Three polarization operators: Averaging of
the net magnetic phase factor over realizations of random magnetic
field}

\vspace{3mm}

To derive analytical expressions for $\delta\nu^{(+)}(\omega)$ and
$\delta\nu^{(-)}(\omega)$ one has to perform  averaging of
Eqs.~(\ref{nuPLUS}), (\ref{nuMINUS}) over realizations of the
random field. Such an averaging has already been carried out for
the Friedel oscillations. In the latter case we had averaged
$\langle\exp\left(2i\delta\varphi_{0\rightarrow {\bf
r}}\right)\rangle$. In the case of the density of states, the
exponents to be averaged are
$\langle\exp\left(2i\delta\varphi_{\Sigma}^{(\pm)}\right)\rangle$,
defined by  Eqs.~(\ref{product1}), (\ref{product2}). Our most
important observation is that the {\em net} phase
$\delta\varphi_{\Sigma}^{(-)}=\delta\varphi_{0\rightarrow {\bf
r}_1}+\delta\varphi_{{\bf r}_1\rightarrow {\bf r}_2}+
\delta\varphi_{{\bf r}_2\rightarrow 0}$ {\em does not} contain
integrals of $\Lambda ^2(x)$, since they {\em cancel out}. This
can be clearly seen from Eq.~(\ref{dphi}). Instead,
$\delta\varphi_{\Sigma}^{(-)}$ is expressed via integrals of
$\Lambda(x)$ {\em in the first power} as follows
\begin{eqnarray}
\label{combination}
\delta\varphi_{\Sigma}^{(-)}=\frac{1}{\Phi_0^2k_{\text{\tiny{F}}}}\Biggl[\frac{1}{r_1}
\left(\int_0^{r_1}dx\Lambda(x)\right)^2
+\frac{1}{r_2-r_1}\\
\times\left(\int_{r_1}^{r_2}dx\Lambda(x)\right)^2
-\frac{1}{r_2}\left(\int_0^{r_2}dx\Lambda
(x)\right)^2\Biggr].\nonumber
\end{eqnarray}
This cancellation, as we demonstrate below, has a dramatic
consequence for the average
$\langle\exp(i\delta\varphi_{\Sigma})\rangle$. It turns out that,
while $\langle\exp(i\delta\varphi_{0\rightarrow {\bf r}})\rangle$
decays with $r$ {\em exponentially}, the average
$\langle\exp(i\delta\varphi_{\Sigma})\rangle$ falls off only as
{\em a power law}. This, in turn, leads to a slow decay of a
zero-bias anomaly, $\delta\nu(\omega/\omega_1)$, with $\omega$.

On the technical level, cancellation of $\int dx \Lambda^2(x)$
terms
 leads to a drastic simplification of the disorder
averaging of Eqs.~(\ref{nuPLUS}), (\ref{nuMINUS}) in the regime
II, as compared to the averaging of the Friedel oscillations in
Section~\ref{FOII}, since the averaging of
$\exp(2i\delta\varphi_{\Sigma})$ can be performed with the help of
the Hubbard-Stratonovich transformation. For the purpose of
functional averaging, it is convenient to rewrite
Eq.~(\ref{combination}) in a slightly different form
\begin{eqnarray}
\label{different}
\delta\varphi_{\Sigma}^{(-)}=\frac{1}{\Phi_0^2k_{\text{\tiny{F}}}(r_2-r_1)}
\Biggl[\sqrt{\frac{r_2}{r_1}}\int_0^{r_1}dx\Lambda(x)\\-
\sqrt{\frac{r_1}{r_2}}\int_0^{r_2}dx\Lambda(x)\Biggr]^2.\nonumber
\end{eqnarray}
Subsequent integration by parts yields the further simplification
of Eq.~(\ref{different})
\begin{eqnarray}
\label{further}
\delta\varphi_{\Sigma}^{(-)}=\frac{1}{\Phi_0^2k_{\text{\tiny{F}}}(r_2-r_1)}
\Biggl[\sqrt{\frac{r_2}{r_1}}\int_0^{r_1}dx\;(r_1-x)h(x,0)\nonumber\\
-\sqrt{\frac{r_1}{r_2}}\int_0^{r_2}dx\;(r_2-x)h(x,0)\Biggr]^2.\qquad\qquad
\end{eqnarray}
Now the averaging over realizations of $h(x,y)$ can be performed
by a sequence of standard steps outlined below.

\subsubsection{Averaging procedure}

Using Eq.~(\ref{combination}) we rewrite the definition of average
$\bigl\langle\exp(2i\delta\varphi_{\Sigma})\bigr\rangle$ by
introducing the auxiliary integration variable, $c$
\begin{widetext}
\begin{eqnarray}
\label{cc}
\Bigl\langle\exp\{2i\delta\varphi_{\Sigma}^{(-)}\}\Bigr\rangle=
\int_{-\infty}^{\infty}\!dc\;\exp\left(-ic^2\right)
\Biggl\langle\delta\Biggl(c-\frac{\sqrt{2}}{\Phi_0k_{\text{\tiny{F}}}^{1/2}\sqrt{r_2-r_1}}
\left[\sqrt{\frac{r_2}{r_1}}\int_0^{r_1}dx\;(r_1-x)h(x,0)
\right.\nonumber\\
\left.-\sqrt{\frac{r_1}{r_2}}\int_0^{r_2}dx\;(r_2-x)h(x,0)\right]\Biggr)\Biggr\rangle_{h(x,y)},
\end{eqnarray}
where the averaging $\langle\dots\rangle_{h(x,y)}$ is defined by
Eq.~(\ref{functional}).
Next we use the following integral representation of the
$\delta$-function in Eq.~(\ref{cc})
\begin{eqnarray}
\label{delta} \Bigl\langle
\exp\{2i\delta\varphi_{\Sigma}^{(-)}\}\Bigr\rangle=
\int_{-\infty}^{\infty}\!dc\;\exp\left(-ic^2\right)
\int_{-\infty}^{\infty}\!\frac{dt}{2\pi}e^{ict}
\Biggl\langle\exp\Biggl\{-it\sqrt{2}\left[\sqrt{\frac{r_2}{r_1}}\int_0^{r_1}\frac{dx\;(r_1-x)
h(x,0)}
{\Phi_0k_{\text{\tiny{F}}}\sqrt{r_2-r_1}}\right.\nonumber\\
\left.
-\sqrt{\frac{r_1}{r_2}}\int_0^{r_2}\frac{dx\;(r_2-x)h(x,0)}{\Phi_0k_{\text{\tiny{F}}}\sqrt{r_2-r_1}}
\right]\Biggr\}\Biggr\rangle_{h(x,y)}.
\end{eqnarray}
Now the integration over $c$ can be performed explicitly, yielding
\begin{eqnarray}
\label{overc} \Bigl\langle
\exp\{2i\delta\varphi_{\Sigma}^{(-)}\}\Bigr\rangle=
\sqrt{\frac{\pi}{2}}e^{-i\pi/4}
\int_{-\infty}^{\infty}\!\frac{dt}{2\pi}e^{it^2/4}
\Biggl\langle\exp\Biggl\{-it\sqrt{2}\left[\sqrt{\frac{r_2}{r_1}}\int_0^{r_1}\frac{dx\;(r_1-x)
h(x,0)}
{\Phi_0k_{\text{\tiny{F}}}\sqrt{r_2-r_1}}\right.\nonumber\\
\left.
-\sqrt{\frac{r_1}{r_2}}\int_0^{r_2}\frac{dx\;(r_2-x)h(x,0)}{\Phi_0k_{\text{\tiny{F}}}\sqrt{r_2-r_1}}
\right]\Biggr\}\Biggr\rangle_{h(x,y)}.
\end{eqnarray}
\end{widetext}
It follows from Eq.~(\ref{overc}) that evaluation of $\Bigl\langle
\exp\{2i\delta\varphi_{\Sigma}^{(-)}\}\Bigr\rangle$ reduces to the
Gaussian averaging of the exponent of a  {\em linear} in $h(x)$
functional, which is standard
\begin{eqnarray}
\label{standard}
&&\!\!\!\!\!\!\!\Bigg\langle\!\exp\left\{-it\int_0^{r_2}dx\int_{-\infty}^{\infty}dy
h(x,y)f(x)\delta(y)\right\}\!\Bigg\rangle_{h(x,y)}\!\!\!\!\!\!=\\
&&\!\!\!\!\exp\left\{-\frac{t^2}{4}\int_0^{r_2}dx_1\int_0^{r_2}dx_2f(x_1)K(x_1,0,x_2,0)f(x_2)\right\},\nonumber
\end{eqnarray}
where  $K(x_1,0,x_2,0)$  is related to the correlator of the
random field Eq.~(\ref{correlator}) as follows
$K(x_1,0,x_2,0)=h_0^2\text{\large K}(|x_1-x_2|/\xi)$. Subsequent
integration over $t$ yields the final result
\begin{eqnarray}
\label{overt1} \Bigl\langle \exp\{2i\delta\varphi_{\Sigma}^{(-)}\}\Bigr\rangle\qquad\qquad\qquad\qquad\qquad\qquad\qquad\;\;\;\\
=\frac{1}{\sqrt{1+i\int_0^{r_2} \int_0^{r_2} dx_1dx_2
f(x_1)\text{K}(x_1,0,x_2,0)f(x_2)}}.\nonumber
\end{eqnarray}
As seen from Eq.~(\ref{overc}) the  function $f(x)$ in
Eq.~(\ref{standard}) has the form
\begin{eqnarray}
\label{f-}
f_{-}(x)&=&\frac{\sqrt{2}}{\Phi_0k_{\text{\tiny{F}}}^{1/2}\sqrt{r_2-r_1}}\Biggl[\sqrt{\frac{r_2}{r_1}}(r_1-x)\theta(r_1-x)\nonumber\\
&-&\sqrt{\frac{r_1}{r_2}}(r_2-x)\Biggr].
\end{eqnarray}
Averaging of $\exp\left\{i\delta\varphi_{\Sigma}^{(+)}\right\}$ is
performed similarly, and also yields Eq.~(\ref{standard}) with
$f(x)$ having the form
\begin{eqnarray}
\label{f+}
f_{+}(x)&=&\frac{\sqrt{2}}{\Phi_0k_{\text{\tiny{F}}}^{1/2}\sqrt{r_2}}\Biggl[\sqrt{\frac{r_1+r_2}{r_1}}(r_1-x)\theta(r_1-x)\nonumber\\
&-&\sqrt{\frac{r_1}{r_1+r_2}}(r_1+r_2-x)\Biggr].
\end{eqnarray}
We emphasize that expression Eq.~(\ref{overt1}) is {\em general},
and is valid for arbitrary, $h_0$ and $\xi$, i.e., in both regimes
I and II. For the regime I, we had already performed the averaging
over realizations of the random field. With regard to
Eq.~(\ref{overt1}), regime I corresponds to replacement of the
correlator {\em by unity}. In regime II, the distances $r_1, r_2$
are much larger than $\xi$.
For this reason, in regime II, the correlator in
Eq.~(\ref{overt1}) can be replaced by
$\sqrt{2\pi}\gamma\xi\delta(x_1-x_2)$, with $\gamma$ defined by
Eq.~(\ref{replace}). Then the averages
$\langle\exp\{2i\delta\varphi_{\Sigma}^{(-)}\}\rangle$ and
$\langle\exp\{2i\delta\varphi_{\Sigma}^{(+)}\}\rangle$ can be
expressed in terms of dimensionless ratios
\begin{eqnarray}
\label{RATIOS}
\varrho_1=\frac{r_1}{\sqrt{6}r_{\text{\tiny{II}}}},~~~~\varrho_2=\frac{r_2}{\sqrt{6}r_{\text{\tiny{II}}}},
\end{eqnarray}
where the characteristic length, $r_{\text{\tiny{II}}}$, is
defined by Eq.~(\ref{r0}).

Eq.~(\ref{overt1}) and analogous expression for
$\langle\exp\{2i\delta\varphi_{\Sigma}^{(+)}\}\rangle$ are
sufficient to perform the averaging over realizations of random
magnetic field in Eqs.~(\ref{nuPLUS}),  (\ref{nuMINUS}). However,
{\em averaged} Eqs.~(\ref{nuPLUS}),  (\ref{nuMINUS}) contain the
real and imaginary parts
\begin{eqnarray}
\label{REIM}
\langle\exp\{2i\delta\varphi_{\Sigma}^{(\pm)}\}\rangle={\mathcal
U}_{1}^{\pm}(\varrho_1,\varrho_2) +i~{\mathcal
U}_{2}^{\pm}(\varrho_1,\varrho_2)
\end{eqnarray}
of the average exponents, {\em separately}. The expressions for
${\mathcal U}_{1}^{\pm}$ and ${\mathcal U}_{2}^{\pm}$ readily
follow
after replacing correlator by delta-function and performing
integrations over $x_1$ and $x_2$ in  Eq.~(\ref{overt1})
\begin{eqnarray}
\label{DIMENSIONLESS1-} {\mathcal
U}_{1}^{-}\!=\!\frac{\sqrt{\sqrt{\varrho_1^2(\varrho_2\!-\!\varrho_1)^2\!+\!1}\!+\!1}\!+\!
\sqrt{\sqrt{\varrho_1^2(\varrho_2\!-\!\varrho_1)^2\!+\!1}\!-\!1}}{\sqrt{2}\sqrt{\varrho_1^2(\varrho_2-\varrho_1)^2+1}}\!,
\nonumber\\
\end{eqnarray}
\begin{eqnarray}
\label{DIMENSIONLESS2-} {\mathcal
U}_{2}^{-}\!=\!\frac{\sqrt{\sqrt{\varrho_1^2(\varrho_2\!-\!\varrho_1)^2\!+\!1}\!+\!1}\!-\!
\sqrt{\sqrt{\varrho_1^2(\varrho_2\!-\!\varrho_1)^2\!+\!1}\!-\!1}}{\sqrt{2}\sqrt{\varrho_1^2
(\varrho_2-\varrho_1)^2+1}}\!,
\nonumber\\
\end{eqnarray}
\begin{eqnarray}
\label{DIMENSIONLESS1+} {\mathcal
U}_{1}^{+}=\frac{\sqrt{\sqrt{\varrho_1^2\varrho_2^2+1}+1}+
\sqrt{\sqrt{\varrho_1^2\varrho_2^2+1}-1}}{\sqrt{2}\sqrt{\varrho_1^2\varrho_2^2+1}}\!,
\nonumber\\
\end{eqnarray}

\begin{eqnarray}
\label{DIMENSIONLESS2+} {\mathcal
U}_{2}^{+}=\frac{\sqrt{\sqrt{\varrho_1^2\varrho_2^2+1}+1}-
\sqrt{\sqrt{\varrho_1^2\varrho_2^2+1}-1}}{\sqrt{2}\sqrt{\varrho_1^2\varrho_2^2+1}}.
\nonumber\\
\end{eqnarray}
\begin{figure}[t]
\centerline{\includegraphics[width=85mm,angle=0,clip]{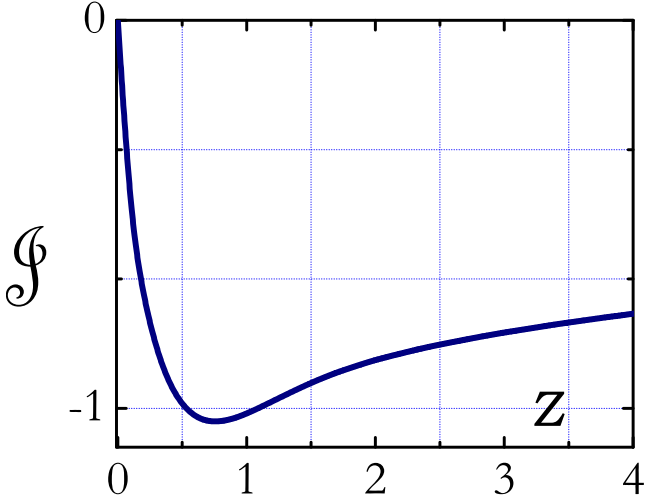}}
\caption{(Color online) Dimensionless density of states in regime
II, ${\mathcal J}(z)={\mathcal J}^{+}(z)+{\mathcal J}^{-}(z)$, is
plotted in the units of $(\nu_0{\mathcal D})$ from
Eqs.~(\ref{cal1}), (\ref{cal2}), (\ref{J1+}), (\ref{J2+}) versus
dimensionless frequency, $z=\omega/\omega_1$, where $\omega_1$ is
defined by Eq.~(\ref{OMEGA1}) and ${\mathcal D}$ is defined by
Eq.~(\ref{B}) .} \label{RegimeII}
\end{figure}
 Final expressions for the contributions $\langle \delta\nu^{-}(\omega)\rangle$
and  $\langle \delta\nu^{+}(\omega)\rangle$ to the averaged
density of states in the second regime are obtained by performing
integration over $\Omega$ in Eqs.~(\ref{nuPLUS}) and
(\ref{nuMINUS}) and using
Eqs.~(\ref{DIMENSIONLESS1-})-(\ref{DIMENSIONLESS2+}). We present
this expression in the form similar to Eq.~(\ref{CALI})
\begin{eqnarray}
\Bigl\langle\frac{\delta \nu^{\pm}(\omega)}{\nu_0}\Bigr\rangle=
{\mathcal D}~{\mathcal
J}^{\pm}\left(\frac{\omega}{\omega_1}\right),
\end{eqnarray}
where the prefactor ${\mathcal D}$ is defined as
\begin{eqnarray}
\label{B} {\mathcal D}=-\frac{(\nu_0V)^3}{6^{3/4}(\pi
k_{\text{\tiny{F}}}r_{\text{\tiny{II}}})^{3/2}},
\end{eqnarray}
and the dimensionless functions ${\mathcal J}^{\pm}$ are the
following integrals over $\varrho_1$, $\varrho_2$
\begin{widetext}
\begin{eqnarray}
\label{cal1} {\mathcal
J}_1^{-}(z)=\frac{1}{4}\int_{\varrho_2>\varrho_1}\frac{d\varrho_1d\varrho_2}
{(\varrho_1\varrho_2)^{3/2}}(\varrho_2-\varrho_1)^{1/2}
\Bigl({\mathcal U}_{1}^{-}(\varrho_1,\varrho_2)-1\Bigr)
\Biggl(\frac{\rho_1+\rho_2}{\rho_1\rho_2}\sin\left[\frac{\pi}{4}+2z(\varrho_2-\varrho_1)\right]\nonumber\\
- \frac{1}{\rho_1}\sin\left[\frac{\pi}{4}+2z\varrho_2\right]
-\frac{1}{\rho_2}\sin\left[\frac{\pi}{4}-2z\varrho_1\right]\Biggr),
\end{eqnarray}
\begin{eqnarray}
\label{cal2} {\mathcal
J}_2^{-}(z)=\frac{1}{4}\int_{\varrho_2>\varrho_1}\frac{d\varrho_1d\varrho_2}
{(\varrho_1\varrho_2)^{3/2}}(\varrho_2-\varrho_1)^{1/2}
\Bigl({\mathcal U}_{2}^{-}(\varrho_1,\varrho_2)-1\Bigr)
\Biggl(\frac{\rho_1+\rho_2}{\rho_1\rho_2}\cos\left[\frac{\pi}{4}+2z(\varrho_2-\varrho_1)\right]\nonumber\\
- \frac{1}{\rho_1}\cos\left[\frac{\pi}{4}+2z\varrho_2\right]
-\frac{1}{\rho_2}\cos\left[\frac{\pi}{4}-2z\varrho_1\right]\Biggr),
\end{eqnarray}

\begin{eqnarray}
\label{J1+} {\mathcal
J}_1^{+}(z)=\frac{1}{4}\int_{\varrho_2>\varrho_1}\frac{d\varrho_1d\varrho_2}
{(\varrho_1\varrho_2)^{3/2} (\varrho_1+\varrho_2)^{1/2}}
\biggl\{\Bigl({\mathcal
U}_{1}^{+}(\varrho_1,\varrho_2)-1\Bigr)\Bigl(\cos\left[\frac{\pi}{4}+2z(\varrho_1+\varrho_2)\right]-
\frac{1}{\sqrt{2}}\Bigr)\nonumber\\
+\Bigl({\mathcal U}_{2}^{+}(\varrho_1,\varrho_2)-1\Bigr)\Bigl(
\sin\left[\frac{\pi}{4}+
2z(\varrho_1+\varrho_2)\right]-\frac{1}{\sqrt{2}}\Bigr)\biggl\},
\end{eqnarray}

\begin{eqnarray}
\label{J2+} {\mathcal
J}_2^{+}(z)=\frac{z}{2}\int_{\varrho_2>\varrho_1}\frac{d\varrho_1d\varrho_2}
{(\varrho_1\varrho_2)^{3/2}} (\varrho_1+\varrho_2)^{1/2}
\biggl\{\Bigl({\mathcal U}_{1}^{+}(\varrho_1,\varrho_2)-1\Bigr)
\sin\left[\frac{\pi}{4}+2z(\varrho_1+\varrho_2)\right]\nonumber\\
-\Bigl({\mathcal U}_{2}^{+}(\varrho_1,\varrho_2)-1\Bigr)
\sin\left[\frac{\pi}{4}-2z(\varrho_1+\varrho_2)\right]\biggl\},
\end{eqnarray}
\end{widetext}
where $z=\omega/\omega_1$ is the dimensionless frequency. The new
energy scale is related to the characteristic length,
$r_{\text{\tiny{II}}}$, in the second regime in a usual way
\begin{eqnarray}
\label{OMEGA1}
\omega_1=\frac{v_{\text{\tiny{F}}}}{\sqrt{6}r_{\text{\tiny{II}}}}.
\end{eqnarray}
The second regime corresponds to  long distances,
$r_{\text{\tiny{II}}}>\xi$, travelled by electron. This is
reflected in the fact that the frequency $\omega_1$ is smaller
than $\omega_0$-the characteristic frequency for the first regime.
Using Eq.~(\ref{r0}), we can establish the relation between
$\omega_0$ and $\omega_1$, namely, $\omega_1\sim
\omega_0\varepsilon^{1/6}$, where  $\varepsilon$ is the small
parameter, defined by Eq.~(\ref{varepsilon}). We emphasize that
the second regime exists only if the condition $\varepsilon \ll 1$
is met.

It is important  to compare the scale $\omega_1$ to the
``diffusive'' energy scale $\omega_{\text{\tiny diff}} \sim
v_{\text{\tiny F}}/l_{\text{\tiny tr}}$, where $l_{\text{\tiny
tr}}$ is the transport mean free path. In the regime II we
have\cite{Mirlin1}
\begin{eqnarray}
\label{ltr} l_{\text{\tiny tr}}\sim v_{\text{\tiny
F}}\left(k_{\text{\tiny F}}\xi\right)^2 \left[\frac{v_{\text{\tiny
F}}h_0^2\xi^3}{\Phi_0^2}\right]^{-1}= \frac{k_{\text{\tiny
F}}^2\Phi_0^2}{h_0^2\xi}.
\end{eqnarray}
In this estimate the combination, $h_0^2\xi^3v_{\text{\tiny
F}}/\Phi_0^2$, stands for a single-particle scattering rate,
calculated from the golden rule, with $h_0^2\xi^2/\Phi_0^2$ coming
from the square of the matrix element; the factor
$\left(k_{\text{\tiny F}}\xi\right)^2$ accounts for the
small-angle scattering. Eq.~(\ref{ltr}) leads to the following
relation between the transport mean free path and $r_{\text{\tiny
II}}$
\begin{equation}
\label{MeanFreePath} \Bigl(\frac{l_{\text{\tiny tr}}}{
k_{\text{\tiny F}}}\Bigr)^{1/2} \sim r_{\text{\tiny II}}\sim
\frac{\xi}{\sqrt{\varepsilon}}.
\end{equation}
As follows from Eq.~(\ref{MeanFreePath}), the distance
$r_{\text{\tiny II}}$, over which the phase of the Friedel
oscillations is preserved,  is intermediate between
$l_{\text{\tiny tr}}$ and $\xi$. Indeed, the ratio $l_{\text{\tiny
tr}}/r_{\text{\tiny II}}$ is $\sim k_{\text{\tiny
F}}r_{\text{\tiny II}}\sim k_{\text{\tiny
F}}\xi/\sqrt{\varepsilon}$. This ratio is large both because
$k_{\text{\tiny F}}\xi \gg 1$ and because $\varepsilon \ll 1$.
Thus we conclude that the energy scale, $\omega_1$, is much larger
than $\omega_{\text{\tiny diff}}$, since $\omega_{\text{\tiny
diff}}/\omega_1$ is $\sim  r_{\text{\tiny II}}/l_{\text{\tiny tr}}
\ll 1$, i.e., the conventional diffusive zero-bias anomaly
develops at frequencies much smaller than the width of the
zero-bias anomaly in regime II.

\subsection{Discussion}
Dimensionless density of states, ${\mathcal J}={\mathcal
J}^{-}+{\mathcal J}^{+}$, is plotted in Fig.~\ref{RegimeII}. It is
seen that the function ${\mathcal J}(z)$ exhibits pronounced
minimum at $z\approx 0.75$, which is followed by a {\em
monotonous} decay. This behavior should be contrasted to the
dimensionless density of states in the regime I, plotted in
Fig.~\ref{RegimeI}. The difference is that the function $\mathcal
I$ exhibits damped oscillations with alternating maxima and
minima, while ${\mathcal J}$ contains only a single minimum. This
difference is not unexpected on qualitative grounds. Indeed, the
distance $\sim r_{\text{\tiny I}}$, at which the oscillations are
formed in regime I, is much smaller than the correlation radius,
$\xi$, while the characteristic distance, $\sim r_{\text{\tiny
II}}$, in regime II is much bigger than $\xi$. Therefore, it is
remarkable that ${\mathcal J}(z)$ exhibits even a single minimum.
However, qualitative difference between ${\mathcal I(z)}$ and
${\mathcal J}(z)$ at large $z$ is much harder to trace from their
respective representations as double integrals over $\rho_1$ and
$\rho_2$ [see Eqs.~(\ref{averaged}), (\ref{cal1}), (\ref{cal2}),
(\ref{J1+}), (\ref{J2+})]. The structure of one of several
contributions to ${\mathcal I}(z)$ and ${\mathcal J}(z)$ can be
loosely rewritten as
\begin{eqnarray}
\label{IandJ1} {\text{I}}\!\!\!&:&\!\!\!\quad
\int_0^{\infty}\int_0^{\infty}\frac{d\rho_1d\rho_2}{(\rho_1\rho_2)^{3/2}}\sqrt{\rho_1+\rho_2}\;\frac{\sin
z\left(\rho_1+\rho_2\right)}{\sqrt{1+\rho_1^2\rho_2^2(\rho_1+\rho_2)^2}},\nonumber\\
\\
\label{IandJ2} {\text{II}}\!\!\!&:&\!\!\!\quad
\int_0^{\infty}\int_0^{\infty}\frac{d\rho_1d\rho_2}{(\rho_1\rho_2)^{3/2}}\sqrt{\rho_1+\rho_2}\;\frac{\sin
z\left(\rho_1+\rho_2\right)}{\sqrt{1+\rho_1^2\rho_2^2}}.\nonumber\\
\end{eqnarray}
The integrands in Eq.~(\ref{IandJ1}) and Eq.~(\ref{IandJ2}) differ
only by the structure of the denominators. This difference can be
traced to Eq.~(\ref{overt1}) in which the correlator is set either
constant (regime I) or a $\delta$-function (regime II).
 From the form of the contribution Eq.~(\ref{IandJ1}),
it is not obvious at all that the large-$z$ behavior is determined
by well-defined values $\rho_1=\rho_2=\rho_0$ in the complex
plane, with $\rho_0$ satisfying $1+\rho_0^6=0$, so that the
contribution is {\em oscillatory} Eq.~(\ref{oscillation}). This
fact was established above by taking the large-$z$ asymptote prior
to the averaging over realizations. It is also supported by
numerics in Fig.~\ref{RegimeI}.

Monotonous behavior of ${\mathcal J}(z)$ at large $z$ implies that
the integral Eq.~(\ref{IandJ2}) is not dominated by distinct
complex $\rho_1=\rho_2=\tilde\rho_0$, such that
$1+\tilde\rho_0^4=0$. The only vague explanation of this is that
denominator, $\sqrt{1+\rho_1^2\rho_2^2(\rho_1+\rho_2)^2}$, in
Eq.~(\ref{IandJ1}) fixes $\rho_1\approx\rho_2\approx\rho_0$ much
more efficiently that the denominator,
$\sqrt{1+\rho_1^2\rho_2^2}$, in Eq.~(\ref{IandJ2}) fixes $\rho_1$,
$\rho_2$ near $\tilde\rho_0$.

\section{Implications}
\label{Implications}
\subsection{Half-filled Landau level}

Experimental situation of a two-dimensional electron gas placed in
inhomogeneous magnetic field can be created artificially, see,
{\em e.g.},
Refs.~\onlinecite{Geim90,bending90,Geim92,Geim94,smith94,mancoff95,gusev96,gusev00,rushforth04}.
This situation also emerges in electron gas in a strong {\em
constant} magnetic field, when the filling factor of the lowest
Landau level is close to $1/2$. In the latter case, constant field
transforms electrons into composite fermions \cite{Jain,HLR}, with
well defined Fermi surface
\cite{composite0,composite1,composite2,composite3,composite4},
 while the randomness
of magnetic field is due to spatial inhomogeneity of the electron
density. Transport properties of {\em noninteracting} gas of
composite fermions under these conditions were considered
theoretically in
Refs.~\onlinecite{chklovskii94,chklovskii95,Falko94,Khveshchenko96,Simons99,Shelankov00,Mirlin1,Mirlin2,Mirlin3}.

With regard to the tunnel density of states near the half-filling,
for the case of homogeneous gas, it was addressed theoretically in
Refs.~\onlinecite{He93,Shytov98,Shytov01} both for tunneling into
the bulk and into the edge. Unlike interacting {\em homogeneous}
electron gas,\cite{mishchenko02} composite fermions are expected
to exhibit a zero-bias anomaly {\em even without
inhomogeneity}\cite{He93,Shytov98,Shytov01}. This difference
between composite fermions and free electrons can be traced to the
form of density-density correlator of composite fermions at small
momenta\cite{HLR}. Namely, the pole of this correlator defines the
mode of neutral excitations with dispersion $\omega \propto iq^3$,
even slower than the diffusive mode in the presence of disorder.
Resulting suppression of tunneling into the edge of homogeneous
electron gas at half filling, predicted in
Refs.~\onlinecite{Shytov98},~\onlinecite{Shytov01}, turned out to
be stronger than in the experiment\cite{Chang1,Chang2}.

It is convenient to express random static magnetic field
originating from spatial inhomogeneity with magnitude $\delta n$,
in the units of the cyclotron frequency
\begin{eqnarray}
\frac{\delta\omega_c}{\Omega_{1/2}} =\frac{2\delta n}{n_{1/2}},
\end{eqnarray}
where $n_{1/2}$ is concentration of electrons at which the filling
factor in the field, $\Omega_{1/2}$, is equal to $1/2$. Density
fluctuations not only smear out the ``intrinsic'' zero-bias
anomaly, but also give rise to the smooth-disorder-induced
zero-bias  anomaly, studied in the present paper. Quantitatively,
we predict the following relation between the width of zero-bias
anomaly and the magnitude, $\delta n$ of the density fluctuations
\begin{eqnarray}
\label{RELATION1} \omega_0\sim \Omega_{1/2}\Biggl(\frac{\delta
n}{n_{1/2}}\Biggr)^{2/3}.
\end{eqnarray}
This relation follows directly from Eq.~(\ref{energies}) and
applies for smooth fluctuations with spatial scale, $\xi$,
satisfying the condition
\begin{eqnarray}
n_{1/2}\xi^2>\Biggl(\frac{n_{1/2}}{\delta n}\Biggr)^{4/3}.
\end{eqnarray}
This condition is equivalent to the condition $\varepsilon >1$,
where the parameter $\varepsilon$ is defined by
Eq.~(\ref{varepsilon}). In the opposite case of ``fast''
fluctuations the width, $\omega_1$, is given by
\begin{eqnarray}
\label{RELATION2} \omega_1\sim
\Omega_{1/2}\left[n_{1/2}\xi^2\right]^{1/4} \left(\frac{\delta
n}{n_{1/2}}\right),
\end{eqnarray}
as follows from Eq.~(\ref{energies}). Concerning the magnitude of
the anomaly, Eqs.~(\ref{A}) and (\ref{B}) predict $\delta
\nu/\nu_0 \sim \left(\delta n/n_{1/2}\right)$ for slow
fluctuations Eq.~(\ref{RELATION1}), and $\delta \nu/\nu_0 \sim
\left(\delta n/n_{1/2}\right)^{3/2}
\Bigl[n_{1/2}\xi^2\Bigr]^{3/8}$ for the fast fluctuations
Eq.~(\ref{RELATION2}), respectively.

Qualitative difference between the ``intrinsic'' zero-bias
anomaly\cite{He93,Shytov98,Shytov01} and inhomogeneity-induced
zero-bias anomaly, considered in the present paper, is that the
latter necessarily involves electron-electron scattering processes
with momentum transfer $\approx 2k_{\text{\tiny F}}$. As was
mentioned above, the intrinsic anomaly gets stronger towards the
edge\cite{Shytov98,Shytov01}. We would like to emphasize that the
anomaly due to the $2k_{\text{\tiny F}}$-processes also gets
stronger towards the edge. The reason is that the average electron
concentration decreases monotonically upon approaching the edge.
This decrease translates into a {\em non-fluctuating} magnetic
field, acting on composite fermions\cite{chklovskii95}, which {\em
increases} towards the edge. Correction, $\delta\nu(\omega)$, to
the density of states in this case is given by Eq.~(\ref{calB}),
and is plotted in Fig.~\ref{Oscillations}. Then we conclude that
the ratio of magnitudes, $\delta\nu_{\text{\tiny
bulk}}/\delta\nu_{\text{\tiny edge}}$, is simply $\sim
\left(\delta n_{\text{\tiny bulk}}/\delta n_{\text{\tiny
edge}}\right)\ll 1$, where $\delta n_{\text{\tiny bulk}}$ and
$\delta n_{\text{\tiny edge}}$ are the deviations of electron
density from $n_{1/2}$ in the bulk and at the edge, respectively.
The widths of $\delta\nu_{\text{\tiny bulk}}(\omega)$ and
$\delta\nu_{\text{\tiny edge}}(\omega)$ are related as $\sim
\left(\delta n_{\text{\tiny edge}}/\delta n_{\text{\tiny
bulk}}\right)^{2/3}\ll 1$.

\subsection{Spin-fermion model}

Similarly to composite fermions, the dispersion of neutral
excitations right at the critical point in the spin-fermion model
is dominated by a slow mode,\cite{hertz76,millis93} $\omega
\propto iq^3$. Outside the critical region, the propagator of the
neutral excitations (bosons) in the spin-disordered phase has a
conventional Ornstein-Zernike form $\chi(q)\propto
1/\left(q^2+\xi^{-2}\right)$, where $\xi$ is the correlation
radius, which diverges at the critical point.
Interaction of electrons with slow critical fluctuations can be
viewed as scattering by the smooth disorder. The question that we
will discuss below is how the growth of $\xi$, upon approaching
the critical point, manifests itself in the behavior of the
averaged (over the fluctuations of the order parameter) density of
states. Our calculations demonstrate that the dimensionless
parameter $\varepsilon$, defined by Eq.~(\ref{varepsilon}), plays
a crucial role.

Traditionally, in the studies of the response functions, like spin
susceptibility, of two-dimensional electrons near the quantum
critical point, see, {\em e.g.}, Refs.
\onlinecite{hertz76,millis93,belitz02,Chubukov04,Chubukov06,vojta07},
electrons are treated as ballistic. More specifically, they
interact only with critical fluctuations, but {\em not with each
other}. Transport at the quantum critical point was also
considered for non-interacting ballistic\cite{Narozhny06} or
diffusive\cite{Paul07} electrons that are scattered by  bosonic
excitations.

In all theoretical treatments of the spin-fermion model,
modification of the response of the electron gas due to
interaction with bosons was governed by the processes with {\em
small momentum transfer}. Our main point is that incorporating
{\em direct} electron-electron interactions into the spin-fermion
model gives rise to a novel feature in the response of the
electron gas near the critical point in spin-disordered phase. The
underlying reason is that, while critical bosonic fluctuations are
``smooth'', so that their momenta are $\ll k_{\text{\tiny F}}$,
electron-electron interactions allow $2k_{\text{\tiny
F}}$-processes. Then the physics, discussed in the present paper,
emerges in the following way:

(i) interaction with slow bosonic fluctuations, curves slightly
the electron trajectories;

(ii) interaction between the electrons, moving along slightly
curved trajectories, generates a small energy scale, which
reflects the ``degree'' of curving;

(iii) the degree of curving grows with correlation radius, $\xi$,
of the bosonic excitations.

As a result, the character of critical fluctuations is reflected
in
the density of states, $\delta\nu(\omega)$, in a very nontrivial
fashion. Namely, they give rise to the lively low-frequency
feature and even aperiodic oscillations in $\delta\nu(\omega)$, as
was demonstrated above. This suggests that information about
proximity to the critical point can be inferred from tunneling
experiments.

To quantify the above scenario, we will assume for simplicity
\cite{Aharonov92} that bosonic critical fluctuations of
magnetization, ${\bf S}({\bf r})$, interact with electron spins
not as $\bm{\sigma}\cdot{\bf S}$, where $\bm{\sigma}$ are the
Pauli matrices, but via the {\em position-dependent} Zeeman
energy, $E_{\text{\tiny Z}}({\bf r})$, with characteristic
magnitude, $E_0$. Assuming that the fluctuations, ${\bf S}({\bf
r})$, are {\em static}, we get for correlator of random Zeeman
energy, $E_{\text{\tiny Z}}({\bf r})$, the standard expression
\begin{eqnarray}
\label{ZEEMAN} \left\langle E_{\text{\tiny Z}}({\bf
r})E_{\text{\tiny Z}}({\bf r}^{\prime})\right\rangle &=&E_0^2\int
\frac{d{\bf q}}{2\pi}\;\frac{e^{i{\bf q}({\bf r}-{\bf
r}^{\prime})}}
{q^2+\xi^{-2}}\nonumber\\&=&E_0^2\;\text{K}_0\left(\vert
x_1-x_2\vert/\xi\right),
\end{eqnarray}
where $\text{K}_0$ is the Macdonald function.

As a next step, we notice that the {\em force}, $\nabla
E_{\text{\tiny Z}}({\bf r})$, curves the electron trajectories
{\em in the same way} as random magnetic field, $h(x,y)$. This
allows us to use  general expressions Eqs.~(\ref{nuPLUS}),
(\ref{nuMINUS}) for the interaction correction to the density of
states.  We can also employ the result Eq.~(\ref{overt1}) for the
general averaging procedure, i.e., to treat critical fluctuations
as a disorder. With the help of Eq.~(\ref{ZEEMAN}) the result
Eq.~(\ref{overt1}) assumes the form
\begin{eqnarray}
\label{overt2} \Bigl\langle \exp\{2i\delta\varphi_{\Sigma}^{(-)}\}
\Bigr\rangle=\Bigl[1+iE_0^2
\int_0^{r_2}\!\int_0^{r_2} dx_1dx_2\\
\times f_{-}(x_1)f_{-}(x_2)\partial_{x_1}\partial_{x_2}
\text{K}_0\left(\frac{\vert
x_1-x_2\vert}{\xi}\right)\Bigr]^{-1/2},\nonumber
\end{eqnarray}
where the function $f_{-}$ is defined by Eq.~(\ref{f-}) for the
case of random magnetic field. For the case of random Zeeman
energy, the prefactor, $1/\Phi_0k_{\text{\tiny F}}^{1/2}$, should
be replaced by $k_{\text{\tiny F}}^{1/2}/E_{\text{\tiny F}}$.
Characteristic energy scales can be now inferred from
Eq.~(\ref{overt2}) on the basis of the following reasoning.
Characteristic distances $r_1$, $r_2$ in Eq.~(\ref{overt2}) are
determined by the condition
\begin{eqnarray}
\label{CONDITION} &&\int_0^{r_2}\!\int_0^{r_2} dx_1dx_2
 \text{K}_0\left(\frac{\vert
x_1-x_2\vert}{\xi}\right)\frac{\partial}{\partial x_1} f_{-}(x_1)
\frac{\partial}{\partial x_2} f_{-}(x_2)\nonumber\\
&&\sim \frac{1}{E_0^2},
\end{eqnarray}
where we performed integration by parts in Eq.~(\ref{overt2}).
Then the characteristic width of a zero-bias anomaly is equal to
$\omega \sim  v_{\text{\tiny F}}/r_1\sim v_{\text{\tiny F}}/r_2$.

Recall now, that in the case of random magnetic field, double
integral in the left-hand side of Eq.~(\ref{CONDITION}) did not
contain derivatives and was $\propto r_2^3$ in the regime I, and
$\propto r_2^2\xi$ in regime II, respectively. This is because the
function, $f_{-}(x_1)$, is $\sim r_2^{1/2}$ at $x_1 \sim r_2$, see
Eq.~(\ref{f-}). Due to the fact that the effective ``force'' in
the spin-fermion model is $\propto \nabla E_{\text{\tiny Z}}({\bf
r})$, the left-hand side in Eq.~(\ref{CONDITION}) is $\sim
k_{\text{\tiny F}}r_2/E_{\text{\tiny F}}^2$ for $\xi \gg r_2$. In
this limit, Eq.~(\ref{CONDITION}) yields (with logarithmic in
$\xi/r_2$ accuracy)
\begin{eqnarray}
\label{xiC} r_2 \sim k_{\text{\tiny
F}}^{-1}\left(\frac{E_{\text{\tiny F}}}{E_0}\right)^2=\xi_c ,~~
\omega \sim \frac{E_0^2}{E_{\text{\tiny F}}}=E_c.
\end{eqnarray}
Note that $E_c$ is {\em independent} of $\xi$. We conclude that,
upon approaching the critical point, as the correlation radius
exceeds the value $\xi_c$, the zero-bias anomaly ``freezes''.
 Its form is shown in Fig.~\ref{RegimeI}, and its magnitude is $\sim
\left(E_0/E_{\text{\tiny F}}\right)^3$. An alternative way to
recover the scales Eq.~(\ref{xiC}) is to notice that parameter
$\varepsilon$, which is defined by Eq.~(\ref{varepsilon}) in
context of random magnetic field, in the situation with random
Zeeman energy acquires the form $\varepsilon=(k_{\text{\tiny
F}}\xi)\left(E_0/E_{\text{\tiny F}}\right)^2$. Then $\xi_c$ given
by Eq.~(\ref{xiC}) corresponds to $\varepsilon=1$, i.e., to the
boundary of the regime I.

For $\xi<\xi_c$  the integral in the left-hand side of
Eq.~(\ref{CONDITION}) is proportional to $\xi$ and is {\em
independent} of $r_2$. Then  Eq.~(\ref{CONDITION}) does not have a
solution. Therefore, characteristic $r_1$ and $r_2$ in the
expression for the density of states are $\sim \xi$, and  the
width of the anomaly is simply $\sim v_{\text{\tiny F}}/\xi
=E_c(\xi_c/\xi)$. Concerning the magnitude of the anomaly at $\xi
<\xi_c$, it should be estimated with the account that the integral
in right-hand side of  Eq.~(\ref{overt2}) is smaller than $1$ {\em
for all} $r_2$. Therefore, $\Bigl\langle
\exp\{2i\delta\varphi_{\Sigma}^{(-)}\} \Bigr\rangle$ in
Eq.~(\ref{overt2}) can be approximately replaced by
$\bigl\{1-(i/2\xi_c)\left[r_2\Theta(\xi-r_2)+\xi\Theta(r_2-\xi)\right]\bigr\}$,
where the second term is a small correction. However, only this
correction causes a zero-bias anomaly. Substituting this
correction into Eq.~(\ref{nuMINUS}), we find the estimate for the
magnitude,
\begin{eqnarray}
\label{dnu} \frac{\delta\nu}{\nu_0} \sim
\left(\frac{E_0}{E_{\text{\tiny
F}}}\right)^3\left(\frac{\xi_c}{\xi}\right)^{1/2} \sim
\left(\frac{E_0}{E_{\text{\tiny
F}}}\right)^2\frac{1}{(k_{\text{\tiny F}}\xi)^{1/2}}.
\end{eqnarray}
We conclude that, as $\xi$ grows and approaches $\xi_c$, the
magnitude of the anomaly falls off as $1/\sqrt{\xi}$, and the
anomaly narrows as $1/\xi$.

The remaining issue to discuss is whether or not the assumption
that fluctuating Zeeman energy, $E_{\text{\tiny Z}}({\bf r})$, is
{\em static} applies at relevant frequency and spatial scales,
$E_c$ and $\xi_c$. For this purpose, we recall the correlator of
Zeeman energies in the momentum space does not have a simple
Ornstein-Zernike form but is rather $\left\langle \vert
E_{\text{\tiny Z}}({\bf q})\vert^2\right\rangle \propto
1/\left(q^2+\xi^{-2}+\varsigma \omega/q\right)$, where the dynamic
term, $\varsigma \omega/q$, describes the damping of bosons due to
creation of electron-hole pairs. The prefactor $\varsigma$ (the
Landau damping coefficient) is thus quadratic in  coupling of
electrons to the spin density fluctuations, i.e.,  $\varsigma
\propto E_0^2$. For characteristic frequencies the dynamic term,
$\varsigma(\omega/q)\sim\varsigma E_c\xi_c\sim\varsigma
v_{\text{\tiny F}}$. Therefore, it is negligible only if the
condition, $\xi_c^{-2}=k_{\text{\tiny
F}}^2\left(E_0/E_{\text{\tiny F}}\right)^4 \gg\varsigma
v_{\text{\tiny F}}$, holds. With $\varsigma$ being proportional to
$E_0^2$, the above condition is met for large enough coupling,
$E_0$. In the opposite case, when the dynamic part of correlator
dominates at $\omega \sim E_c$ and $q\sim \xi_c^{-1}$, the
zero-bias anomaly develops only away from the critical point when
$\xi$ becomes smaller than $(\varsigma v_{\text{\tiny
F}})^{-1/2}$. Upon further departure from the critical point, our
prediction $\delta\nu/\nu_0 \propto \xi^{-1/2}$ and $\omega \sim
v_{\text{\tiny F}}/\xi$ should apply. Note finally, that, {\em
directly at the critical point}, the slow mode $\omega \approx
iq^3/\varsigma$ gives rise to the ``intrinsic'' zero-bias
anomaly,\cite{Chubukov06} similar to the composite fermions.




\acknowledgments The authors acknowledge the support of NSF (Grant
No. DMR-0503172) and of the Petroleum Research Fund (Grant No.
43966-AC10). We are grateful to E.~G.~Mishchenko and O.~A.~Starykh
for numerous discussions.

\appendix

\section{Polarization operator
in the coordinate space} \label{AppendixA} Here we derive
Eqs.~(\ref{operator0}) and (\ref{operator2kF}) for polarization
operator in coordinate space using the known expression
\cite{Stern67} for $\Pi({\bf q},\omega)$ in the momentum space.
Since we are interested in behavior of $\Pi({\bf r},\omega)$ at
distances $\vert {\bf r}\vert \gg k_{\text{\tiny{F}}}^{-1}$, it is
sufficient to perform the Fourier transform
\begin{eqnarray}
\label{decomposition} \Pi({\bf
r},\omega)=\frac{1}{2\pi}\int\!d{\bf q}\;e^{i{\bf q r}}\;\Pi({\bf
q},\omega),
\end{eqnarray}
using the asymptotes of $\Pi({\bf q},\omega)$ at small $q\ll
k_{\text{\tiny{F}}}$ and at $q$ close to $2k_{\text{\tiny{F}}}$.
The small-$q$ asymptote of $\Pi({\bf q},\omega)$ has the form
\begin{eqnarray}
\label{omega} \Pi_0({\bf q},\omega)=-\nu_0\left[1+\frac{i\omega\;
\Theta(qv_{\text{\tiny{F}}}-\omega)}
{\sqrt{q^2v_{\text{\tiny{F}}}^2-\omega^2}} \right. \nonumber\\
\left.~~~~~~~~~~~~~~~~
 +\frac{\omega\; \Theta(\omega-qv_{\text{\tiny{F}}})}{\sqrt{\omega^2-q^2v_{\text{\tiny{F}}}^2}} \right],
\end{eqnarray}
where $\Theta(x)$ is the step-function. The easiest way to perform
the integration Eq.~(\ref{decomposition}) is to first Fourier
transform Eq.~(\ref{omega}) with respect to {\em frequency}
\begin{eqnarray}
\label{relation} -\frac{\Theta(qv_{\text{\tiny{F}}}-\omega)}
{\sqrt{q^2-\left(\omega/v_{\text{\tiny{F}}}\right)^2}}+\frac{i\Theta(\omega-qv_{\text{\tiny{F}}})}
{\sqrt{\left(\omega/v_{\text{\tiny{F}}}\right)^2-q^2}}\qquad\nonumber\\
\qquad=\int_0^{\infty}\!\!ds\;J_0(qs)\;\exp\left\{\frac{i\omega\;s}{v_{\text{\tiny{F}}}}\right\}.
\end{eqnarray}
Substituting Eq.~(\ref{omega}) into Eq.~(\ref{relation}) and using
the orthogonality relation
$\int_0^{\infty}dq\;qJ_0(qs)J_0(qr)=\delta(r-s)/r$, we readily
obtain
\begin{eqnarray}
\label{P0} \Pi_0({\bf
r},\omega)=-\frac{i\nu_0\omega}{v_{\text{\tiny{F}}}r}\exp\left\{\frac{i\omega\;r}{v_{\text{\tiny{F}}}}\right\}.\\\nonumber
\end{eqnarray}
 In order to calculate $\Pi_{2k_{\text{\tiny{F}}}}({\bf r},\omega)$ we use
the form of polarization operator in momentum space for $\vert
q-2k_{\text{\tiny{F}}}\vert \ll k_{\text{\tiny{F}}}$ and $\omega
\ll E_{\text{\tiny{F}}}$
\begin{eqnarray}
\label{P2kF}
\Pi_{2k_{\text{\tiny{F}}}}({\bf q},\omega)=\nu_0\Biggl[1-\frac{1}{\sqrt{4k_{\text{\tiny{F}}}}}\qquad\qquad\qquad\qquad\qquad\\
\times\left(\sqrt{q-2k_{\text{\tiny{F}}}+\omega/v_{\text{\tiny{F}}}}+\sqrt{q-2k_{\text{\tiny{F}}}-\omega/v_{\text{\tiny{F}}}}\right)\Biggr],\nonumber
\end{eqnarray}
where the square roots should be understood as
$\sqrt{x}\rightarrow \text{sign}(x)\sqrt{x}$. Then the integral
over ${\bf q}$ in Eq.~(\ref{decomposition}) assumes the form
\begin{eqnarray}
\label{intermediate} \Pi_{2k_{\text{\tiny{F}}}}({\bf
r},\omega)=-\nu_0\int_0^{\infty}dq \;q J_0(qr)\Bigl
[\sqrt{q-2k_{\text{\tiny{F}}}+\omega/v_{\text{\tiny{F}}}}\;\;\;\;\;\;\;\;\nonumber\\
+\sqrt{q-2k_{\text{\tiny{F}}}-\omega/v_{\text{\tiny{F}}}}\Bigr]\approx
\sqrt{\frac{4k_{\text{\tiny{F}}}}{\pi r}}\int_0^{\infty}dq\cos\left(qr-\frac{\pi}{4}\right)\quad\quad\;\;\nonumber\\
\times
\Bigl[\sqrt{q-2k_{\text{\tiny{F}}}+\omega/v_{\text{\tiny{F}}}}+\sqrt{q-2k_{\text{\tiny{F}}}-\omega/v_{\text{\tiny{F}}}}\Bigr]\qquad\;\;\;
\end{eqnarray}
where we used that fact that $k_{\text{\tiny{F}}}r\gg 1$ and
replaced the Bessel function by its large-$q$ asymptotics.
Integration over variable $q$ in Eq.~(\ref{intermediate}) is
performed with the use of the identity
\begin{eqnarray}
\int_a^{\infty}dz\;\cos
z\sqrt{z-a}={\frac{\sqrt{\pi}}{2}}\sin\left(a+\frac{\pi}{4}\right),
\end{eqnarray}
and yields the zero-temperature limit of Eq.~(\ref{operator2kF}).

\section{Polarization operator in a constant magnetic
field} \label{AppendixB} We start from the general expression
\cite{aleiner95} for the polarizability in arbitrary magnetic
field
\begin{eqnarray}
\label{general} {\large{\Pi}}\;
(q)=-\frac{2m}{\pi}\sum_{n_1=0}^{\infty}\sum_{n_2=0}^{\infty}
\frac{(-1)^{(n_2-n_1)}\bigl(f_{n_1}-f_{n_2}\bigr)}{{n_2-n_1}}\nonumber\\
\times \exp(-q^2l^2/2)\;{\text{\large
L}}_{n_1}^{n_2-n_1}\left(\frac{q^2l^2}{2}\right) {\text{\large
L}}_{n_2}^{n_1-n_2}\left(\frac{q^2l^2}{2}\right),
\end{eqnarray}
where ${\text{\large L}}_{n_1}^{n_2-n_1}(x)$ and ${\text{\large
L}}_{n_2}^{n_1-n_2}(x)$ are the Laguerre polynomials, and
$f_n=\bigl\{\exp\bigl[(n-N_{\mbox{\tiny
F}})\hbar\omega_c/T\bigr]+1\bigr\}^{-1}$ is the Fermi
distribution. At small $q\ll k_{\mbox{\tiny F}}$
Eq.~(\ref{general}) yields \cite{aleiner95}
$\Pi\;\!(q)=-(m/\pi)\bigl[1-J_0^2(qR_{\mbox{\tiny L}})\bigr]$,
i.e., the characteristic scale is $q\sim R_{\mbox{\tiny L}}^{-1}$.
For $(q-2k_{\mbox{\tiny F}})\ll k_{\mbox{\tiny F}}$ it is
convenient to perform the summation over the Landau levels with
the help of the following integral representation of the Laguerre
polynomial

\begin{eqnarray}
\label{representation} {\text{\large
L}}_m^n(x)=\frac{1}{2\pi}\int_0^{2\pi}\!\!\frac{d\theta}{\left(1-e^{i\theta}\right)^{n+1}}
\exp\left\{\frac{xe^{i\theta}}{e^{i\theta}-1}-im\theta\right\}.\nonumber\\
\end{eqnarray}
In the vicinity $q=2k_{\mbox{\tiny F}}$ Eq.~(\ref{representation})
contains a small factor $\exp(-q^2l^2/2)$. This factor is
compensated by the product of Laguerre polynomials, since each of
them is $\propto\exp(x/2)$, which comes from the exponent in
Eq.~(\ref{representation}) taken at $\theta=\pi$. With
contribution from the vicinity $\theta=\pi$ dominating the
integral (\ref{representation}), we can expand the integrand
around this point as $\exp\bigl[x/2+i\pi m
+i\phi(\psi)\bigr]/2^{n+1}$, where $\psi=(\theta - \pi)$, and the
phase, $\phi(\psi)$, is equal to
\begin{eqnarray}
\label{expansion} {\large\phi}(\psi)= \left({x\over
4}-m-\frac{n+1}{2} \right)\psi+\frac{x\psi ^3}{48}.
\end{eqnarray}
Now we make use of the fact that only relatively small number
$\sim (k_{\mbox{\tiny F}}l)^{2/3}\ll N_{\mbox{\tiny F}}$
 of Landau levels around
$E_{\mbox{\tiny F}}$ contribute to the sum
Eq.~(\ref{representation}). This suggests that we can present
$n_1$ and $n_2$ as $n_1=N_{\mbox{\tiny F}}+m_1$ and
$n_2=N_{\mbox{\tiny F}}-m_2$, respectively,  and extend the sum
over $m_1$, $m_2$ from $-\infty$ to $+\infty$. After that the
summation over Landau levels can be easily carried out with the
help of the following identity
\begin{eqnarray}
\label{identity2} \sum\limits _{m_1,m_2=-\infty}^{\infty}
\frac{f_{N_{\mbox{\tiny F}}-m_1}-f_{N_{\mbox{\tiny F}}+m_2}}{m_1+m_2}\qquad\qquad\nonumber\\
\times\cos\bigl[(m_1-m_2)\alpha+\beta\bigr]=\frac{2 \pi
^2T\cos\beta}{\hbar \omega _c\sinh \bigl(2\pi\vert \alpha\vert T/
\hbar\omega _c)}.
\end{eqnarray}
As a next step, we substitute the representation
Eq.~(\ref{representation}) of Laguerre polynomials with integrand
expanded according to Eq.~(\ref{expansion}), into
Eq.~(\ref{general}). Upon this substitution, we perform the
summation over Landau levels using the relation
Eq.~(\ref{identity2}). Then the double integral, which emerges in
Eq.~(\ref{general}) as a result of representing the two laguerre
polynomials Eq.~(\ref{representation}), assumes the form
\begin{eqnarray}
\label{doubleintegral}
\int_{-\infty}^{\infty}\int_{-\infty}^{\infty}\frac{d\psi_1d\psi_2}{
\vert\psi_1+\psi_2\vert}\qquad\qquad\qquad\qquad\qquad\nonumber\\
\times\cos\left[\left(\psi_1^3+\psi_2^3\right)\frac{N_{\mbox{\tiny
F}}}{12}-\left(\psi_1+\psi_2\right)\frac{\delta q R_{\mbox{\tiny
L}}}{2}\right],
\end{eqnarray}
where $\delta q=q-2k_{\text{\tiny{F}}}$. Note, that integration
over the difference, $(\psi_1-\psi_2)$, in
Eq.~(\ref{doubleintegral}) can be performed explicitly. It is
convenient to present the final result not for $\Pi(q)$, but
rather for the derivative,
$\Pi^{\prime}(q,T)=\partial\Pi(q,T)/\partial q$. Knowledge of
$\Pi^{\prime}(q,T)$ is sufficient for finding the large-distance
behavior of the potential, created by the short-range impurity.
Indeed, this potential can be expressed directly through
$\Pi^{\prime} (2k_{\mbox{\tiny F}}+ Q)$ as follows
\begin{eqnarray}
\label{calculation}
 V_H(r)=\frac{V(2k_{\mbox{\tiny F}})g}{2(\pi k_{\mbox{\tiny F}}r)^{3/2}}
\int_{-\infty}^{\infty}\!\!dQ \sin\! \left[(2k_{\mbox{\tiny
F}}+Q)r-\!{\pi\over
 4}\right]\times\nonumber\\
\Pi ^\prime (2k_{\mbox{\tiny F}}+Q,T).\qquad
\end{eqnarray}
At zero temperature and in a zero magnetic field we have
$\Pi^{\prime}(q,0)\propto \theta (\delta q)/\sqrt{\delta q}$. At
finite magnetic field and finate temperature, taking derivative of
Eq.~(\ref{doubleintegral}) with respect to $\delta q$, we arrive
to the result
\begin{eqnarray}
\label{finiteT} \Pi^{\prime}(q,T)=-\frac{2^{1/3}mT}{(\pi
k_{\mbox{\tiny F}}{p_0})^{1/2}\epsilon_0}
\int_0^{\infty}\frac{dx\; x^{1/2}}{\sinh(2\pi xT/\epsilon_0)}\nonumber\\
\times \sin\left(2^{2/3}\frac{\delta
q}{p_0}\;x+\frac{1}{3}x^3+\frac{\pi}{4}\right).
\end{eqnarray}
In the limit $T\rightarrow 0$, substitution of Eq. (\ref{finiteT})
into Eq.~(\ref{calculation}) and integration over $Q$ reproduces
Eq.~(\ref{modified1}).

Interestingly, for $T=0$, the integral Eq.~(\ref{finiteT}) can be
evaluated analytically
\begin{equation}
\label{AiBi}
\Pi^{\prime}(q) =-\frac{m}{(k_{\mbox{\tiny
F}}{p_0})^{1/2}}\;{\text{\large \it Ai}}\left(\frac{\delta
q}{p_0}\right) \;{\text{\large \it Bi}}\left(\frac{\delta
q}{p_0}\right),
\end{equation}
where $Ai(z)$ is the Airy function, and  $Bi(z)$ is another
solution of the Airy equation defined, {\em e.g.}, in Ref.
\onlinecite{book}. It is seen that the singularity at
$q=2k_{\mbox{\tiny F}}$ is smeared by the magnetic field in a
rather peculiar way: for positive $\delta q \gg p_0$ the $(\delta
q)^{-1/2}$ zero-field behavior [see Eq.~(\ref{intermediate})] is
restored. However, for large negative $\delta q/p_0$, the
derivative $\Pi^{\prime}(q)$ approaches zero {\em with
oscillations}, namely, as $\cos\bigl[4(\vert \delta
q\vert/p_0)^{3/2}/3\bigr]/(\vert\delta q\vert)^{1/2}$.
 As the difference
$2k_{\mbox{\tiny F}}-q$ increases and becomes comparable to
$k_{\mbox{\tiny F}}$, these oscillations cross over to the
``classical'' oscillations\cite{aleiner95} $\Pi^{\prime}(q)
\propto J_0(qR_{\mbox{\tiny L}})J_1(qR_{\mbox{\tiny L}})\propto
\cos(2qR_{\mbox{\tiny L}})$.

\section{Evaluation of the functional integral}
\label{AppendixC}

Upon combining Eqs.~(\ref{W1}) and (\ref{h2}) the quadratic form
in the exponent in the numerator of the functional integral
Eq.~(\ref{functional}) assumes the form
\begin{widetext}
\begin{eqnarray}
\label{EXP1} 2i\delta\varphi(r)-W\{h\}=\frac{2i\varepsilon
r^3}{\xi^3}\Biggl\{\frac{1}{12}\left[\int dq\; {\mathcal
A}_{0,q}\right]^2 +\sum_{n>0}c_n\Big\vert \int dq\; {\mathcal
A}_{n,q}\Big\vert^2 +
\int dq\; {\mathcal A}_{0,q}{\mathcal G}\{{\mathcal
A}_n\}\Biggr\}\\
-\frac{2r}{\gamma\xi}\sum_{n>0}\int \!dq\;\frac{\vert{\mathcal
A}_{n,q}\vert^2} {\tilde{\mathcal K}(q)}-\frac{r}{\gamma\xi}\int
\!dq\;\frac{\vert{\mathcal A}_{0,q}\vert^2} {\tilde{\mathcal
K}(q)},\nonumber
\end{eqnarray}
with numerical coefficients $c_n=1/2\pi^2n^2$ and $b_n=-c_n
+i/2\pi n$ defined by Eq.~(\ref{bc}). In the above expression we
had introduced a short-hand notation
\begin{eqnarray}
\label{shorthand} {\mathcal G}\{{\mathcal
A}_{n,q}\}=\sum_{n>0}\Bigl[b_n\int dq\; {\mathcal A}_{n,q} +
b_n^{\ast}\int dq\; {\mathcal A}_{n,q}^{\ast}\Bigr].
\end{eqnarray}
We adopt the following sequence of integration over the variables
${\mathcal A}_{n,q}$. First we integrate over ${\mathcal A}_{0,q}$
using the following decoupling
\begin{eqnarray}
\label{exponent1} {\text{\large H}}\{ {\mathcal G}\}=\int
\prod_qd{\mathcal A}_{0,q}\exp\Biggl\{-\frac{r}{\gamma\xi}\int
\!dq\;\frac{\vert{\mathcal A}_{0,q}\vert^2} {\tilde{\mathcal
K}(q)}+\frac{i\varepsilon r^3}{6\xi^3}\left[\int dq\; {\mathcal
A}_{0,q}\right]^2 +
i{\mathcal G}\int dq\; {\mathcal A}_{0,q}\Biggr\}=\\
e^{-i\pi/4}\sqrt{\frac{3r^3}{2\pi\varepsilon\xi^3}} \int
\prod_qd{\mathcal A}_{0,q} d{\mathcal B}_0\exp\Biggl\{
-\frac{3ir^3{\mathcal B}_0^2}{2\varepsilon\xi^3} +i{\mathcal
B}_0\int dq\; {\mathcal A}_{0,q}-\frac{r}{\gamma\xi}\int
\!dq\;\frac{\vert{\mathcal A}_{0,q}\vert^2} {\tilde{\mathcal
K}(q)}+i{\mathcal G}\int dq\; {\mathcal A}_{0,q}\Biggr\},\nonumber
\end{eqnarray}
\end{widetext}
where we had introduced an auxiliary variable ${\mathcal B}_0$.
Function ${\text{\large H}}\{ {\mathcal G}\}$ combines all
integrals in Eq.~(\ref{EXP1}) containing ${\mathcal A}_{0,q}$.
Subsequent integration first over the variables  ${\mathcal
A}_{0,q}$ and then over the auxiliary variable ${\mathcal B}_0$
yields
\begin{eqnarray}
\label{exponent2}
{\text{\large H}}\{ {\mathcal
G}\}=\sqrt{\frac{\pi\xi\gamma}{ir\int dq\;\tilde{\mathcal K}(q)}}
\frac{\exp\bigl\{i{\mathcal F}(r){\mathcal G}^2\bigr\}}{\sqrt{
1-\frac{2i}{3}\left(\frac{r}{r_{\text{\tiny{II}}}}\right)^2 }},
\nonumber
\end{eqnarray}
where we had used the definition
$r_{\text{\tiny{II}}}=2\xi/\left(\sqrt{2\pi}\gamma\varepsilon\right)^{1/2}$.
In Eq.~(\ref{exponent2}) the complex  function ${\mathcal F}(r)$
is defined as
\begin{eqnarray}
\label{auxiliaryf} {\mathcal
F}(r)=-\frac{3\varepsilon^{1/2}\left(\gamma\int\!dq\;\tilde{\mathcal
K}(q)\right)^{3/2}}{16\left(\frac{r}{r_{\text{\tiny{II}}}}\right)^3
+24i\left(\frac{r}{r_{\text{\tiny{II}}}}\right)}.
\nonumber\\
\end{eqnarray}
As a result of integration over ${\mathcal A}_{0,q}$ the exponent
in the functional integral Eq.~(\ref{EXP1}) assumes the form
\begin{eqnarray}
\label{exponent3}
i\sum_{n>0}{\tilde c}_n\Big\vert \int \!dq\; {\mathcal A}_{n,q}\Big\vert^2\qquad\qquad\qquad\\
\qquad\qquad\qquad- \frac{2r}{\gamma\xi}\sum_{n>0}\int
\!dq\;\frac{\vert{\mathcal A}_{n,q}\vert^2} {\tilde{\mathcal
K}(q)}+i{\mathcal F}(r){\mathcal G}^2,\nonumber
\end{eqnarray}
where ${\tilde c}_n$ is related to $c_n$ via a dimensionless
factor
\begin{eqnarray}
\label{factor} {\tilde c}_n=\frac{2\varepsilon r^3}{\xi^3}c_n.
 \end{eqnarray}
The first and the third terms in Eq.~(\ref{exponent3}) contain
squares of the linear combinations of ${\mathcal A}_{n,q}$. To
decouple these squares,
 we introduce a set of auxiliary variables, $\alpha_n, \alpha_n^{\ast}$ for the first
term, and one auxiliary variable, $\alpha_0$ for the third term as
follows
\begin{widetext}
\begin{eqnarray}
\label{HS2} e^{i{\mathcal V}\left\vert\int\!dq\;{\mathcal
A}_{n,q}\right\vert^2}=\frac{1}{2\pi}\int\!d\alpha_nd\alpha_n^{\ast}
\exp\Bigl\{-i\vert\alpha_n\vert^2 +{\mathcal
V}^{1/2}\alpha_n^{\ast}\int\!dq\;{\mathcal A}_{n,q}-{\mathcal
V}^{1/2}\alpha_n\int\!dq\;{\mathcal A}_{n,q}^{\ast}\Bigr\},
\end{eqnarray}
\begin{eqnarray}
\label{HS1} e^{i{\mathcal F}(r){\mathcal G}^2}=\frac{1}{\sqrt{4\pi
{\mathcal
F}(r)}}\int_{-\infty}^{\infty}d\alpha_0\exp\left\{-\frac{i\alpha_0^2}{4{\mathcal
F}(r)}+i\alpha_0{\mathcal
G}\right\}.\nonumber\\
\end{eqnarray}
Note that $\text{Im}\left[1/{\mathcal F}(r)\right]<0$, so that the
decoupling Eq.~(\ref{HS1}) of the quadratic in ${\mathcal G}$ term
in the exponent of Eq.~(\ref{auxiliaryf}) is justified.

As a next step, we perform gaussian integration over the infinite
set of variables, $\{{\mathcal A}_{n,q}\}$
\begin{eqnarray}
\label{Anq} \int d{\mathcal A}_{n,q}d{\mathcal
A}_{n,q}^{\ast}\exp\Biggl\{\int\!dq\Bigl[-\frac{2r\vert{\mathcal
A}_{n,q}\vert^2}{\gamma\xi{\tilde{\mathcal K}(q)}}+{\mathcal
A}_{n,q} \left({\tilde
c}_n^{1/2}\alpha_n^{\ast}+i\alpha_0b_n\right)\!+\! {\mathcal
A}_{n,q}^{\ast}\left(-{\tilde c}_n^{1/2}\alpha_n+i\alpha_0b_n^{\ast}\right)\Bigr]\!\Biggr\}\nonumber\\
\qquad=
\frac{2i\pi}{(2r/\gamma\xi)\int\!dq\;\left[{\tilde{\mathcal
K}(q)}\right]^{-1}} \exp\Bigl\{-\left\vert-i{\tilde
c}_n^{1/2}\alpha_n^{\ast}+\alpha_0b_n\right\vert^{2}
\frac{\gamma\xi}{2r}\int\!dq\;{\tilde{\mathcal K}(q)}\Bigr\}.
\end{eqnarray}
As follows from Eqs.~(\ref{HS2}) and (\ref{Anq}), the integrals
over all $\alpha_n$ are gaussian and can be easily evaluated
\begin{eqnarray}
\label{alpha-n} &&\int\frac{d\alpha_nd\alpha_n^{\ast}}{2\pi}
\exp\Bigl\{-i\vert\alpha_n\vert^2- \left\vert-i{\tilde
c}_n^{1/2}\alpha_n^{\ast}+\alpha_0b_n\right\vert^{2}
\frac{\gamma\xi}{2r}\int\!dq\;{\tilde{\mathcal K}(q)}\Bigr\}\nonumber\\
&&=\frac{2r}{2r-i\gamma\xi{\tilde c}_n\int\!dq{\tilde{\mathcal
K}(q)}} \exp\Biggl\{\!-\frac{\gamma\xi\alpha_0\vert
b_n\vert^2}{2r}\!\int\!dq{\tilde{\mathcal K}(q)}\!-\!\frac{{\tilde
c}_n\left[\alpha_0\vert b_n\vert\gamma\xi\int\!dq{\tilde{\mathcal
K}(q)}\right]^2}{4ir^2+2r\gamma\xi{\tilde
c}_n\int\!dq{\tilde{\mathcal K}(q)}}\!\Biggr\}.
\end{eqnarray}
The remaining integral over $\alpha_0$ is also gaussian. Note now,
that the denominator in Eq.~(\ref{functional}), responsible for
the normalization, can be evaluated by performing the same steps
as above. This evaluation amounts to setting ${\tilde c}_n =0$ in
Eq.~(\ref{alpha-n}) and taking the limit
$r_{\text{\tiny{II}}}\rightarrow \infty$ in
Eq.~(\ref{auxiliaryf}). As a result, the functional integral
reduces to the ratio of the
ordinary integrals
\begin{eqnarray}
\label{aver2iphi} &&\langle e^{2i\delta\varphi(r)}\rangle=
\frac{1}
{\sqrt{1-\frac{2i}{3}\left(\frac{r}{r_{\text{\tiny{II}}}}\right)^2
}}\left[\prod_{n=1}^{\infty}\frac{n^2}{n^2-2i(r/r_{\text{\tiny{II}}})^{2}/\pi^2}\right]
\;\frac{\int_{-\infty}^{\infty}\!\!d\alpha_0\exp\left\{-w\alpha_0-u_1\alpha_0^2\right\}}
{\int_{-\infty}^{\infty}\!\!d\alpha_0\exp\left\{-w\alpha_0-u_0\alpha_0^2\right\}},
\end{eqnarray}
\end{widetext}
where the coefficients $w$, and $u_0$ are defined as
\begin{eqnarray}
\label{wu1u0} &&w=\frac{\gamma\xi}{2r}\left[\sum_{n>0}\vert
b_n\vert^2\right]\!\int\!dq{\tilde{\mathcal K}(q)},\nonumber\\
&&u_0=\frac{r}{\xi}\Bigl(\int\!dq{\tilde{\mathcal
K}(q)}\Bigr)^{-3/2},
\end{eqnarray}
while the definition of the coefficient $u_1$ is the following
\begin{eqnarray}
\label{u1} &&u_1=\frac{i}{4{\mathcal F}(r)}+
\left[\gamma\xi\int\!dq{\tilde{\mathcal K}(q)}\right]^2\nonumber\\
&&\times\sum_{n>1}\frac{{\tilde c}_n\vert
b_n\vert^2}{4ir^2+2r\gamma\xi{\tilde c}_n\int\!dq{\tilde{\mathcal
K}(q)}}.
\end{eqnarray}
For characteristic $r\sim r_{\text{\tiny{II}}}$ the first term in
Eq.~(\ref{u1}) is $\sim \varepsilon^{-1/2}$, as follows from
Eq.~(\ref{auxiliaryf}). On the other hand, the product
$r\xi{\tilde c}_n$ in the denominator of the second term in
Eq.~(\ref{u1}) is $\sim \varepsilon r^4/\xi^2 \sim
r^4/r_{\text{\tiny{II}}}^2$. Thus, for $r\sim
r_{\text{\tiny{II}}}$ both terms in the denominator of the sum in
the second term are $\sim r_{\text{\tiny{II}}}^2$. The numerator
in the sum over $n$ is $\sim \varepsilon^{-1/2}$ for $r \sim
r_{\text{\tiny{II}}}$. Then the estimate for the second term in
Eq.~(\ref{u1}) is $\xi^2/r_{\text{\tiny{II}}}^2\varepsilon^{1/2}$,
so that the second term is smaller than the first term in
parameter $\xi^2/r_{\text{\tiny{II}}}^2\sim \varepsilon$. Next we
notice that, for  $r\sim r_{\text{\tiny{II}}}$ both $u_0$ and
$u_1$ are of the same order and are $\sim \varepsilon^{-1/2}$. On
the other hand, as seen from Eq.~(\ref{wu1u0}), the parameter $w$
for $r\sim r_{\text{\tiny{II}}}$  is small, $w \sim
\varepsilon^{1/2}$. This allows to disregard $w$ both in numerator
and denominator in Eq.~(\ref{aver2iphi}), so that the ratio of
integrals reduces to $\left(u_0/u_1\right)^{1/2}$. Using
Eq.~(\ref{auxiliaryf}), this ratio can be rewritten as
$\left[1-(2i/3)(r/r_{\text{\tiny{II}}})^2\right]^{-1/2}$.
Substituting it into Eq.~(\ref{aver2iphi}), we arrive at
Eq.~(\ref{after}) in  Section~\ref{FO}.


\section{Analysis of the integrals Eq.~(\ref{averaged})}

\label{AppendixD}
The dimensionless function ${\mathcal I}(z)$ defined by
Eq.~(\ref{averaged}) can be naturally divided into two parts
${\mathcal I}(z)={\mathcal I}_{+}+{\mathcal I}_{-}$, where
\begin{eqnarray}
\label{minus} {\mathcal
I}_{-}(z)=\int_{\rho_2>\rho_1}\frac{d\rho_2d\rho_1}{(\rho_2\rho_1)^{3/2}}\int_0^{z}dz^{\prime}
\sin \bigl[(z-z^{\prime})(\rho_2+\rho_1)\bigr]\nonumber\\
\times\sqrt{\rho_2-\rho_1}\Biggl\{ \sin\bigl[\pi
/4+(z+z^{\prime})(\rho_2-\rho_1)\bigr]\qquad\qquad\nonumber\\
\times\sqrt{\frac{1+\sqrt{1+\rho_2^2\rho_1^2(\rho_2-\rho_1)^2}}
{1+\rho_2^2\rho_1^2(\rho_2-\rho_1)^2}}\nonumber\\
+ \cos\bigl[\pi
/4+(z+z^{\prime})(\rho_2-\rho_1)\bigr]\qquad\qquad\qquad\qquad\qquad\nonumber\\
\times\sqrt{\frac{\sqrt{1+\rho_2^2\rho_1^2(\rho_2-\rho_1)^2}-1}{1+\rho_2^2\rho_1^2(\rho_2-\rho_1)^2}}
\Biggr\},\;\;\;\;
\end{eqnarray}
and
\begin{eqnarray}
\label{plus} {\mathcal I}_{+}(z)=
\int_{\rho_2>\rho_1}\frac{d\rho_2d\rho_1}{(\rho_2\rho_1)^{3/2}}\int_0^{z}dz^{\prime}
\sin \bigl[(z-z^{\prime})(\rho_2+\rho_1)\bigr]\nonumber\\
\times\sqrt {\rho_2+\rho_1} \Biggl\{ \sin\bigl[\pi
/4-(z+z^{\prime})(\rho_2+\rho_1)\bigr]\qquad\qquad\\
\times\sqrt{\frac{1+\sqrt{1+\rho_2^2\rho_1^2(\rho_2+\rho_1)^2}}
{1+\rho_2^2\rho_1^2(\rho_2+\rho_1)^2}}\nonumber\\
 + \cos\bigl[\pi /4-(z+z^{\prime})(\rho_2+\rho_1)\bigr]\qquad\qquad\qquad\qquad\qquad\nonumber\\
\times\sqrt{\frac{\sqrt{1+\rho_2^2\rho_1^2(\rho_2+\rho_1)^2}-1}{1+\rho_2^2\rho_1^2(\rho_2+\rho_1)^2}}
\Biggr\}.\;\;\;\;\nonumber
\end{eqnarray}
The complexity in numerical evaluation of ${\mathcal I}_{+}$ and
${\mathcal I}_{-}$ stems from the fact that, upon integration over
$z^{\prime}$, both integrals turn into the sums of two
contributions, each of which is {\em divergent} in the limit
$z\rightarrow 0$. Therefore, it is necessary to rewrite the result
of integration over $z^{\prime}$ in ${\mathcal I}_{+}$ and in
${\mathcal I}_{-}$ in such a way that  cancellation of the
divergent contributions is explicit.

We start with ${\mathcal I}_{-}$. Integration over $z^{\prime}$
generates the combination of three terms
\begin{eqnarray}
\label{COMBINATION}
&&\frac{\rho_1+\rho_2}{\rho_1\rho_2}\cos\left[\frac{\pi}{4}+2z(\rho_1-\rho_2)\right]\nonumber\\
&&-
\frac{1}{\rho_2}\cos\left(\frac{\pi}{4}+2z\rho_1\right)-\frac{1}{\rho_1}\cos\left(\frac{\pi}{4}-2z\rho_2\right)
\end{eqnarray}
In order to treat all these three terms on the equal footing, in
the first term of Eq.~(\ref{COMBINATION}) we introduce the
following new variables
\begin{eqnarray}
\label{zbarminus}
\tilde{z}&=&z(\rho_2-\rho_1),\nonumber\\
x&=&\frac{\rho_2\rho_1}{z^3(\rho_2-\rho_1)^2}.
\end{eqnarray}
In the second term we introduce $\tilde{z}=z\rho_1$, and finally,
in the third term, $\tilde{z}=z\rho_2$. After that, the expression
for ${\mathcal I}_{-}$ assumes the form
\begin{eqnarray}
\label{IMINUS} {\mathcal
I}_{-}(z)&=&\frac{1}{\sqrt{2}z^3}\int_0^{\infty}\frac{dx}{x^{5/2}}
\Biggl[F_1(x)-F_1(0)\Biggr]
\nonumber\\
&+&\frac{1}{\sqrt{2}z^3}\int_0^{\infty}\frac{dx}{x^{5/2}}
F_2(x,z)\\
&+&\frac{1}{\sqrt{2}z^3}\int_0^{1/4z^3}\frac{dx}{x^{5/2}}\Biggl[F_3(x,z)-F_3(0,0)\Biggr],
\nonumber
\end{eqnarray}
where the functions $F_1$, $F_2$, and $F_3$ are defined as
\begin{eqnarray}
\label{FONE}
F_1(x)&=&\int_0^{\infty}\frac{d\tilde{z}}{\tilde{z}^{5/2}}\Biggl\{
\bigl( \cos 2\tilde{z}-\sin 2\tilde{z} \bigr)
\sqrt{\frac{\sqrt{1+\tilde{z}^6x^2}-1}{1+\tilde{z}^6x^2}}
\nonumber\\
&+& \bigl(\cos 2\tilde{z}+\sin
2\tilde{z}\bigr)\sqrt{\frac{1+\sqrt{1+\tilde{z}^6x^2}}
{1+\tilde{z}^6x^2}}\Biggr\},
\end{eqnarray}
\begin{eqnarray}
\label{FTWO} F_2(x,z)=
\frac{\sqrt{\frac{1}{4}+xz^3}-\frac{1}{2}}{2\sqrt{\frac{1}{4}+xz^3}}
\\
\times\int_0^{\infty}\frac{d\tilde{z}}{\tilde{z}^{5/2}}\Biggl\{
\bigl( \sin 2\tilde{z}\!&-&\!\cos 2\tilde{z}
\bigr) \sqrt{\frac{\sqrt{1+\tilde{z}^6x^2}+1}{1+\tilde{z}^6x^2}}\nonumber\\
- \bigl(\cos 2\tilde{z}\!&+&\!\sin
2\tilde{z}\bigr)\sqrt{\frac{\sqrt{1+\tilde{z}^6x^2}-1}
{1+\tilde{z}^6x^2}}\Biggr\},\nonumber
\end{eqnarray}

\begin{eqnarray}
\label{FTHREE}
F_3(x,z)=-\frac{\left(\sqrt{\frac{1}{4}-xz^3}+\frac{1}{2}\right)^{3}+
\left(\sqrt{\frac{1}{4}-xz^3}-\frac{1}{2}\right)^{3}}{2\sqrt{\frac{1}{4}-xz ^3}}\nonumber\\
\times\int_0^{\infty}\frac{d\tilde{z}}{\tilde{z}^{5/2}}\Biggl\{
\bigl( \cos 2\tilde{z}+\sin 2\tilde{z} \bigr)
\sqrt{\frac{\sqrt{1+\tilde{z}^6x^2}+1}{1+\tilde{z}^6x^2}}\;\nonumber\\
+ \bigl(\cos 2\tilde{z}-\sin
2\tilde{z}\bigr)\sqrt{\frac{\sqrt{1+\tilde{z}^6x^2}-1}
{1+\tilde{z}^6x^2}}\Biggr\}.\;\;\;\;\;\;\;\;
\end{eqnarray}
Subtraction of $x=0$ values from $F_1(x)$ and $F_3(x,z)$ in
Eq.~(\ref{IMINUS}) insures the convergence of integrals over
$\tilde{z}$ in Eqs.~(\ref{FONE}) and (\ref{FTHREE}). On the other
hand, this subtraction shifts ${\mathcal I}_{-}$ by
$z$-independent constant.

It is seen that, in the limit $z\rightarrow 0$, the difference
$F_2(x,z)-F_2(0,z)$ behaves  as $z^3$, so that the contribution
from $F_2$ to ${\mathcal I}_{-}(z)$ remains finite in this limit.
On the other hand, the contributions from $F_1$ and $F_3$ both
behave as $1/z^3$. To demonstrate that the two divergent
contributions cancel out, we divide the integration domain in the
first term of ${\mathcal I}_{-}$ into the intervals $\{0,1/4z^3\}$
and $\{1/4z^3,\infty\}$. We then combine the two integrals from
$0$ to $1/4z^3$ to obtain

\begin{eqnarray}
\label{combined} &&\!\!\!\!{\mathcal
I}_{-}(z)=\frac{1}{\sqrt{2}z^3}\int_0^{1/4z^3}\frac{dx}{x^{5/2}}
\Biggl[\Bigl(F_1(x)-F_1(0)\Bigr)\nonumber\\
&&\!\!\!\!+\Bigl(F_3(x,z)-F_3(0,z)\Bigr)\Biggr]
+\frac{1}{\sqrt{2}z^3}\int_{1/4z^3}^{\infty}\frac{dx}{x^{5/2}}\nonumber\\
&&\!\!\!\!\times\Biggl[F_1(x)-F_1(0)\Biggr]+\frac{1}{\sqrt{2}z^3}
\int_0^{\infty}\frac{dx}{x^{5/2}} F_2(x,z).\nonumber\\
\end{eqnarray}
The second and the third terms in Eq.~(\ref{combined}) are
convergent in the limit $z\rightarrow 0$. The integrand in the
first term has the form
\begin{eqnarray}
&&\!\!\!F_1(x)-F_1(0)+F_3(x,z)-F_3(0,z)\\
&&\!\!\!=\Biggl\{1-\frac{\left(\sqrt{\frac{1}{4}-xz^3}+\frac{1}{2}\right)^{3}+
\left(\sqrt{\frac{1}{4}-xz^3}-\frac{1}{2}\right)^{3}}{2\sqrt{\frac{1}{4}-xz
^3}}\Biggr\}\nonumber\\
&&\!\!\!\times\int_0^{\infty}\frac{d\tilde{z}}{\tilde{z}^{5/2}}\Biggl\{
\bigl( \cos 2\tilde{z}+\sin 2\tilde{z} \bigr)
\Biggl[\sqrt{\frac{\sqrt{1+\tilde{z}^6x^2}+1}{1+\tilde{z}^6x^2}}-\sqrt{2}\Biggr]\nonumber\\
&&\!\!\!+ \bigl(\cos 2\tilde{z}-\sin
2\tilde{z}\bigr)\sqrt{\frac{\sqrt{1+\tilde{z}^6x^2}-1}
{1+\tilde{z}^6x^2}}\Biggr\}.\nonumber
\end{eqnarray}
We see that in the limit $z\rightarrow 0$ expression in the curly
brackets behaves as $\propto z^3$, and thus cancels the divergent
prefactor. Now  all three terms in Eq.~(\ref{combined}) yield a
finite contribution at $z \rightarrow 0$. Our numerical results
for ${\mathcal I}_{-}(z)$ were obtained from Eq.~(\ref{combined}).

We now turn to ${\mathcal I}_{+}(z)$. In order to deal with
small-$z$ behavior in the integral Eq.~(\ref{plus}), we introduce,
after performing  integration over $z^{\prime}$, the following new
variables
\begin{eqnarray}
\label{zbarplus}
\tilde{z}&=&z(\rho_2+\rho_1),\nonumber\\
x&=&\frac{\rho_2\rho_1}{z^3(\rho_2+\rho_1)^2}.
\end{eqnarray}
Then one obtains ${\mathcal I}_{+}={\mathcal I}_{+}^{1}+{\mathcal
I}_{+}^{2}$, where the two contributions are given by
\begin{eqnarray}
&&{\mathcal
I}_{+}^{(1)}=\frac{1}{\sqrt{2}}\int_0^{1/4z^3}\frac{dx}{x^{3/2}\sqrt{\frac{1}{4}-xz^3}}
\int_0^{\infty}\frac{d\tilde{z}}{\tilde{z}^{5/2}}\\
&&\Biggl\{
 \bigl(1-\sin
2\tilde{z}-\cos 2\tilde{z}\bigr)\sqrt{\frac{\sqrt{1+\tilde{z}^6x^2}-1} {1+\tilde{z}^6x^2}}\nonumber\\
&&-\bigl( 1-\cos 2\tilde{z}+\sin 2\tilde{z} \bigr)
\Biggl[\sqrt{\frac{\sqrt{1+\tilde{z}^6x^2}+1}{1+\tilde{z}^6x^2}}-\sqrt{2}\Biggr]\Biggr\}\nonumber
\end{eqnarray}
and
\begin{eqnarray}
&&\!\!\!\!\!\!\!{\mathcal I}_{+}^{(2)}=\frac{2}{\sqrt{2}}
\int_0^{1/4z^3}\frac{dx}{x^{3/2}\sqrt{\frac{1}{4}-xz^3}}
\int_0^{\infty}\frac{d\tilde{z}}{\tilde{z}^{3/2}}\\
&&\!\!\!\!\!\!\!\Biggl\{
 \bigl(\sin
2\tilde{z}-\cos 2\tilde{z}\bigr)\sqrt{\frac{\sqrt{1+\tilde{z}^6x^2}-1} {1+\tilde{z}^6x^2}}\nonumber\\
&&\!\!\!\!\!\!\!+\bigl(\cos 2\tilde{z}+\sin 2\tilde{z} \bigr)
\Biggl[\sqrt{\frac{\sqrt{1+\tilde{z}^6x^2}+1}{1+\tilde{z}^6x^2}}-\sqrt{2}\Biggr]\Biggr\}.\;\;\;\nonumber
\end{eqnarray}
Both these contributions are finite in the limit  $z \rightarrow
0$.

\section{Analysis of the integrals Eqs.~(\ref{formulaP+}), (\ref{formulaP-})}

\label{AppendixD1}

In the integral Eq.~(\ref{formulaP+}) we perform the following
change of variables
\begin{eqnarray}
\label{changeP+}
  \rho_1=\frac{z}{2}\left(1-\sqrt{\frac{v}{v+4}}\right),
\nonumber\\
\rho_2=\frac{z}{2}\left(1+\sqrt{\frac{v}{v+4}}\right),
\end{eqnarray}
after which it acquires the form
\begin{eqnarray}
\label{AFTER1}
 P_1^{+}(x)=
\frac{3\cdot 2^{13/6}}{\pi
^{3/2}}\int_0^{\infty}\frac{dv}{v^{1/2}}
\int_0^{\infty}\frac{dz}{z^{3/2}}\\
\Biggl\{
\cos\left[xz-\frac{\pi}{4}-\frac{z^3}{v+4}\right]-\cos\left[xz-\frac{\pi}{4}\right]
\Biggr\}.\nonumber
\end{eqnarray}

In the integral Eq.~(\ref{formulaP-}) we perform the following
change of variables
\begin{eqnarray}
\label{changeP-}
 \rho_1=\frac{z}{2}\left(1+\sqrt{\frac{v+4}{v}}\right),\nonumber\\
 \rho_2=\frac{z}{2}\left(\sqrt{\frac{v+4}{v}}-1\right),
\end{eqnarray}
after which it acquires the form
\begin{eqnarray}
\label{AFTER2}
 P_1^{-}(x)
=\frac{3\cdot 2^{13/6}}{\pi
^{3/2}}\int_0^{\infty}\frac{dv}{(v+4)^{1/2}}
\int_0^{\infty}\frac{dz}{z^{3/2}}\\
\Biggl\{
\cos\left[xz+\frac{\pi}{4}+\frac{z^3}{v}\right]-\cos\left[xz+\frac{\pi}{4}\right]
\Biggr\}.\nonumber
\end{eqnarray}
It is convenient to present $\int_0^{\infty}dv$ in
Eq.~(\ref{AFTER2}) as the following  difference of integrals
\begin{eqnarray}
\label{AFTER3}
 P_1^{-}(x)=\frac{3\cdot 2^{13/6}}{\pi ^{3/2}}\int_0^{\infty}\frac{dz}{z^{3/2}}\\
\times\Biggl(-\int_{-4}^0\frac{dv}{\sqrt{v+4}}+\int_{-4}^\infty\frac{dv}{\sqrt{v+4}}\Biggr)\nonumber\\
\times\Biggl(\cos\left[xz+\frac{\pi}{4}+\frac{z^3}{v}\right]-\cos\left[xz+\frac{\pi}{4}\right]\Biggr).\nonumber
\end{eqnarray}
We now observe that the the second term cancels {\em identically}
the function $P_{+}$. Then we readily arrive to Eq.~(\ref{last}).


\section{Asymptotics of the density of states}
\label{AppendixE}

The idea of derivation of Eq.~(\ref{oscillation}) from
Eq.~(\ref{FORM}) is that the major contribution to the integral
Eq.~(\ref{FORM}) comes from the domain $\vert \rho_2-\rho_1 \vert
\ll \rho_1,\rho_2$, i.e., from the domain where $\rho_1$ and
$\rho_2$ are close to each other. To make use of this
simplification we rewrite the argument of cosine in
Eq.~(\ref{FORM}) as
\begin{eqnarray}
\label{REWRITTEN}
\frac{(\rho_1+\rho_2)^3}{4}+\frac{\pi}{4}-\frac{2^{7/3}\omega}{\omega_h}(\rho_1+\rho_2)\nonumber\\-
\frac{(\rho_1+\rho_2)(\rho_2-\rho_1)^2}{4},\qquad
\end{eqnarray}
where we had introduced $\omega_h=\omega_0(h/h_0)^{2/3}$. It is
seen from Eq.~(\ref{REWRITTEN}) that the typical value of
$(\rho_2+\rho_1)$ is $\left(\omega/\omega_h\right)^{1/2}\gg 1$,
while the typical value of $(\rho_2-\rho_1)$ is
$(\rho_2+\rho_1)^{-1/2}\sim \left(\omega/\omega_h\right)^{-1/4}$,
i.e., the relevant difference $\rho_2-\rho_1$ is small indeed.
This allows to extend the integration over $\rho_2-\rho_1$ from
zero to infinity and perform the integral. This yields
\begin{eqnarray}
\label{OVERSUM} \left\langle\frac{\delta\nu
(\omega)}{\nu_0}\right\rangle\!=\!- \frac{
(\nu_0V)^3\omega\omega_h^{1/2}}{\pi^{1/2}E_{\text{\tiny{F}}}^{3/2}}
\int_0^{\infty}\!\frac{d\rho}{\rho^3}\nonumber\\
\times\left\langle\cos\left[\frac{\rho^3}{4}-2^{7/3}\rho\frac{\omega}{\omega_h}
\right]\right\rangle.
\end{eqnarray}
The argument of cosine in Eq.~(\ref{OVERSUM}) has a sharp minimum
at $\rho
=\rho_0=\left(2^{13/6}/3^{1/2}\right)\sqrt{\omega/\omega_h}$,
which allows to perform the integration over $\rho$ by introducing
$\delta\rho=\rho-\rho_0$ and extending the integration over
$\delta\rho$ from minus to plus infinity. This yields the
following asymptote of $\delta\nu(\omega)$
\begin{eqnarray}
\label{OVERRHO} \left\langle\frac{\delta\nu
(\omega)}{\nu_0}\right\rangle\!=\!-\frac{1}{64\cdot
2^{7/12}\sqrt{\pi}}
\frac{(\nu_0V)^3\omega_h^{9/4}}{E_{\text{\tiny{F}}}^{3/2}\omega^{3/4}}\nonumber\\
\times
\left\langle\sin\left[\frac{32\sqrt{2}}{3\sqrt{3}}\left(\frac{\omega}{\omega_h}\right)^{3/2}+
\frac{\pi}{4}\right]\right\rangle_{h(x,y)},
\end{eqnarray}
in which the random magnetic field enters through $\omega_h$. The
argument of sine contains the term $\propto\omega_h^{-3/2}$, which
can be presented as $sh_0/h$, where the constant $s$ is equal to
$16(2\omega/3\omega_0)^{3/2}$. The factor in front of sine
contains $\omega_h^{9/4}\propto h^{3/2}$. Then the gaussian
averaging over $h$ can be carried out analytically using the fact
that $s\gg 1$. This yields
\begin{eqnarray}
\label{LASTSTEP} \left\langle h^{3/2}
\sin\left[\frac{sh_0}{h}+\frac{\pi}{4}\right]\right\rangle_{h(x,y)}=
\frac{h_0^{3/2}s^{1/2}}{\sqrt{6}}\sin\Bigl(\frac{3^{3/2}s^{2/3}}{2^{5/3}}\Bigr)\nonumber\\
\times\exp\left\{-\frac{3s^{2/3}}{2^{5/3}}\right\}.\qquad
\end{eqnarray}
Combining Eqs.~(\ref{LASTSTEP}) and (\ref{OVERRHO}), one
reproduces Eq.~(\ref{oscillation}) of the main text.

\end{document}